\RequirePackage{lineno} 

\documentclass[twocolumn,showpacs,preprintnumbers,amsmath,amssymb,nofootinbib]{revtex4}

\usepackage{graphicx}
\usepackage{dcolumn}
\usepackage{bm}
 
\usepackage{epstopdf}

\def\bea{\begin{eqnarray}}
\def\eea{\end{eqnarray}}

\def\pp{\mbox{$p$-$p$}}

\def\pa{\mbox{$p$-A}}

\def\auau{\mbox{Au-Au}}

\def\pbpb{\mbox{Pb-Pb}}
\def\ppb{\mbox{$p$-Pb}}
\def\pau{\mbox{$p$-Au}}
\def\dau{\mbox{$d$-Au}}
\def\hau{\mbox{$h$-Au}}

\def\aa{\mbox{A-A}}
\def\nn{\mbox{N-N}}

\def\ppbar{\mbox{$p$-$\bar p$}}

\def\pt{$p_t$}
\def\mt{$m_t$}
\def\yt{$y_t$}

\def\nch{$n_{ch}$}
\def\mmpt{$\bar p_t$}

\begin{document} 

\setlength{\pdfpagewidth}{8.5in}
\setlength{\pdfpageheight}{11in}

\setpagewiselinenumbers
\modulolinenumbers[5]

\preprint{version 2.6}

\title{Questioning quark-gluon plasma formation in small collision systems
}

\author{Thomas A.\ Trainor}\affiliation{CENPA 354290, University of Washington, Seattle, WA 98195}


\date{\today}

\begin{abstract}

A recent letter published in the journal Nature reports observation at the relativistic heavy ion collider (RHIC) of quark-gluon plasma (QGP) formation in small asymmetric collision systems denoted as $p$-Au, $d$-Au and $^3$He-Au. The claimed phenomenon, described as ``short-lived QGP droplets,'' is inferred from a combination of Glauber Monte Carlo simulations, measurement of two azimuth Fourier amplitudes $v_2$ and  $v_3$ and hydro theory calculations. While that claim follows a trend in recent years to report ``signals'' conventionally attributed to QGP as appearing also in smaller collision systems, the new result remains surprising in the context of expectations before first RHIC operation that small systems, e.g.\ $d$-Au collisions, would provide {\em control experiments} in which a QGP was unlikely to appear. An alternative interpretation of the recent RHIC result is that small-system control experiments do convey an important message: The ``signals'' conventionally attributed to QGP formation in larger A-A collisions do not actually represent that phenomenon in any system. The present study reviews a broad array of experimental evidence for or against QGP formation in several collision systems. It examines in particular hydro theory descriptions of spectra and correlations usually interpreted to support QGP formation. Available evidence suggests that data features conventionally attributed to QGP formation represent either minimum-bias jet production or a nonjet azimuth quadrupole with properties inconsistent with a hydro hypothesis.

\end{abstract}


\maketitle

 \section{Introduction}

A recent publication in Nature~\cite{nature} presents  a claim that quark-gluon plasma (QGP) ``droplets'' have been created in small asymmetric collision systems \pau, \dau\ and \hau%
\footnote{The expression $^3$He + Au appearing in Ref.~\cite{nature} includes the symbol for the neutral atom $^3$He whereas the symbol $h$ in this text refers to the bare {\em helion} nucleus that is actually accelerated.
} at the relativistic heavy ion collider (RHIC). The claim is based on certain assumptions: (a) that a strongly-coupled sQGP is formed in more-central \auau\ collisions, (b) that the main evidence for sQGP in \auau\ is description of measured azimuthal asymmetries (e.g.\ $v_2$ and $v_3$ data) by viscous-hydro theory assuming a very low fluid viscosity, (c) that similar azimuthal asymmetries have been observed recently in \dau\ collisions at the RHIC and \ppb\ collisions at the large hadron collider (LHC) and  (d) that recent RHIC results from \pau, \dau\ and \hau\ collisions include  $v_2$ and $v_3$ data that correspond to certain viscous-hydro model descriptions derived from a Glauber model of initial collision geometry.

The claim can be challenged on the basis of several issues considered in the present article: 
(i)  $v_2$ and $v_3$ data represent a small fraction of the total information carried by particle data, and their interpretation in terms of flows is questionable -- other data characteristics conflict with a flow interpretation. 
(ii) The conventional Monte Carlo (MC) Glauber model of A-B collision geometry or initial conditions conflicts with LHC \ppb\ data. 
(iii) In some instances hydro models have been applied to data features subsequently associated with minimum-bias (MB) dijets. Thus, claims of QGP formation in small systems should be confronted by a more complete data representation.

The two-component (soft + hard) model (TCM) of hadron production near midrapidity has been successfully applied to a broad array of particle data from several A-B collision systems over a range of collision energies from SPS (17 GeV) to LHC (13 TeV) within their published uncertainties~\cite{alicetomspec}. The TCM is applied to yield, \pt\ spectrum and two-particle correlation data. 2D angular correlations do require a significant third component -- a {\em nonjet} (NJ) azimuth quadrupole~\cite{v2ptb,njquad} (acronym NJ will be applied only to the term ``quadrupole''). However, systematics of the NJ quadrupole in \auau~\cite{quadspec,davidhq,davidhq2,anomalous,v2ptb}, \pbpb~\cite{multipoles,sextupole,njquad} and \pp~\cite{ppquad} collisions are inconsistent with hydro expectations~\cite{quadspec,njquad}. The TCM and related differential techniques enable accurate distinctions between jet-related and nonjet data features. The TCM provides a reference against which QGP claims may be tested.

The TCM soft component is associated with data manifestations in spectra and angular correlations representing longitudinal projectile-nucleon fragmentation. Similarly, the hard component is associated with transverse fragmentation of large-angle scattered partons (low-$x$ gluons) to MB dijets. Those processes are observed to dominate hadron production near midrapidity in all A-B collision systems and should provide a context for claims of QGP production in small (or any) collision systems.

For example, quantities $v_2$ and $v_3$ are Fourier amplitudes inferred from 1D azimuth correlations to which jet structure must make substantial contributions. Interpretation of Fourier amplitudes as representing flows must compete with alternative interpretation in terms of MB dijets (referred to as ``nonflow''). The distinction depends critically on the choice of analysis method. If interpreted as flows,  $v_2$ and $v_3$ should represent {\em azimuthal modulation} of transverse or radial flow. Evidence for radial flow should then be demonstrated by {\em differential} \pt\ spectrum analysis.
Arguing by analogy with reported results for \aa\ collisions, if a flowing dense medium (QGP) plays a significant role then jet formation should be altered measurably (``jet quenching'') as indicated by changes to single-particle \pt\ spectra. Evolution of jet modification with collision centrality or charge density $\bar \rho_0$ should correspond to evolution of $v_2$ and $v_3$ interpreted as flows.

More generally, any proposed analysis method should establish a clear, comprehensive and accurate distinction between MB dijet contributions to yields, spectra and two-particle correlations~\cite{mbdijets} and nonjet contributions, some of which might be related to flows. No data manifestation should be assumed {\em a priori} to represent flows. For instance, a valid \pt\ spectrum model should identify the full MB dijet contribution over the entire \pt\ acceptance, not just the ``high-\pt'' contribution above some imposed \pt\ threshold. Ratio measures that by definition discard critical information should be abandoned in favor of differential modeling of isolated spectra. 2D angular correlations should be modeled intact rather than as 1D projections that also discard critical information, and the MB dijet contribution should be accurately isolated before attempting to identify nonjet (possibly flow) contributions within the remainder.

The present study compares the arguments presented in Ref.~\cite{nature} for QGP droplets in  small collision systems, based on certain critical assumptions and a limited set of data, with evidence from several collision systems and energies in the broader context of the TCM. Topics include validity of claims for perfect liquid (sQGP) formation in \aa\ collisions at the RHIC and LHC, applicability of the MC Glauber model to small asymmetric collision systems, relevance of hydro models to any high-energy A-B collisions, comparison of $v_n$ measurement methods and flow interpretations, and correspondence (or lack thereof) between jet modifications attributed to a dense medium and radial flow as inferred (or not) from \pt\ spectrum structure. This study concludes that formation of QGP droplets in small collision systems is unlikely based on comparisons of data trends from several collision systems. 

This article is arranged as follows:
Section~\ref{droplets} summarizes arguments supporting claims of QGP droplet formation in small $x$-A collision systems.
Section~\ref{perfect} considers evidence for and against so-called perfect liquid formation in \aa\ collisions.
Section~\ref{glauber} reviews application of the Monte Carlo Glauber model to \aa, \pa\ and $x$-Au collision systems.
Section~\ref{vnmeasures} summarizes methods and results for inference of centrality dependence of azimuth multipoles as measured by Fourier amplitudes $v_n(b)$.
Section~\ref{quadrupole} reviews results for measurements of the \pt\ dependence of quadrupole $v_2(p_t)$.
Section~\ref{radial} summarizes evidence for and against radial flow as inferred from single-particle hadron spectra.
Section~\ref{hydro} considers hydro theory models of high-energy collisions, supporting assumptions and their role as fit models.
Section~\ref{predict} reviews specific examples of hydro models as applied to data, especially whether models respond to nonjet or jet-related data features. Section~\ref{summ} presents a summary.

\section{$\bf s$QGP droplets in small systems} \label{droplets}

This section reviews  assumptions, arguments and experimental data introduced in Ref.~\cite{nature} to support a claim of QGP formation in small collision systems at the RHIC. The main subject of this study is asymmetric small collision systems. It is convenient to describe the smaller partner as the projectile and the larger partner as the target in reference to earlier fixed-target experiments.

\subsection{Underlying assumptions} \label{assumptions}

Several assumptions are invoked in Ref.~\cite{nature} explicitly or implicitly as follows:  
(a) Claims for sQGP formation in \aa\ collisions as in Ref.~\cite{perfect} are valid. 
(b) The Glauber model (based on the eikonal approximation) correctly estimates an initial-state geometry of participant nucleons N or \nn\ binary collisions that relates directly to energy densities and pressure gradients within a locally-thermalized medium. 
(c) Fourier amplitudes $v_n$ as conventionally defined actually measure final-state flows as opposed to some other phenomenon (e.g.\ jets). 
(d) Hydrodynamic models assuming a low-viscosity dense medium, including ``plasma droplets'' in small systems, have some relation to actual collision dynamics. 
(e) An alternative theory (aside from color flux tubes~\cite{fluxtubes}) or additional information derived from particle data cannot better explain the $v_n(p_t)$ data. 
The present study examines each of those assumptions in the context of the TCM and an assortment of particle data and analysis methods.

\subsection{Arguments and experimental results}

The conclusion that a near-perfect (``inviscid'') fluid is formed in \aa\ collisions at the RHIC is based primarily on the apparent ability of viscous-hydro models to describe Fourier coefficients $v_n(p_t)$ in the expression for single-particle density (modulo a constant offset) $\bar \rho_0(p_t,\phi) \approx \sum_{n=1}^\infty 2 V_n(p_t) \cos[n(\phi - \psi_n)]$, with $v_n = V_n / V_0$ and $\psi_n$ the event-plane angle, and with $n = 2,~3$ receiving primary attention. White papers from four RHIC collaborations published in 2005 are cited in support. Properties of the $v_n(p_t)$ in concert with successful hydro descriptions are  interpreted to signal formation of a sQGP in any \aa\ collision system. Observation of similar $v_n(p_t)$ trends in \pp~\cite{ppridge}, \ppb~\cite{ppbridge} and (via updated analysis) \dau~\cite{dauridge} collisions then raises the question: does sQGP appear in {\em any} high-energy collision~\cite{dusling}?

Since identification of sQGP in \aa\ collisions relies on viscous-hydro model descriptions of $v_n$ data, and hydro models effectively map initial-state collision geometries to final-state momentum distributions, Ref.~\cite{nature} describes a strategy by which $v_n$ data from three asymmetric collision systems with nominally different initial-state azimuth distributions are compared to relevant hydro predictions. If systematic differences among $v_n$ data for three systems correspond to hydro predictions derived from specific initial-state geometries sQGP formation in the small systems is considered likely, otherwise not.

A-B initial-state transverse geometry is measured by spatial eccentricities $\epsilon_n$ (representing here mean values for simplicity) estimated by a Glauber Monte Carlo. It is assumed that $\epsilon_2$ is a measure of {\em ellipticity} corresponding to elliptic flow as measured by $v_2$, and $\epsilon_3$ is a measure of {\em triangularity} corresponding to triangular flow as measured by $v_3$ according to Ref.~\cite{alver}. While $\epsilon_n$ variation in \aa\ collisions is dominated by centrality (i.e.\ impact parameter $b$) it is assumed in Ref.~\cite{nature} that {\em projectile shape} has a dominant influence in asymmetric $x$-Au collisions, with deuterons $d$ emphasizing $\epsilon_2$ (quadrupole) and helions $h$ emphasizing $\epsilon_3$ (sextupole) referring to cylindrical multipoles with pole number $2n$. $\epsilon_n$ estimates for three collision systems are presented in Fig.~1 (a) of Ref.~\cite{nature}.

The Glauber Monte Carlo model estimates the distribution of participant nucleons and \nn\ binary collisions on the polar coordinate system $(r,\phi)$ in the plane perpendicular to the beams. Certain assumptions are required to relate a simulated participant or binary-collision distribution to an energy density or temperature distribution from which a flow field may be estimated via a hydro theory model. Examples are provide in Fig.~1 (b) of Ref.~\cite{nature}.

The principal experimental result is $v_n(p_t)$ data for three collision systems with similar shapes on \pt\ but different overall amplitudes. $v_n\{\text{EP}\}(p_t) \approx v_n\{2\}(p_t)$ data%
\footnote{The ``method'' in $v_n\{\text{method}\}$ refers to the specific analysis method as summarized in Sec.~\ref{vnmeasures}.}
 for three 0-5\% central 200 GeV $x$-Au collision systems are shown in Fig.~2 of Ref.~\cite{nature}. For each particle pair, and for $v_2$ data, one particle is selected from central detectors with $|\eta| < 0.35$ and the second particle is selected within $\eta \in [-3,-1]$ for $p$ and $d$ projectiles and within $\eta \in [-3.9,-3.1]$ for $h$ projectiles (i.e.\ within the Au hemisphere in either case).%
\footnote{Text in Ref.~\cite{nature} refers to {\em event-plane} estimation within e.g.\ $\eta \in [-3,-1]$, but because $v_n\{\text{EP}\}(p_t) \approx v_n\{2\}(p_t)$~\cite{njquad} the statement in the text above is more relevant to the issues for this study.
} For $v_3$ data $\eta \in [-3.9,-3.1]$ is used for all projectiles. The resulting $v_n$ data are Fourier amplitudes inferred from hadron-pair azimuth distributions. Reference~\cite{nature} reports that overall amplitudes of $v_n(p_t)$ trends follow the ordering of MC Glauber $\epsilon_n$, suggesting that initial-state (IS) geometry does control final-state (FS) azimuth asymmetries as expected for hydro evolution, consistent with QGP formation. 

The  $v_n(p_t)$ data are in turn compared to hydro theory in  Fig.~3 of Ref.~\cite{nature}. Variation of hydro theory curves is observed to be similar to data, but the theory curves must by definition correspond to initial-state $\epsilon_n$ estimates. It is concluded that 
``The simultaneous constraints of $v_2$ and $v_3$ in p/d/$^3$He+Au collisions definitively demonstrate that the $v_n$ coefficients are correlated with the initial geometry. ... Hydrodynamical models, which include the formation of a short-lived QGP droplet, provide the best simultaneous description of these measurements.'' 

\subsection{Initial responding comments}

As described in the present study, alternative analysis methods reveal phenomena and data trends that  conflict with the arguments and conclusions reported in Ref.~\cite{nature}. It is notable that in Ref.~\cite{nature} and related publications the role of MB dijets in high-energy  collisions is not explicitly acknowledged. Jet production is referred to only indirectly as one possible source of ``nonflow,'' and jet-related features in collision data are not explicitly identified. The kinematic scope of jet fragmentation is assumed to be relatively small, and almost all aspects of high-energy nuclear collisions are assumed to be flow related.

Descriptions of selected data features by hydro theory are seen as confirming the dominant role of flows in \aa\ collisions: ``A multitude  of measurements of the Fourier coefficients, utilizing a variety of techniques, have been well described by hydrodynamical models, thereby establishing the fluid nature of the QGP in large-ion collisions''~\cite{nature}.  But hydro theory is a complex system with multiple components, versions and parametrizations that are {\em selected by comparisons with data}. An evolving theory system is in effect matched to an evolving system of analysis methods until a best fit is achieved. The theory is then not predictive and cannot be falsified.

Acknowledged data features and associated analysis methods are specifically selected for compatibility with a hydro description. Only limited intervals of \pt\ spectra are considered, and specifically for inference of radial flow. Instead of considering the full information conveyed by 2D angular correlations only 1D projections onto azimuth (``azimuthal asymmetries,'' ``momentum anisotropies'') are considered, and then only modeled by Fourier series. The Fourier terms are by assumption interpreted as flows, whereas alternative data models are more efficient~\cite{tombayes} and lead to different physical interpretations~\cite{njquad}.  The remainder of this article considers in detail a number of such issues related to the claims in Ref.~\cite{nature}.

\section{Is a perfect liquid formed in A-A?} \label{perfect}

This section responds to assumption (a) as noted in Sec.~\ref{assumptions}: Claims for sQGP formation in \aa\ collisions as in Ref.~\cite{perfect} are valid. Claims of perfect-liquid formation in \aa\ collisions are largely based on Fourier decomposition of the projection onto 1D azimuth $\phi$ of 2D angular correlations, thus greatly reducing information in correlation data actually utilized and ignoring \pt\ spectrum data and evidence therein (or not) for radial flow. 

The assumption in Ref.~\cite{nature} that a QCD perfect liquid is formed in more-central RHIC \auau\ collisions is justified by evidence presented in white papers by the four RHIC collaborations~\cite{whitebrahms,whitephob,whitestar,whitephen} as summarized in Ref.~\cite{perfect}, where the term ``perfect liquid'' is first used to describe a sQGP. One basis for such a claim would be evidence that a locally-thermalized flowing dense medium is described by a QGP equation of state (EoS). Of central importance is measurement of elliptic flow in the form $v_2(p_t)$ for identified hadrons.  Measured $v_2(p_t)$ distributions for central \auau\ collisions are said to be compatible with nonviscous ideal hydrodynamics below $p_t \approx 2$ GeV/c, leading to inference of low-viscosity sQGP or perfect liquid. A QGP EoS is said to be confirmed by the extent of {\em mass ordering} of $v_2(p_t)$ for identified hadrons below 2 GeV/c.

Another basis for inferring formation of a dense QGP medium is  strong suppression of hadron \pt\ spectra above 4 GeV/c in more-central \auau\ collisions. Since jet fragments are expected to dominate that $p_t$ interval the suppression effect is referred to as ``jet quenching.''  The principal mechanism is thought to be parton energy loss within a dense colored medium via gluon bremsstrahlung, associated with the sQGP inferred from $v_2(p_t)$ data.

Those assumptions and results can be questioned as follows: (i) Are $v_2(p_t)$ data for identified hadrons actually consistent with a flowing dense QCD medium as the main source for final-state hadrons? (ii) Are \pt\ spectrum modifications consistent with parton energy loss for higher-energy partons and full absorption for lower-energy partons? (iii) Are $v_2(p_t)$ data trends, interpreted as indicators for a flowing dense medium, actually correlated with \pt-spectrum modification trends interpreted as indicators for parton energy loss in the same medium?

Replies to such questions are offered in a recent RHIC program review~\cite{rhicreview}. Results differ markedly depending on preferred analysis methods as discussed in Sec.\ II of Ref.~\cite{rhicreview}. Certain preferred methods, selected from a range of possibilities, tend to support a sought-after physical mechanism whereas alternative methods may yield contradicting results as demonstrated in Sec.\ X of Ref.~\cite{rhicreview}. Examples are provided in following subsections.

\subsection{PID $\bf v_2(p_t)$ data and interpretations} \label{interpretations}

\begin{figure}[h]
	\includegraphics[width=1.65in,height=1.68in]{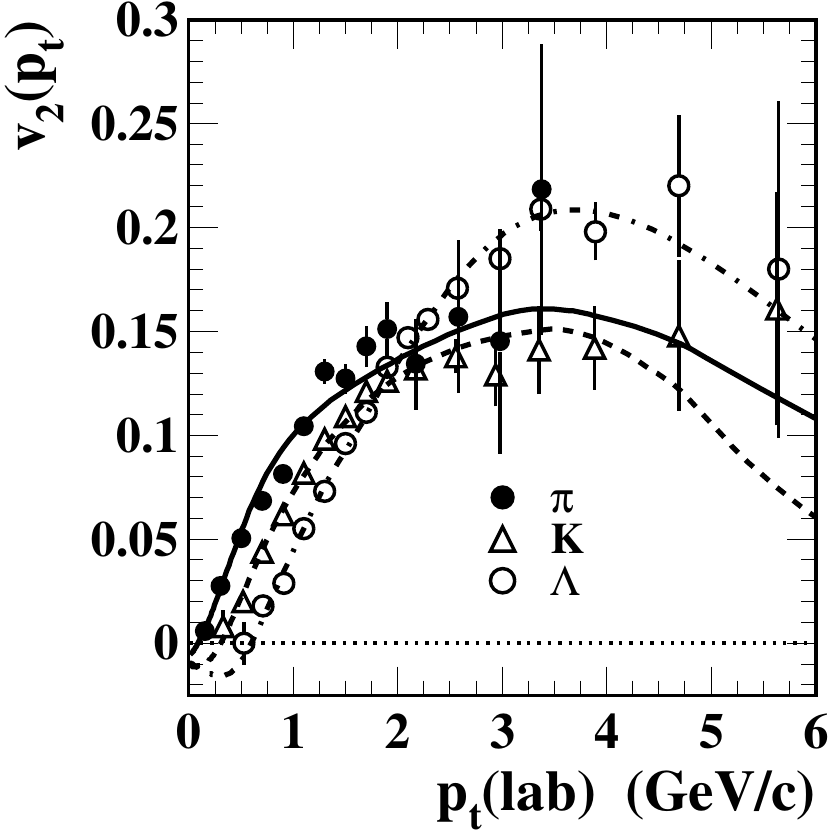}
	\includegraphics[width=1.65in,height=1.65in]{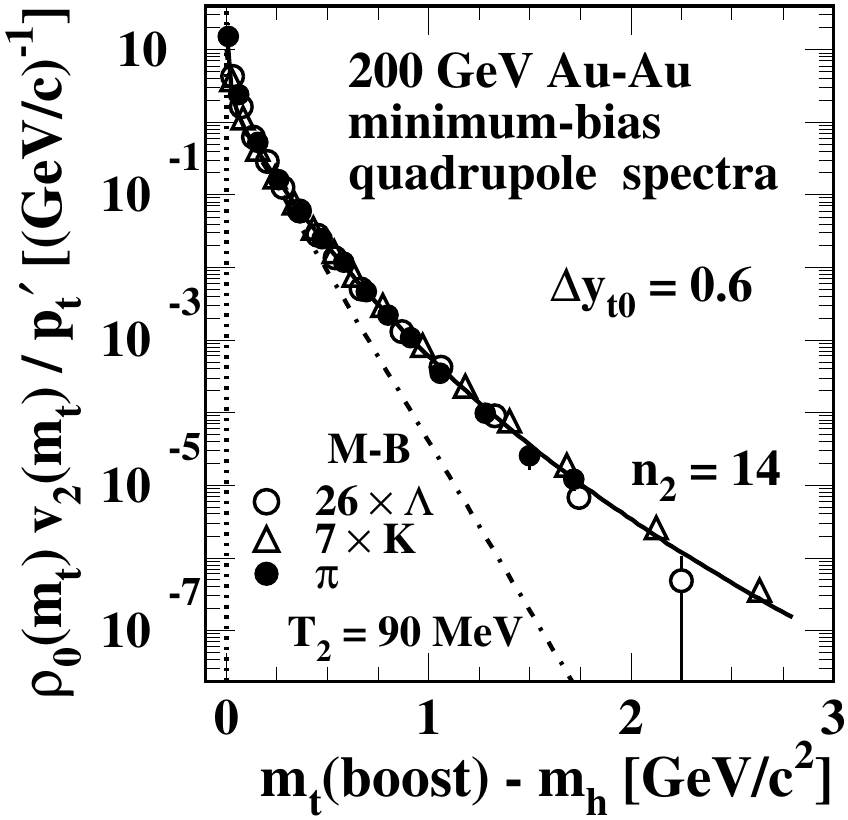}
	\caption{\label{quad3}
		Left: Published PID $v_2(p_t)$ data for three hadron species from Refs.~\cite{v2pions,v2strange} plotted in the conventional $v_2(p_t)$ vs \pt\ format. The curves are transformed from a common quadrupole spectrum described in the text.
		Right: Data from the left panel transformed to the boost frame (shifted left on $y_{th}$ by $\Delta y_{t0}$), transformed to $m_{th} - m_h$ and scaled by expected hadron abundances (statistical model) to reveal a common {\em quadrupole spectrum} in the boost frame (solid curve)~\cite{quadspec}. $p_t'$ represents $p_t(\text{boost})$ or \pt\ in the boost frame.
	} 
\end{figure}

Figure~\ref{quad3} responds to question (i) above and the relation of $v_2(p_t)$ to a flowing dense medium or sQGP. The left panel shows published $v_2(p_t)$ data for minimum-bias 200 GeV \auau\ collisions and identified-hadron (PID) $\pi$, K and $\Lambda$ (points) plotted in the conventional $v_2(p_t)$ vs $p_t$ format~\cite{v2pions,v2strange,quadspec}. The three curves through data are obtained by transformations of a single universal {\em quadrupole spectrum model} described below and shown in the right panel (solid curve). Note that data in the left panel are consistent with model zero crossings below 1 GeV/c. The mass trend for different hadron species below 2 GeV/c (described as {\em mass ordering}) is said to confirm a hydrodynamic interpretation of $v_2$ data as representing elliptic flow. These $v_2(p_t)$ data correspond to a minimum-bias data sample averaged over \auau\ centrality. Although the data were obtained early in the RHIC program they remain illustrative. Analysis of more recent $v_2(p_t,b)$ data~\cite{v2ptb} is consistent with these early results.

Figure~\ref{quad3} (right) shows the result of a sequence of transformations of the $v_2(p_t)$ data on the left as described in Ref.~\cite{quadspec}. The first step is based on definition of $v_2(p_t)$ as a {\em ratio}~\cite{galehydro2} in which the denominator is the single-particle \pt\ spectrum $\bar \rho_0(p_t)$ (from which jet quenching is inferred)
\bea
v_2(p_t) &\equiv& \frac{\int d\phi f(p_t,\phi) \cos[n(\phi - \psi_2)]}{\int d\phi f(p_t,\phi) \equiv \bar \rho_0(p_t)},
\eea
where $\psi_2$ is the event-plane angle.
The numerator contains information specifically relevant to a quadrupole correlation feature which may or may not relate to flows.

The sequence of transformations is as follows:
(a) For each hadron species $h$ multiply $v_2(p_t)$ data by the corresponding single-particle spectrum $\bar \rho_0(p_t)$ and divide by $p_t(\text{lab})$ (\pt\ in the lab frame).
(b) Replot those data on transverse rapidity $y_{th} \equiv \ln[(p_t + m_{th})/m_h]$ where $m_h$ is the proper hadron mass and $m_{th}^2 = p_t^2 + m_h^2$.  Plotted in that format the data reveal a {\em common fixed source boost} $\Delta y_{t0} \approx 0.6$ (common zero intercept) for all hadron species.
(c) Transform from lab frame to boost frame simply by shifting all $y_{th}$ spectra to the left by the single inferred source boost $\Delta y_{t0}$. 
(d) For each hadron species transform spectra now in the boost frame from $y_{th}$ to $m_{th}$.
(e) Multiply spectra by ratio  $p_t(\text{lab})/p_t(\text{boost})$ determined exactly by $\Delta y_{t0}$.
(f) Rescale the \mt\ spectra by constant factors (1,7,26) as indicated in the right panel, consistent with statistical-model values for respective hadron species abundances at $T_{chem} \approx 150$ MeV. 

All quadrupole spectra for three hadron species are then quantitatively described by a single L\'evy distribution (solid curve) within published data uncertainties. The $v_2(p_t)$ data are described up to 5-6 GeV/c. The quadrupole spectra are cold ($T_2 \approx 90$ MeV) and do not correspond to single-particle spectra $\bar \rho_0(y_t)$ representing most hadrons ($T_0 \approx 145$ MeV). The dash-dotted curve is an exponential with $T_2 \approx 90$ MeV as determined by data below 0.5 GeV/$c^2$ demonstrating the large difference from a L\'evy distribution with $n_2 \approx 14$ (solid).

While the data in Fig.~\ref{quad3} represent a minimum-bias average over \auau\ centrality subsequent analysis has established that quadrupole source boost $\Delta y_{t0}$ is independent of \auau\ centrality to the uncertainty limits of those data~\cite{davidhq2,v2ptb} (see Sec.~\ref{quadspecc}). These results imply that from all 200 GeV \auau\ $v_2(p_t,b)$ data for identified hadrons two numbers are obtained: (a) NJ quadrupole {\em amplitude} $\Delta y_{t2}(b,\sqrt{s})$ depending on centrality and energy as reported in Refs.~\cite{quadspec,davidhq2} and (b) {\em fixed} quadrupole source boost $\Delta y_{t0}$ common to all collision centralities. The quadrupole spectrum inferred from these \auau\ data appears to be universal for all collision conditions. The solid curve in Fig.~\ref{quad3} (right) predicts all $v_2(p_t,b)$ data for any hadron species as in the left panel. 

Of major importance is the implication of a fixed source boost $\Delta y_{t0} \approx 0.6$. A Hubble-expanding dense medium should be represented, at least in more-central \aa\ collisions, by a {\em broad} boost distribution leading to a more complex $v_2(p_t)$ data configuration at low \pt, and the boost distribution should depend strongly on collision centrality. Also, the quadrupole spectrum shape should correspond to that for almost all hadrons emerging from the flowing dense medium [i.e.\ single-particle spectrum $\bar \rho_0(p_t)$]. Neither condition is met by these $v_2(p_t)$ data, casting doubt on claims for a dense medium or sQGP.


\subsection{$\bf p_t$ spectrum modifications and jet quenching}

\begin{figure}[h]
	\includegraphics[width=1.65in,height=1.65in]{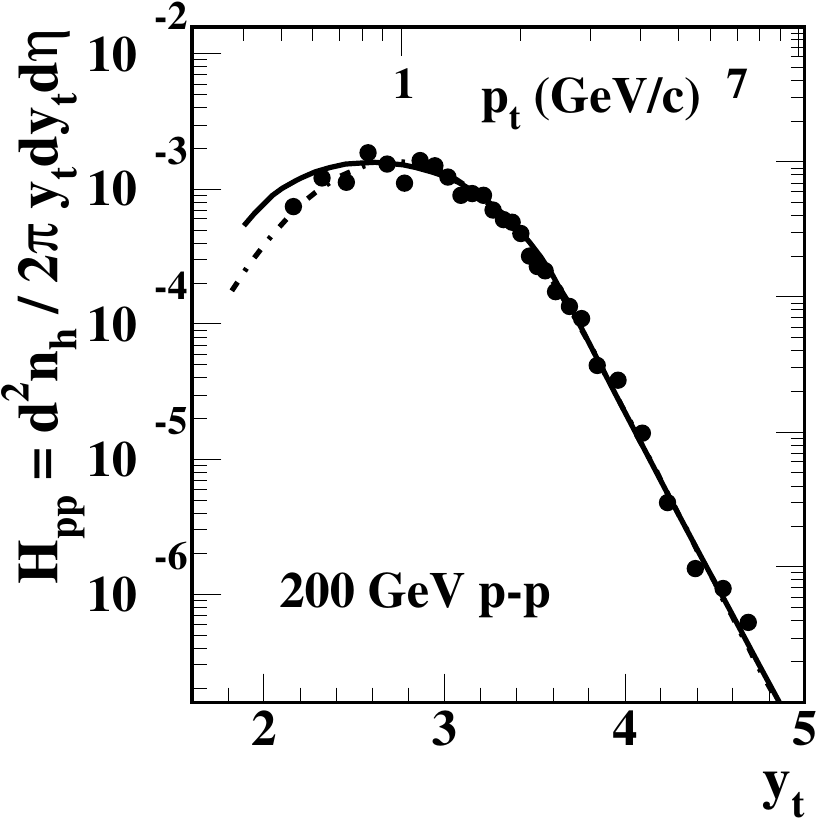}
	\includegraphics[width=1.65in,height=1.65in]{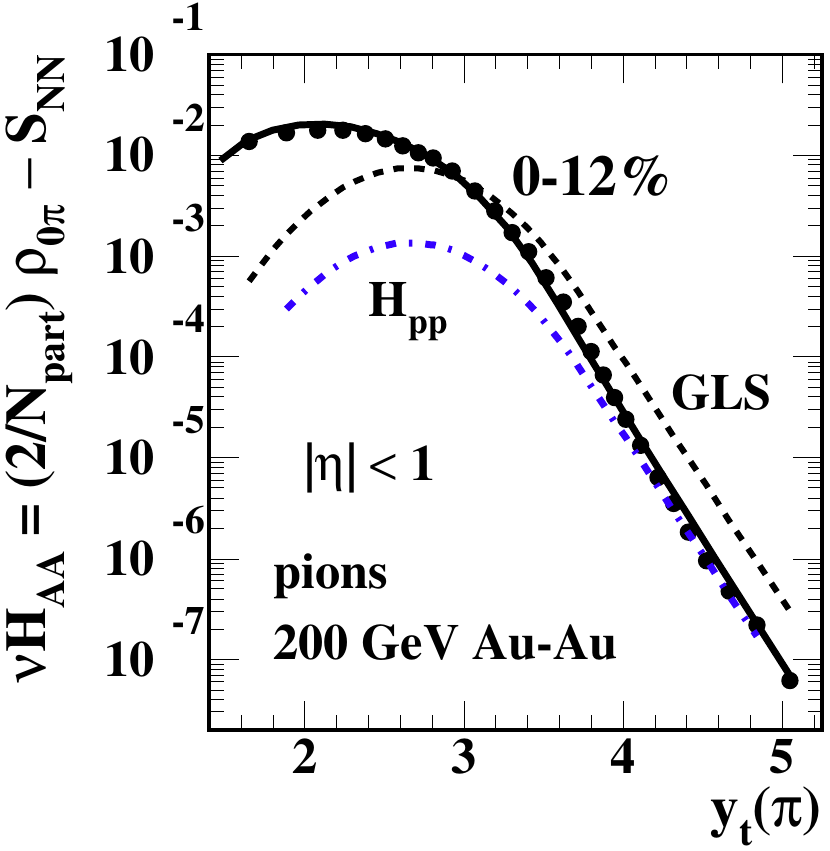}
	\caption{\label{specfrag}
		Left:  Spectrum hard component representing 200 GeV NSD \pp\ collisions (points)~\cite{ppprd,ppquad}. The solid curve is a pQCD prediction for the corresponding fragment distribution derived from measured fragmentation functions and 200 GeV \pp\ jet energy spectrum~\cite{fragevo}.
		Right: Spectrum hard component for 0-12\% central 200 GeV \auau\ collisions (points)~\cite{hardspec}. The solid curve is a pQCD prediction based on a simple modification of fragmentation functions~\cite{fragevo}. The dashed curve is a TCM prediction (GLS) extrapolated from \pp\ data.
	} 
\end{figure}

Question (ii) above deals with parton energy loss vs \pt\ spectrum modification. Spectrum modification attributed to jet quenching is conventionally determined via spectrum ratio $R_{AA}$ relating a \pt\ spectrum for central \aa\ to a spectrum for \pp\ collisions. But ratio $R_{AA}$ of {\em full} (soft + hard) \pt\ spectra conceals the jet contribution below 4 GeV/c,  whereas the mode of the MB jet fragment distribution is near 1 GeV/c [see Fig.~\ref{specfrag} (left)]. Thus, $R_{AA}$ {\em conceals almost all jet information} whereas direct comparison of isolated spectrum hard components conveys all available information as demonstrated below.

Figure~\ref{specfrag} (left) shows a non-single-diffractive (NSD) average of 200 GeV \pp\ \pt\ spectrum hard components (points) from  Ref.~\cite{ppquad}. The solid curve is a prediction~\cite{fragevo} for the \pp\ spectrum hard component based on a measured MB jet energy spectrum (bounded below at 3 GeV and integrating to 2.5 mb)~\cite{jetspec2} and measured \ppbar\ fragmentation functions (FFs)~\cite{fragevo}. The dash-dotted curve  is a Gaussian plus exponential tail from Ref.~\cite{ppprd}. 

Figure~\ref{specfrag} (right) shows the spectrum hard component inferred from 0-12\% central 200 GeV \auau\ collisions (points)~\cite{hardspec}. The dash-dotted curve is the \pp\ Gaussian with exponential tail $H_{pp}$ from the left panel. The dashed curve is a Glauber linear-superposition (GLS) prediction $\nu H_{pp}$ (with $\nu \equiv 2N_{bin} / N_{part} \approx 6$) for  central \auau\ (no jet quenching). 
The solid curve is a pQCD description for central \auau\ $\nu H_{AA}$ data based on a {\em single modification} of measured FFs (single-parameter change in a gluon splitting function compared to \pp\ FFs) and no change in the underlying jet spectrum (no jets are lost to a dense medium)~\cite{fragevo}. Fragment reduction at larger $y_t$ as revealed by $R_{AA}$ is balanced by much larger fragment enhancement at smaller $y_t$ that approximately {\em conserves the parton energy within resolved jets} but is concealed by $R_{AA}$. It is notable that the spectrum soft component $S_{NN} \rightarrow S_{pp}$ is held fixed for all centralities, consistent with {\em no radial flow} in \auau\ collisions~\cite{hardspec}.

\subsection{Relation between $\bf v_2$ data and jet modification}

Question (iii) above deals with the existence (or not) of a flowing dense medium common to both $v_2(p_t)$ data trends at lower \pt\ (interpret as demonstrating flows and local thermalization  early in collisions) and spectrum modification per ratio $R_{AA}$ (sensitive to jet modification only at higher \pt). A key issue is the centrality dependence of each phenomenon. Do reported elliptic-flow trends and jet-quenching trends correspond, as should be expected if a dense medium is the common element?

\begin{figure}[h]
	\includegraphics[width=1.65in,height=1.65in]{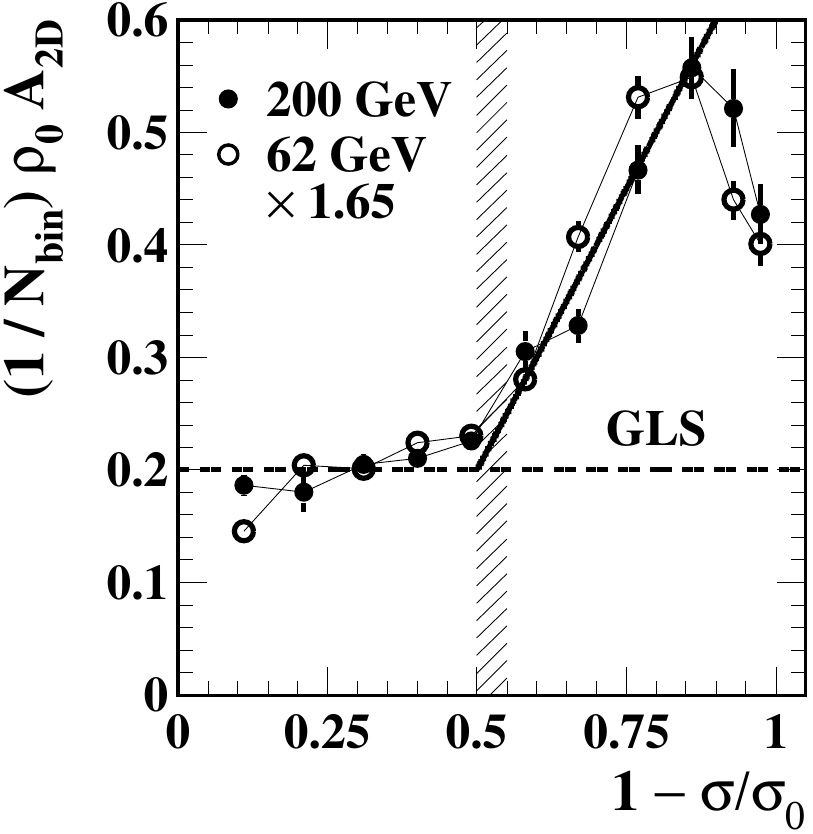}
	\includegraphics[width=1.65in,height=1.65in]{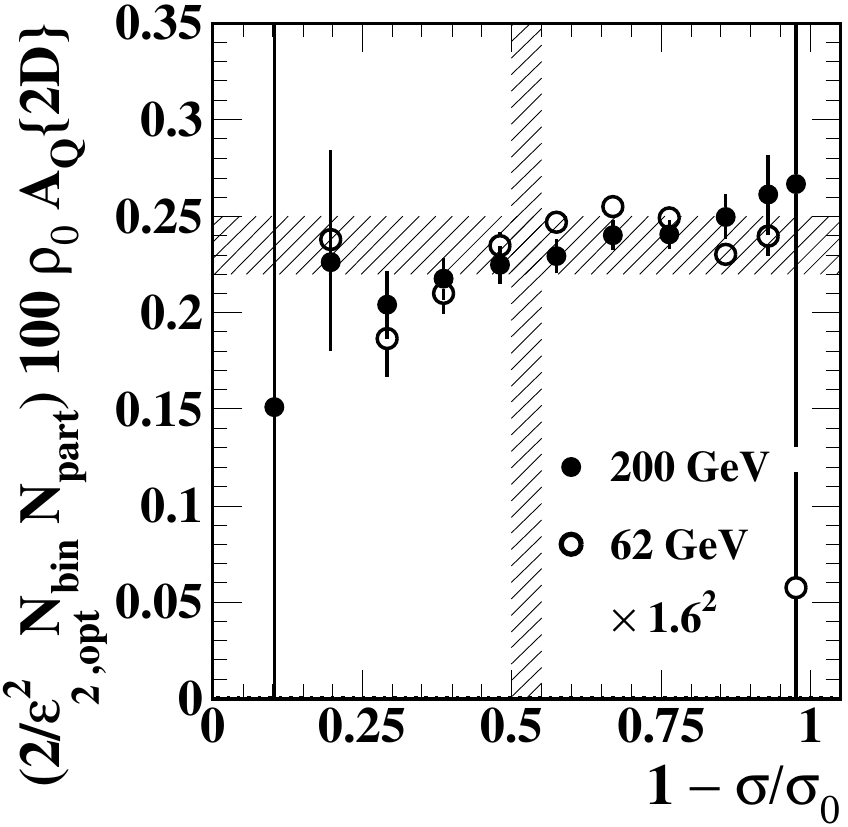}
	\caption{\label{quad2}
		Fractional cross section in the form $1 - \sigma/ \sigma_0$ proceeds from peripheral to central, left to right.
		Left: SS 2D peak amplitude in the form $\bar \rho_0 A_{\text{2D}}$ (left) scaled by the number of \nn\ binary collisions $N_{bin}$ showing good agreement with the expectation for transparent \auau\ collisions (GLS) over the peripheral 50\% of the total cross section~\cite{anomalous,nov2}. The vertical hatched band indicates a {\em sharp transition} in jet properties.
		Right: NJ quadrupole data from Ref.~\cite{v2ptb} scaled as indicated in the axis label. Relative to the scaling function the  data remain constant within data uncertainties for all \auau\ centralities.
	} 
\end{figure}

Figure~\ref{quad2} (left) shows amplitude $A_{\text{2D}}$ of the same-side (SS) 2D jet peak inferred by model fits to 2D angular correlations from 62 and 200 GeV \auau\ collisions~\cite{anomalous}. Symbol $A_{\text{2D}}$, referring to the SS 2D jet peak, is denoted by $A_1$ in Ref.~\cite{anomalous}. 
Because $A_{\text{2D}}$ is defined as a {\em per-particle} quantity the product $\bar \rho_0 A_{\text{2D}}$ represents {\em all correlated pairs} associated with MB dijets within the detector acceptance. Division by number of \nn\ binary collisions $N_{bin}$ then assesses any changes in jet formation. If the ratio is constant the product of number of jets per \nn\ collision and mean number of fragments per jet (integral of FFs) remains constant: there is no jet modification.

Below a {\em sharp transition} (hatched band) the data are consistent with linear superposition and no jet modification. Above the transition point the ratio increases rapidly (solid line) and then saturates. The product of jet number and mean fragment number {\em increases}, suggesting that no jets are lost by absorption in a dense medium. The fall-off for most-central collisions is due to strong elongation of the SS peak along pseudorapidity $\eta$ relative to the finite $\eta$ acceptance of the STAR detector~\cite{anomalous}. The 2D jet-peak centrality trend in the left panel is quantitatively consistent with centrality evolution of the \pt\ spectrum hard component $H_{AA}(y_t)$~\cite{hardspec} (only the most-central $H_{AA}$ distribution is shown in Fig.~\ref{specfrag}, right).

Figure~\ref{quad2} (right) shows the number of  correlated pairs  associated with the NJ quadrupole proportional to the product $\bar \rho_0 A_Q\{\text{2D}\}$, where 2D reminds that amplitude $A_Q\{\text{2D}\}$ is derived from model fits to 2D angular correlations as opposed to conventional Fourier amplitudes derived from 1D azimuth projections. 
The correlated pairs are scaled by quantity $(N_{part}/2) N_{bin} \epsilon_{opt}^2$ which describes the quadrupole trend on centrality within data uncertainties as first reported in Refs.~\cite{davidhq,davidhq2}. Although the individual factors vary over {\em several orders of magnitude} the ratio remains constant within 10\% down to  the most peripheral \auau\ point, which in turn is consistent with \pp\ NJ quadrupole systematics (see Sec.~\ref{ppquadd}). Those data are the same that appear in Fig.~\ref{soft2} (right).

Four results are notable: 
(a) While the dijet contribution undergoes a sharp transition near mid-centrality from linear superposition to nearly three-fold increase in correlation amplitude per \nn\ collision the NJ quadrupole data follow the same trend from most-peripheral to most-central \auau\ collisions within point-to-point systematic uncertainties. 
(b) NJ quadrupole data follow an $\epsilon_{2,opt}^2$ trend based on smooth representation of each Au nucleus by an optical-model matter density, not one based on a Monte Carlo model including discrete nucleons. 
(c) The quadrupole amplitude $A_Q\{\text{2D}\}(b)$ increases to 60\% of its maximum over an interval where \auau\ collisions are transparent (no significant particle rescattering)~\cite{davidhq,nov2}.
(d) The trend for quadrupole pairs in \pp\ collisions~\cite{ppquad} is $\bar \rho_0 A_Q\{\text{2D}\} \propto \bar \rho_s^3$ [see Fig.~\ref{soft} (right)]. That relation can also be written $\bar \rho_0 A_Q\{\text{2D}\} \sim N_{part} N_{bin}$ given that soft-hadron charge density $\bar \rho_s$ can be identified with the number of participant low-$x$ gluons near midrapidity and $\bar \rho_s^2$ is identified with dijet production via gluon-gluon binary collisions~\cite{ppquad}. 
The only difference between \pp\ and \auau\ trends is then factor $\epsilon_{opt}^2$ whose absence in the \pp\ trend is consistent with \pp\ \pt\ spectrum and angular-correlation data that suggest centrality is not relevant for \pp\ collisions~\cite{ppquad,pptheory}. The relevance of elliptic flow to high-energy nuclear collisions in the context of NJ quadrupole vs MB dijets is considered in Ref.~\cite{nonjetquad}.

\section{Initial state and Glauber model} \label{glauber}

This section responds to assumption (b) as noted in Sec.~\ref{assumptions}: The Glauber model (based on the eikonal approximation) estimates an initial-state geometry of participant nucleons or binary collisions.
Reference~\cite{nature} contains the following further assumptions regarding initial-state geometry and hydro initial conditions: ``The $v_2$ and $v_3$ values in $d$+Au and $^3$He+Au are driven almost entirely by the intrinsic geometry of the deuteron and $^3$He, while the values in $p$+Au collisions are driven by fluctuations in the configuration of struck nucleons in the Au nucleus, as the proton itself is, on average, circular,'' in reference to Eq.~(2) of that letter and its Fig.~1 presenting MC Glauber-model estimates of eccentricities $\epsilon_n$ for $n = 2,~3$ and the assumed time-evolution of hydro-estimated temperature distributions. 
Those assumptions, which provide the main basis for interpreting the presented $v_n$ data in terms of ``QGP droplet'' formation, can be questioned as follows.

\subsection{MC Glauber as applied to symmetric A-A}

Two versions of the Glauber model have been used to estimate initial-state (IS) geometries for \aa\ collisions. One version is based on continuous-matter distributions represented by the nuclear optical model and denoted by subscript ``opt.'' The IS transverse geometry is represented by the single eccentricity $\epsilon_2$. Another version is based on discrete-nucleon distributions modeled by Monte Carlo simulation and denoted by subscript ``MC.''  The IS geometry is then parametrized by several eccentricities $\epsilon_n$ mapped by one of several possible hydro models to final-state azimuth asymmetries $v_n$. 

The relation between optical and MC eccentricities is reviewed in Sec.~II F of Ref.~\cite{v2ptb}. For the optical Glauber only $\epsilon_2$ is relevant and goes to asymptotic limit $\epsilon_2 \rightarrow 0$ for central ($b = 0$) \aa\ collisions. For the MC Glauber both $\epsilon_2$ and $\epsilon_3$ are relevant and go to substantial {\em nonzero} limits for central \aa\ collisions and to large values $O(1)$ for peripheral collisions. Whether the optical or MC Glauber model is preferred depends on the method chosen for $v_n$ estimation and on interpretations of resulting $v_n$ data. 

For some $v_n$ methods substantial nonzero values are obtained for central \aa\ collisions where zero is expected for a nonfluctuating smooth initial geometry as modeled by the optical Glauber. 
The nonzero $v_n$ values are interpreted by some in terms of a fluctuating initial geometry modeled by the MC Glauber. The main difference between methods is explicit isolation of MB dijet structure vs indirect reduction of jet contributions to $v_n$ by certain numerical recipes or by cuts on $(\eta_1,\eta_2)$. The conventional multiparticle-cumulant or event-plane method has no explicit model elements corresponding to jet structure, so a jet contribution to angular correlations may emerge at some level as a bias to the $v_n$. A method based on model fits to 2D angular correlations returns $v_2 \approx 0$ for central collisions and $v_n \approx 0$ for all centralities for $n \ge 3$, compatible with the optical Glauber as discussed in Sec.~\ref{vnmeasures}.

\subsection{MC Glauber as applied to p-Pb}

A conventional MC Glauber model has been applied to 5 TeV \ppb\ collisions to estimate collision centrality (impact parameter $b$) and participant number $N_{part}$~\cite{aliceppbprod}. The resulting values conflict strongly with centrality trends inferred from TCM analysis of ensemble-mean \mmpt\ data from the same collision system~\cite{tommpt,tomglauber}. An explanation for the discrepancies is proposed in Ref.~\cite{tomexclude}: An assumption implicit in the MC Glauber model that a projectile nucleon may interact {\em simultaneously} with multiple target nucleons conflicts with data. However, if an {\em exclusivity} condition is introduced the resulting MC Glauber results are consistent with \mmpt\ data provided an exclusion time consistent with a nucleon diameter is used. Subsequent analysis of identified-hadron spectra from \ppb\ collisions confirms the TCM centrality trends~\cite{ppbpid}. The same issue must impact MC Glauber simulations applied to $x$-Au collisions leading to incorrect estimation of IS collision geometry as described in the next subsection.

\subsection{MC Glauber as related to x-Au data}

MC Glauber modeling of 200 GeV \pau, \dau\ and \hau\ collisions is a key element of the analysis reported in Ref.~\cite{nature} with its claims of QGP droplet formation in small asymmetric collision systems. Only results for 0-5\% central $x$-Au collisions are reported which greatly limits the ability to interpret experimental results. Results for two eccentricities $\epsilon_2$ and $\epsilon_3$ representing each of three IS geometries are reported in Fig.~1 (a) of Ref.~\cite{nature}. By hypothesis the three collision systems should exhibit substantially different IS geometries, dominated by those of the projectiles $x$. The extent of correspondence with measured FS $v_n$ values is then seen as a test of the role of hydro flows, and therefore QGP, in the small systems.

However, interpretation and even relevance of such a test depends on the validity of MC Glauber estimation. Questions that arise include: (i) Can the MC Glauber model describe asymmetric small collision systems at all? (ii) If yes, is the accuracy of the MC Glauber model as applied to $x$-Au collisions sufficient to test the claims made in Ref.~\cite{nature}? (iii) If the MC Glauber geometry is sufficiently accurate at the projectile {\em nucleon} level is it relevant to initial conditions required by hydro models that assume local equilibration of a dense medium?

Question (i) is answered by the previous subsection: The naive MC Glauber model without exclusivity condition greatly overestimates participant numbers in \ppb\ collisions, and is therefore likely to do so for any \pa\ collision system. It is also likely that the same will happen for $x$-Au collisions generally. What is more important for eccentricity estimates is the effect of exclusivity on composite few-nucleon projectiles: Will the leading nucleon ``shadow'' trailing nucleons? Is there any correlation among projectiles? If not, do MC Glauber eccentricities relate to projectile $x$ geometry at all as assumed? 

Question (ii) relates to Fig.~1 (b) of Ref.~\cite{nature}. The configurations in that panel are extreme cases in that the symmetry planes of projectiles $d$ and $h$ as depicted coincide with the plane of the panel, a {\em low-probability} configuration displaying maximum structure. Averaged over an event ensemble the eccentricities should be much less apparent. According to a MC Glauber model each projectile nucleon interacts with $N_{bin}$ nucleons ($\approx 7$ for central \auau\ {\em with exclusivity}), peripheral \nn\ interactions being most probable. The result should be additional smearing of effective projectile geometry and reduced eccentricity. Given those several factors one can then question the small uncertainties presented in Fig.~1 (b), and especially the large $\epsilon_2$ values for $d$ and $h$ projectiles.

Question (iii) relates to another major assumption, that IS participant-nucleon distributions as inferred from MC Glauber simulations have anything to do with local equilibration and density/temperature distributions serving as initial conditions for hydro calculations.  Figure~1 of Ref.~\cite{nature} assumes that simulated temperatures and energy densities and responding flow fields as in panel (b) are relevant to physical collisions. TCM analysis of \ppb\ \pt\ spectra at much higher energies~\cite{ppbpid} reveals no effect of changing particle densities and no variation of a universal spectrum soft component that should reflect any significant equilibration and {\em responding radial flow} (see Sec.~\ref{ppbpid1}). It is also notable that \nn\ binary collisions would be separated in space-time (on $z$), not simply additive as suggested by $(x,y)$ projections in Fig.~1 (b).

In summary, several strong assumptions that underlie Fig.~1 of Ref.~\cite{nature} are questionable. In particular, evidence from collision systems at higher energies contradicts certain Glauber-related assumptions. Application of a naive MC Glauber model to \pa\ collisions has been shown to be severely biased by overestimation of the number of participants (3-fold excess). Space-time overlap of multiple \nn\ collisions {\em for the same projectile nucleon} is strongly hindered according to \ppb\ data, which may have a major effect on $x$-A collisions with composite projectiles $x$.

\section{Final state and $\bf v_n$ measurements} \label{vnmeasures}

This section responds to assumption (c) as noted in Sec.~\ref{assumptions}: Fourier amplitudes $v_n$ as conventionally defined actually measure final-state flows as opposed to some other phenomenon (e.g.\ jets).  A third major issue for claims of QGP droplet formation is the accuracy of $v_n$ data and their interpretation relative to hydro predictions derived from MC Glauber $\epsilon_n$ estimates. As noted in the previous section there are two main approaches to $v_n$ measurement: (i) $v_n\{m\}$ derived from multiparticle cumulants applied to 1D projections onto azimuth and (ii) $v_2\{\text{2D}\}$ derived from model fits to 2D angular correlations on $(\eta,\phi)$. In the former case the so-called ``event-plane'' (EP) method is equivalent to $v_2\{2\}$~\cite{quadspec,v2ptb,njquad} and all azimuth correlation structure is reduced to a Fourier-series representation. In the latter case $n > 2$ Fourier terms (``higher harmonics'') are excluded from 2D model fits as being negligible once the SS 2D jet peak is represented independently by a 2D Gaussian~\cite{multipoles,sextupole}.

There are several significant issues for $v_n$ measurements and interpretations: 
(a) MB dijets may make a significant contribution to some $v_n$ data that should be explicitly evaluated or effectively excluded from such measurements. 
(b) $v_2\{\text{2D}\}$ data strongly conflict with some $v_2\{2\} \approx v_2\{\text{EP}\}$ data suggesting substantial jet bias in the latter. 
(c) measured data trends from an assortment of collision systems suggest that alternative {\em nonflow} (and nonjet) interpretations for $v_2\{\text{2D}\}$ data may be relevant.

\subsection{$\bf v_n$ analysis methods}

One method for estimating $v_n$ is multiparticle cumulants denoted by $v_n\{m\}$, where $n = 2,~3$ for quadrupole and sextupole cylindrical multipoles and $m$ is, for instance, 2 and 4 for two- and four-particle cumulants. $v_2\{2\}$ simply corresponds to the quadrupole Fourier amplitude for all 2D angular correlations within an acceptance on $(\eta_1,\eta_2)$ projected onto 1D azimuth. As reported in Refs.~\cite{2004,sorensenv2fluct}  difference $v_2\{2\}^2 - v_2\{4\}^2 \approx 2\sigma^2_{v_2} + \delta_2$ includes contributions from flow fluctuations (first term) and nonflow (second term). It is often assumed that $\sigma^2_{v_2} \gg \delta_2$ and flow fluctuations dominate~\cite{flowfluct}. However, evidence from model fits to 2D angular correlations suggests that contribution $\delta_2$ (mainly SS 2D jet peak) dominates and the fluctuation term is negligible~\cite{azcorr1,gluequad,v2ptb}. 

According to conventional flow descriptions, if $v_2$ fluctuations are found to be substantial in the hadronic final state their origin is likely to be IS geometry fluctuations in $\epsilon_2$ as modeled by the MC Glauber model~\cite{alverfluct}.  One consequence of that reasoning is an alternative explanation for what had been interpreted as evidence for {\em Mach cones}: peaks on azimuth at $\pm 2\pi / 3$ in so-called trigger-associated jet correlations~\cite{phenixzyam}.  An unrecognized jet contribution to published $v_2$ data (nonflow) had resulted in oversubtraction of the estimated $v_2$ component of a combinatorial background for trigger-associated jet correlations~\cite{tzyam} leading to claims for Mach cone formation~\cite{mach}. The jet contribution to  $v_2$ data was later reinterpreted as ``flow fluctuations'' leading to preference for the MC Glauber model. It was argued that given IS geometry fluctuations, including those for $\epsilon_3$ (``triangularity''), one should observe {\em triangular flow} in the final state as measured by $v_3$, including the same peaks on azimuth at $\pm 2\pi / 3$~\cite{alver}.  It was further concluded that {\em any} Fourier component of 1D azimuth angular correlations $v_n$ (i.e.\ ``higher harmonics'') may be interpreted as representing a FS flow resulting from IS geometry fluctuations~\cite{luzum}.

Multiparticle cumulant methods focus on projections of angular correlations onto 1D azimuth $\phi$, thus discarding essential information conveyed by $\eta$ dependence. An alternative method for estimating $v_n$ is based on model fits to 2D angular correlations~\cite{axialci,anomalous,ppquad}. A major advantage of 2D model fits is full use of correlation information (no projection onto 1D) and unambiguous isolation of a SS 2D peak representing all {\em intra}\,jet angular correlations. 

2D correlation histograms are plotted on difference variables $x_\Delta \equiv x_1 - x_2$ defined as diagonal coordinates on pair spaces $(\eta_1,\eta_2)$ and $(\phi_1,\phi_2)$ differentiated from intervals $\Delta x$ defined on a single coordinate axis $x$ (see Sec.~\ref{quadaa}).
The following notation is used: If angular correlations are measured by {\em per-particle} quantity $\Delta \rho / \sqrt{\rho_{ref}}$ as in Refs.~\cite{anomalous, quadspec} 2D fit-model amplitudes are denoted by $A_X \equiv \bar \rho_0 B_X$. $V_2^2 \equiv \bar \rho_0^2 v_2^2 \equiv \bar \rho_0^2 B_Q \equiv \bar \rho_0 A_Q$ is the total-quadrupole amplitude (proportional to the total number of quadrupole-correlated pairs). For \pt-differential data as in Ref.~\cite{v2ptb} $B_Q(p_t,b) \equiv v_2(b)v_2(p_t,b)$. 

The 2D fit model as in Refs.~\cite{axialci,anomalous,ppquad,v2ptb} includes a 2D Gaussian to model the SS 2D peak, with amplitude $A_{\text{2D}}$ (intrajet correlations), a cylindrical-dipole term with amplitude $A_{D}$ to model the away-side (AS) 1D peak on azimuth (back-to-back jet contribution) and a cylindrical-quadrupole term with amplitude $A_Q\{\text{2D}\}$  to model the NJ quadrupole contribution also denoted by $v_2\{\text{2D}\}$.  In more-central \auau\ collisions angular correlations are dominated by the jet-related SS 2D peak and AS dipole (see Fig.~\ref{unique}, right).  The 2D fit model typically describes correlation data within statistical uncertainties~\cite{anomalous,ppquad}.

Consequences of model fits to 2D angular correlations include: (a) $v_2\{\text{2D}\}$ data include no jet (nonflow) contribution by construction. (b) The $v_2\{\text{2D}\}$ centrality trend as in Fig.~\ref{quad2} (right) is consistent with $\epsilon_{2,opt}$, not with $\epsilon_{2,MC}$. (c) 2D model fits include no $v_3$ or higher terms~\cite{multipoles,sextupole}. (d) Bayesian analysis of 1D azimuth correlations strongly rejects an isolated Fourier description with terms higher than dipole and quadrupole~\cite{tombayes}. Given (c), (d) and the $v_2\{\text{2D}\}$ centrality trend the MC Glauber model and ``higher harmonics'' are thus falsified.

\subsection{NJ quadrupole as a unique feature in A-A} \label{quadaa}

\begin{figure}[h]
	\includegraphics[width=1.65in,height=1.5in]{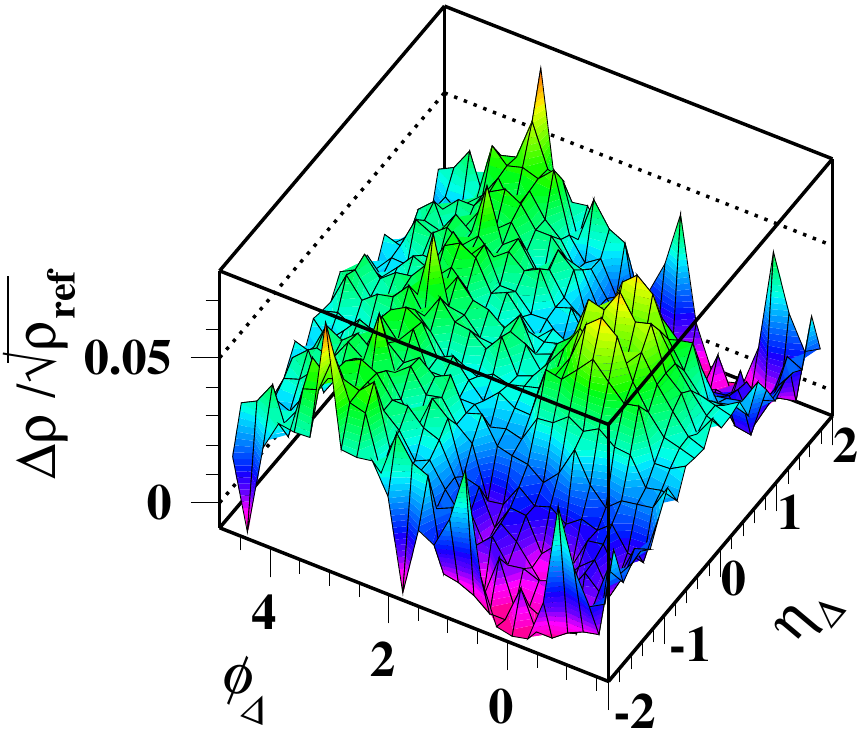}
	\includegraphics[width=1.65in,height=1.5in]{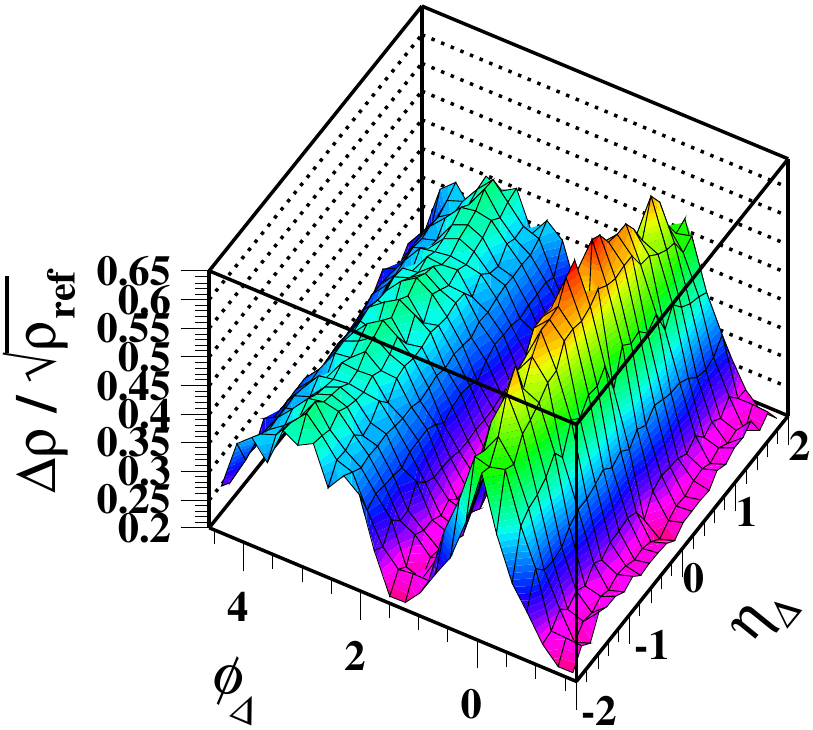}
	\caption{\label{unique} (Color online)
		2D angular correlations on $(\eta_\Delta,\phi_\Delta)$ from 84-93\% central (left) and 0-5\% central (right) 200 GeV \auau\ collisions~\cite{anomalous} showing jet-related features after subtracting several nonjet 2D fit-model elements (see text).
	}  
\end{figure}

Figure~\ref{unique} shows 2D angular correlations from 200 GeV \auau\ collisions for 84-93\% (left, $\bar \rho_0 \approx 4.3$ vs NSD \pp\ $\approx 2.5$) and 0-5\% (right, $\bar \rho_0 \approx 672$) centralities. Statistical errors are largest for edge bins with $|\eta_\Delta| \approx 2$. The published histograms presented in Fig.~1 of Ref.~\cite{anomalous} have been further processed for Fig.~\ref{unique} as follows. Based on 2D model fits reported in Ref.~\cite{anomalous} the following fitted model elements have been subtracted from the basic 2D correlation histograms: (a) a constant offset, (b) a 1D Gaussian on $\eta_\Delta$ representing the TCM soft component, (c) a narrow 2D exponential centered at the origin representing Bose-Einstein correlations and electron pairs, (d) a feature described in Fig.~4 of Ref.~\cite{anomalous} subtracted from the 0-5\% data only, and (e) a separate quadrupole component. Items (a) and (b) have no azimuth structure. Item (c) has narrow widths for all centralities; its Fourier spectrum is thus broad with small amplitude. Item (d) consists of an $\eta_\Delta$-modulated dipole component making no contribution to Fourier coefficients with $n \ge 2$. Item (e) is denoted in Ref.~\cite{anomalous} as the NJ quadrupole component corresponding to reported values of $v_2\{\text{2D}\}$ in Ref.~\cite{v2ptb} ($\approx 0$ for 0-5\%) and discussed in Sec.~\ref{quadacc}. What remains in Fig.~\ref{unique} is accurately described by an AS 1D dipole and a SS 2D Gaussian.

As reported in Ref.~\cite{anomalous} the  two features in Fig.~\ref{unique} are closely correlated for all \auau\ centralities. Over a peripheral interval corresponding to 50\% of the \auau\ total cross section both features closely conform to expectations for contributions from MB dijets, especially precise scaling with number of \nn\ binary collisions $N_{bin}$ (see Fig.~\ref{quad2}, left). In more-central \auau\ collisions both features deviate from simple $N_{bin}$ scaling in related ways: (a) the amplitudes increase {\em together} more rapidly than $N_{bin}$ until saturation is achieved, and (b) the $\eta_\Delta$ width of the SS 2D peak strongly increases while the $\phi_\Delta$ width {\em decreases} from the \pp\ value in such a way that the 2D-peak aspect ratio measured by width ratio $\sigma_{\eta_\Delta} / \sigma_{\phi_\Delta}$ increases smoothly and approximately linearly over the full centrality range of \auau\ collisions, from $\approx 1/2$ for \pp\ to $\approx 3$ for central \auau\ collisions~\cite{anomalous}.

Some implications of those results are as follows: There are two main sources of a quadrupole $n = 2$ amplitude in 2D angular correlations, one being the NJ quadrupole component, the other being the $n = 2$ Fourier component of the SS 2D peak. Evolution of SS 2D peak structure and AS 1D peak amplitude coincide with evolution of 200 GeV \auau\ \pt\ spectra for identified hadrons as described in Ref.~\cite{hardspec}, specifically with the jet-related spectrum hard component. The centrality trend for the NJ quadrupole measured by the number of correlated pairs is $\bar \rho_0 A_Q \propto N_{part} N_{bin} \epsilon^2_{opt}$~\cite{v2ptb}. Thus, inference of $v_n$ by a numerical method must clearly and accurately distinguish the two sources -- NJ quadrupole and MB dijet structure -- before claiming new physics from $v_n$  trends.

In summary,  although its shape on $\eta$ may vary with collision charge density (or \aa\ centrality) the SS 2D peak is a monolith with uniform $\phi$ width independent of $\eta$. The SS 2D peak and the AS dipole approximation to a broad AS 1D peak are closely correlated as to centrality and energy dependence. The trends correspond to expected jet physics in all cases, with close correspondence also to single-particle (SP) \pt\ spectra and the TCM hard component. Arguments to split the SS 2D peak into a ``short-range jet-like'' component narrow on $\eta$ and ``long-range'' tails (ridge) attributed to non-jet processes are not justified by data trends. Projected onto azimuth the SS 2D peak can always be represented by a narrow Gaussian with its own Fourier series decomposition~\cite{tombayes}. But combining individual SS peak Fourier components with other correlation structures leads to an inefficient and confusing data representation. Likewise, the AS dipole has an independent jet-related existence that should be preserved in any data representation.  The NJ azimuth quadrupole thus has its own unique centrality and energy trends for various collision systems and should be distinguished from jet-related structures. The centrality trends of MB dijets and the NJ quadrupole remain dramatically different for all \auau\ centralities (see Fig.~\ref{alice}).

\subsection{NJ quadrupole vs $\bf \bar \rho_s$ in 200 GeV p-p collisions} \label{ppquadd}

While manifestations of elliptic flow in the form of $v_2$ data for \aa\ collisions were sought out and apparently observed~\cite{rhicflow} a similar phenomenon in small collision systems, especially in \pp\ collisions, was not expected. Observation of a SS 1D ``ridge'' in 7 TeV \pp\ collisions was therefore surprising and not immediately understood~\cite{ppridge}. Reference~\cite{tomcmsridge} soon provided an interpretation: the \pp\ 1D ridge is one (SS) lobe of a NJ quadrupole component; the AS lobe increases the magnitude of the (negative) curvature of the AS 1D peak. More recently, a detailed study of 2D angular correlations from 200 GeV \pp\ collisions has revealed the full systematics of the NJ quadrupole in \pp\ collisions~\cite{ppquad}.

\begin{figure}[h]
	\includegraphics[width=.24\textwidth]{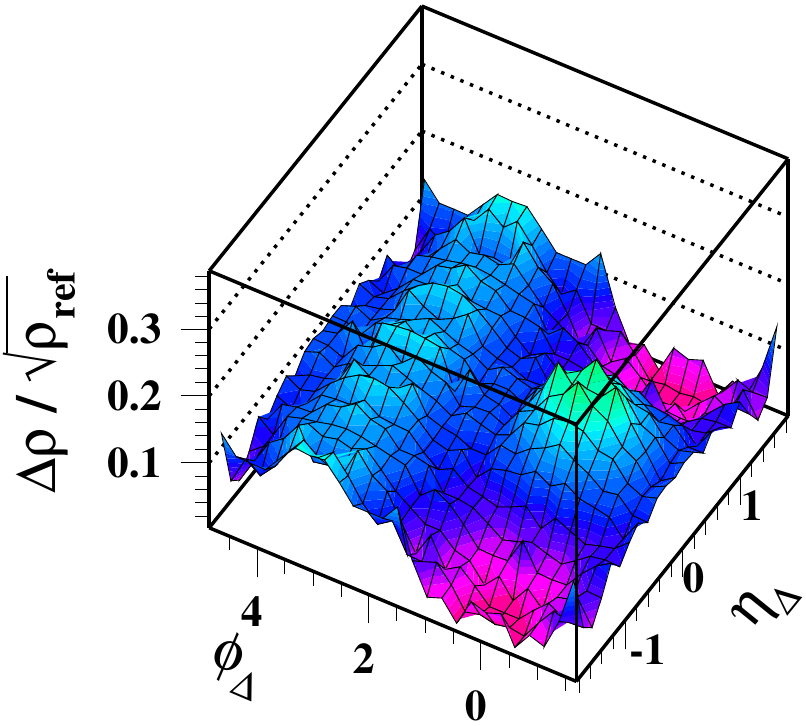}
	\includegraphics[width=1.65in,height=1.6in]{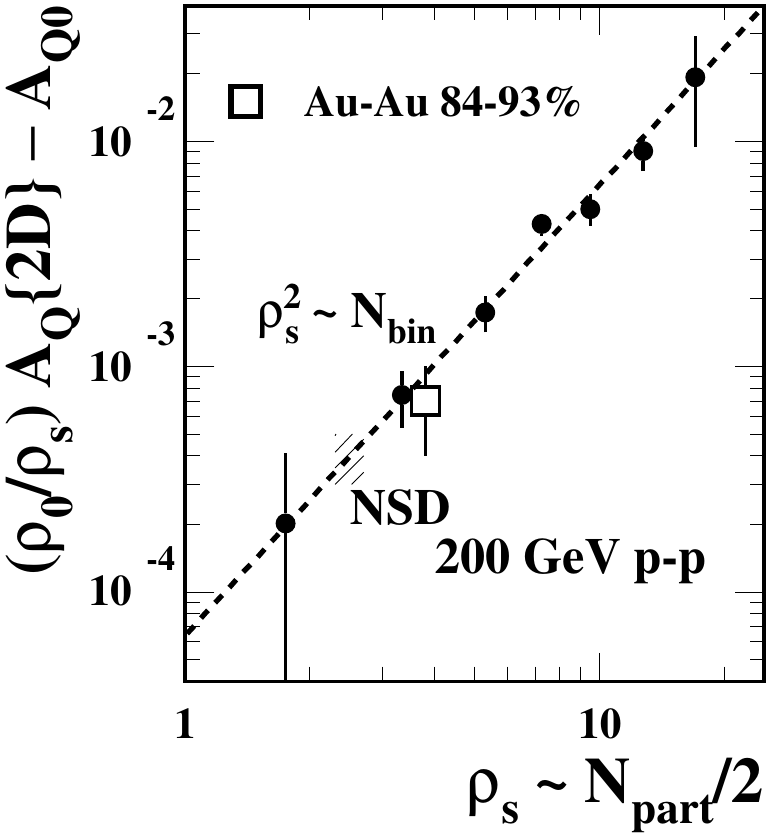}
	\caption{\label{soft} (Color online)
		Left: 
		2D angular correlations from high-multiplicity 200 GeV \pp\ collisions (class 5 of 7 in the right panel) with jet + NJ quadrupole contributions obtained by subtracting fitted offset, soft-component and BEC/electron elements of the fit model from the data histograms~\cite{ppquad}. 
		Right:  
		Rescaled NJ quadrupole amplitude $(\bar \rho_0 / \bar \rho_s)A_Q\{\text{2D}\}(\bar \rho_s) - A_{Q0}$ vs soft-component charge density $\bar \rho_s$ proportional to number of participant parton pairs $N_{part}/2$. The line is $\propto \bar \rho_s^2$ representing number $N_{bin}$ of binary parton-parton (gluon-gluon) collisions. Subtracted offset $A_{Q0}$ may be an aspect of global transverse-momentum conservation. The hatched region represents the $\bar \rho_s$ estimate for NSD \pp\ collisions and the corresponding value from the data trend.
	} 
\end{figure}

Figure~\ref{soft} (left) shows 2D angular correlations for high-multiplicity 200 GeV \pp\ collisions (multiplicity class 5 in Ref.~\cite{ppquad}). The basic correlation data have been modified by subtracting certain 2D fit-model components as follows: a fitted constant offset, a soft-component 1D Gaussian on $\eta_\Delta$ and a narrow 2D exponential representing Bose-Einstein correlations and electron pairs have been subtracted (see the description of Fig.~\ref{unique}).  What remain are a jet-related SS 2D peak at the origin plus an AS 1D peak at $\pi$ and a NJ quadrupole component that is visually manifested by two data features: (a) the SS background on either side of the SS 2D jet peak along $\eta_\Delta$ is flattened on $\phi_\Delta$ and (b) the AS 1D peak curvature is strongly increased compared to the jet-related AS dipole component as noted above. The NJ quadrupole amplitude is accurately determined via 2D model fits.

Figure~\ref{soft} (right) shows fitted NJ quadrupole amplitude $A_Q\{\text{2D}\} = \bar \rho_0 v_2^2\{\text{2D}\}$ rescaled to a per-participant-parton measure by factor $\bar \rho_0 / \bar \rho_s = n_{ch} / n_s$. The dashed line represents the trend
\bea \label{ppquad}
(\bar \rho_0 / \bar \rho_s) A_Q\{\text{2D}\}(\bar \rho_s) &=& (0.008\bar  \rho_s)^2 + A_{Q0},
\eea indicating that the data (modulo the small fixed offset $A_{Q0} \approx 0.0007$) are consistent with a {\em quadratic} dependence on $\bar \rho_s$ in contrast to the linear dependence for jet production reported in Ref.~\cite{ppquad}.
The rescaling strategy $(\bar \rho_0 / \bar \rho_s)  A_Q\{\text{2D}\}$ is preferred because the small contribution to $\bar \rho_0 A_Q\{\text{2D}\}$ from momentum conservation scales with $\bar \rho_s$ (i.e.\ soft component $\sim N_{part}$) which then transforms to the constant term $A_{Q0}$ in Eq.~(\ref{ppquad}). 
That offset for the NJ quadrupole component (and comparable $A_{D0}$ for the AS dipole component) is consistent with TCM soft-component transverse-momentum conservation~\cite{anomalous}. 

The open square corresponds to the most-peripheral 200 GeV \auau\ point in Fig.~\ref{soft2} (right) that is also consistent with the empirical relation $A_Q\{\text{2D}\} = 0.0045 N_{bin} \epsilon_{opt}^2$~\cite{davidhq}.  For 84-93\% \auau\  $\epsilon_{opt} \approx 0.4$ and $N_{bin} \approx 2$ result in $A_Q\{\text{2D}\} \approx 0.0014$, in agreement with the \pp\ data (when $A_{Q0}$ is subtracted). The hatched band represents the \pp\ NSD value for $\bar \rho_s \approx 2.5$ and the corresponding data trend denoted by the dashed line.

The per-participant trend in Fig.~\ref{soft} (right) increases 100-fold over the measured \nch\ interval. The same quadratic trend on $\bar \rho_s$ continues down to negligible hadron density, inconsistent with a collective (flow) phenomenon that might result from final-state particle rescattering. It is notable that the total number of quadrupole-correlated pairs, measured by $\bar \rho_0 A_Q\{\text{2D}\}$, {\em increases 1000-fold} for the plotted \pp\ data. In contrast, 2D model fits including a sextupole $A_S\{\text{2D}\}$ ($\sim v_3$) component return zero amplitude for that component within statistical uncertainties in all cases~\cite{sextupole}. In general, observed features of $v_2\{\text{2D}\}$ data from \pp\ collisions summarized above suggest the presence of a novel QCD phenomenon unrelated to a dense flowing medium or QGP~\cite{gluequad}.

\subsection{NJ quadrupole vs $\bf N_{part}$ in Au-Au collisions} \label{auauquad}

\begin{figure}[h]
	\includegraphics[width=1.65in,height=1.6in]{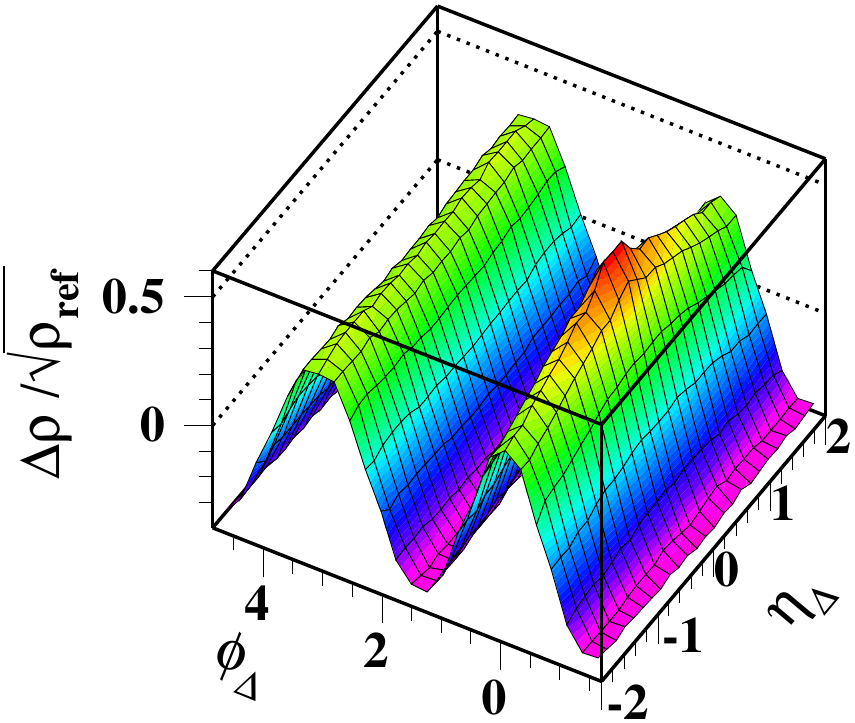}
	\includegraphics[width=1.65in,height=1.6in]{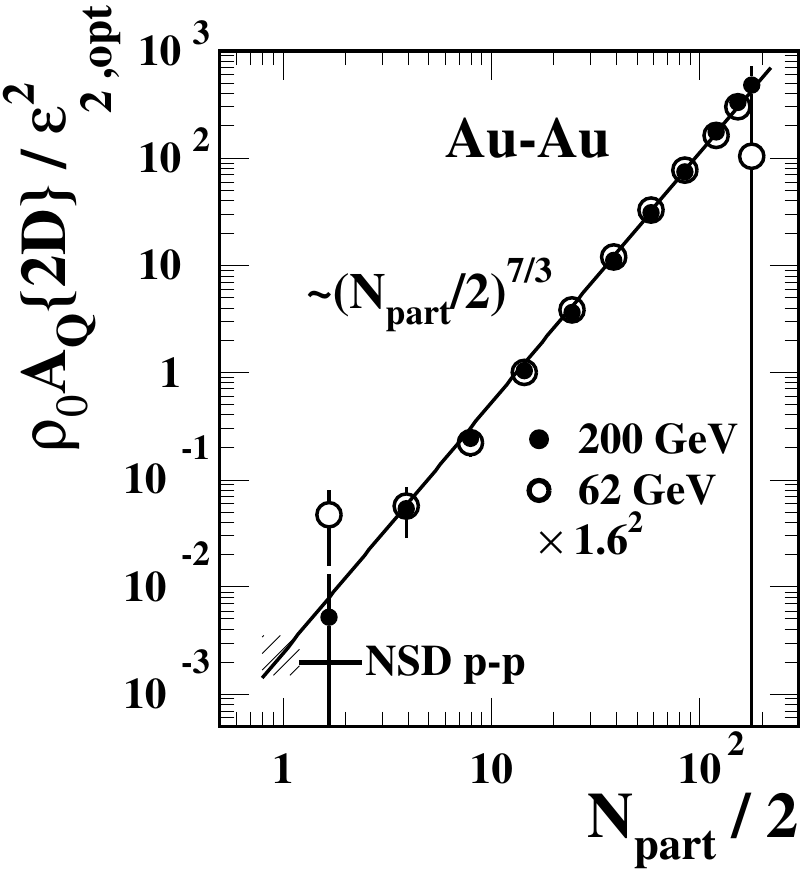}
	\caption{\label{soft2}  (Color online)
		Left: 2D angular correlations from 200 GeV \auau\ collisions for $N_{part}/2 \approx 85$ (18-28\%)~\cite{anomalous}. Correlation data are dominated by the NJ quadrupole and jet-related SS 2D and AS 1D peaks. No fit-model components have been subtracted.
		Right: 	Rescaled NJ quadrupole amplitude $\bar \rho_0 A_Q\{\text{2D}\} / \epsilon^2_{2,opt}$ vs participant-pair number $N_{part} / 2$~\cite{anomalous,v2ptb}. The line is $\propto (N_{part} / 2)^{7/3}$ representing product $N_{bin} N_{part} / 2$ estimated by a Glauber Monte Carlo. The most-peripheral 200 GeV point is the open box in Fig.~\ref{soft} (right).
	}  
\end{figure}

Figure~\ref{soft2} (left) shows 2D angular correlations from 18-28\% central 200 GeV \auau\ collisions. For those mid-central data the two-lobed NJ quadrupole dominates 2D correlations, with the jet-related AS dipole and SS 2D peak as subordinate features in contrast to Fig.~\ref{unique} (right). This is one of eleven centrality classes of 200 GeV \auau\ collisions from which NJ quadrupole $A_Q\{\text{2D}\}$ data were extracted via 2D model fits as reported in Refs.~\cite{anomalous,v2ptb}.

Figure~\ref{soft2} (right) shows the rescaled NJ quadrupole amplitude in the form $\bar \rho_0 A_Q\{\text{2D}\} / \epsilon^2_{2,opt}$ -- the number of quadrupole-correlated pairs divided by the ($n =2$) optical-Glauber eccentricity squared. The rescaled data follow a Glauber-model trend $\propto N_{bin} N_{part} / 2$ (solid line)  over five orders of magnitude -- extending from single \nn\ collisions to central \auau. Based on the Glauber relation $N_{bin} \approx (N_{part}/2)^{4/3}$~\cite{powerlaw} the straight line following $(N_{part}/2)^{7/3}$ describes the data within their uncertainties. The data trend extrapolated to $N_{part} / 2 = 1$ (\nn\ collisions, hatched band) is consistent with the NSD p-p value shown as the hatched band in Fig.~\ref{soft} (right). 

The most-central 0-5\% point for \pt-integral $v_2\{\text{2D}\}(b)$ is consistent with the straight-line trend whereas the \pt-differential $v_2\{\text{2d}\}(p_t)$ data  for the same centrality class in Fig.~\ref{ytcomp1} (left) are consistent with zero (hatched band). The apparent conflict is discussed in Ref.~\cite{v2ptb}.

Whereas ``global'' momentum conservation, scaling as number of participant low-$x$ gluons (i.e.\ soft-component offset $A_{Q0}$ for $(\bar \rho_0 / \bar \rho_s)A_Q\{\text{2D}\}$), is a significant effect for \pp\ collisions as in Fig.~\ref{soft} (right) and Eq.~(\ref{ppquad}), the equivalent contribution to Fig.~\ref{soft2} for \auau\ collisions is small compared to large jet and NJ quadrupole contributions.  

That observation relates to confusion about momentum conservation within the TCM soft component vs momentum conservation within the TCM hard component (parton scattering to dijets). The symbol $v_1$ is sometimes interpreted to represent global transverse-momentum conservation and sometimes to represent a flow manifestation. In  the latter case there is further confusion: ``The first and second coefficient[s]..., $v_1$ and $v_2$, are usually referred to as directed and elliptic flow...''~\cite{hydro} is incorrect. The symbol $v_1$ is employed ambiguously to denote both a {\em spherical} multipole (directed flow, asymmetric on $\eta$) and a {\em cylindrical} multipole (e.g.\ the AS 1D peak or dipole) dominated by back-to-back MB dijet correlations ({\em local} momentum conservation). There is further confusion if the pair distribution on $\phi_\Delta$ is represented solely by a Fourier series, in which case the $v_1$ amplitude represents contributions of opposite sign from the AS 1D jet peak (negative) and the SS 2D jet peak (positive), the combination misinterpreted to represent global momentum conservation or a flow manifestation.

It is notable that based on $v_2\{\text{2D}\}$ data derived from model fits to 2D angular correlations there is a detailed {\em quantitative} correspondence between \pp\ quadrupole \nch\ trends and \aa\ quadrupole centrality (\nch) trends. The correspondence suggests that the same underlying quadrupole production mechanism is responsible in small and large systems, and that the NJ quadrupole is not the result of hydro flows {\em in a dense medium}~\cite{gluequad}. 

\subsection{Relevance of higher harmonics} \label{nonjethigher}

Arguments in Ref.~\cite{nature} supporting QGP droplets in small collision systems are based in part on inferred systematic behavior of ``triangular flow'' as measured by $v_3(p_t)$, which then relates to the general question of ``higher harmonics'' as measured by $v_n(p_t,b)$ with $n > 2$ for data derived from \aa\ collisions at the RHIC and LHC~\cite{multipoles,sextupole}. The  $v_3(p_t)$ measurements reported in Ref.~\cite{nature} are seen as a special case of an established flow phenomenon in \aa\ collisions. However, there is evidence from $v_n(p_t,b)$ data for \aa\ collisions that any reported higher harmonics are entirely MB dijet manifestations which then casts doubt on arguments in Ref.~\cite{nature} based on $v_3(p_t)$ data. 
$v_n$ data from 2.76 TeV \pbpb\ collisions provide an illustrative example. \pbpb\ $v_n(b)$ data trends can be predicted based on a 200 GeV \auau\ data model derived from model fits to 2D angular correlations [see Sec.~\ref{quadacc} and Eqs. (\ref{aqsum}), (\ref{sseq}) and \ref{fm})]. The 200 GeV trends adjusted slightly are able to describe the LHC $v_n$ data accurately and demonstrate the quantitative relation between ``higher harmonics'' and MB dijets.

\begin{figure}[h]
	\includegraphics[width=3.3in]{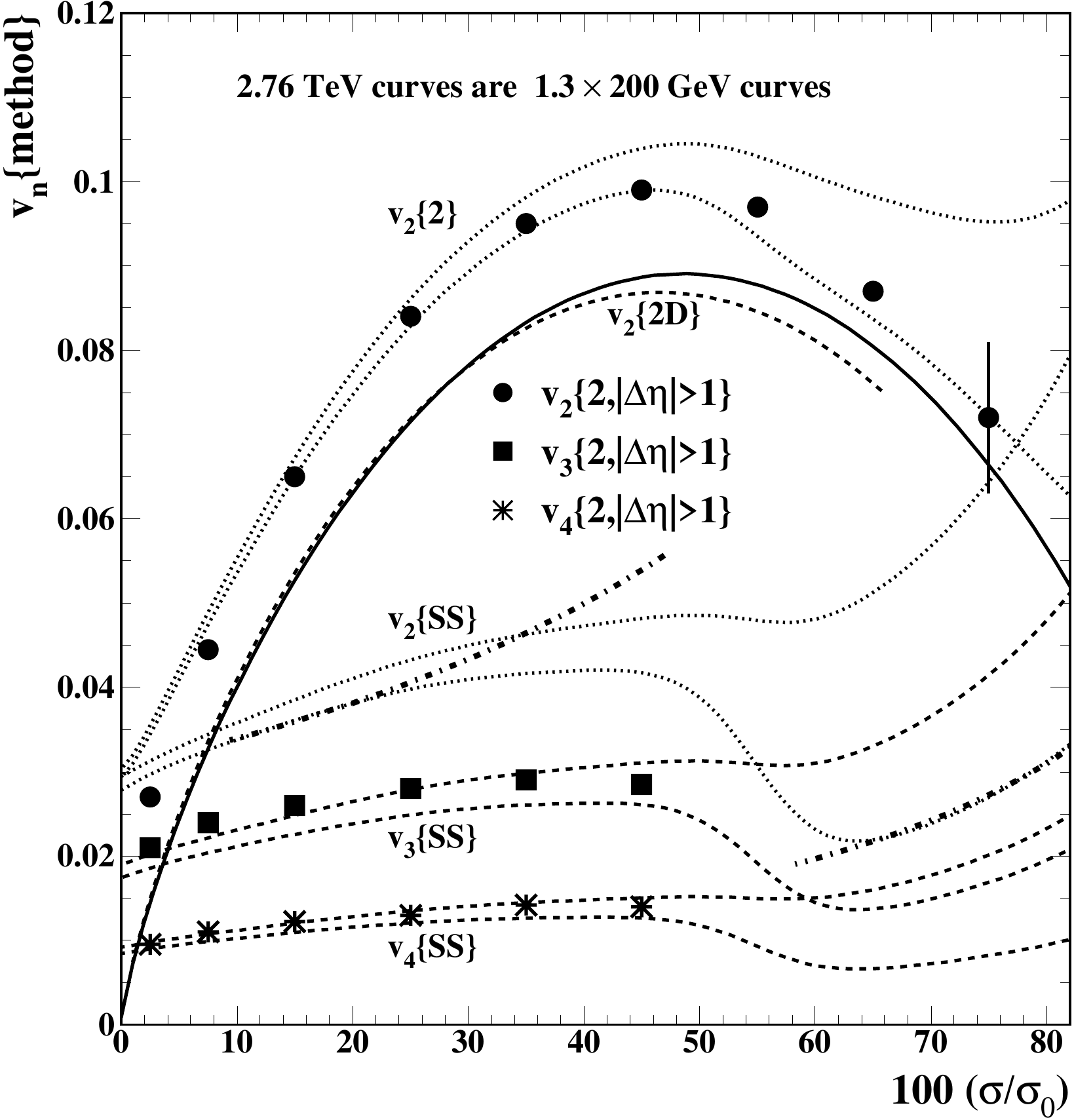}
	\caption{\label{alice}
		The data points are from Ref.~\cite{alice}. The several curves are obtained from Fig.~17 of Ref.~\cite{multipoles} describing 200 GeV \auau\ data, but with $\sigma_{\phi_\Delta}$ reduced from 0.65 to 0.60 and with overall multiplier 1.3. The relative magnitudes and centrality variations agree closely between data and curves, suggesting that the SS 2D peak is the sole source of $v_n\{\text{SS}\}$ at 2.76 TeV and has similar properties to that at 200 GeV, as also noted in Ref.~\cite{ppridge}. The dash-dotted curves illustrate  $v_n\{\text{SS}\}(b)\propto (N_{part}/2)^{-1/3}$ trends expected for MB dijet production.
	}  
\end{figure}

Figure~\ref{alice} shows $v_n\{2\} \approx v_n\{\text{EP}\}$ data (points) for $n = 2,~3,~4$ from 2.76 TeV \pbpb\ collisions (from Fig.~1 of Ref.~\cite{alice}) compared to predicted trends derived from 2D model fits to 200 GeV \auau\ angular correlations on $(\eta,\phi)$ as summarized in Fig.~17 of Ref.~\cite{multipoles} (curves).
The $v_n\{2\}$ data from Ref.~\cite{alice} represent coefficients from a Fourier-series fit to the sum of all two-particle angular correlations projected onto 1D azimuth, with $\eta$ exclusion cut $1 <|\eta_\Delta| <1.6$, denoted $v_n\{2,\text{$\eta$ cut}\}$. The solid curve for $v_2\{\text{2D}\}$ is equivalent to the straight line representing $A_Q\{\text{2D}\}$ in Fig.~\ref{soft2} (right). The $v_n\{\text{SS}\}$ are Fourier components of the SS 2D jet peak as discussed in Sec.~\ref{quadacc}.

All curves in Fig.~\ref{alice} are derived from a description of 200 GeV \auau\ data. For each pair of dashed or dotted curves the upper curve corresponds to a contiguous $0 < |\eta_\Delta| < 1.6$ acceptance (ALICE TPC acceptance $|\eta| < 0.8$) while the lower curve corresponds to a ``nonflow'' exclusion acceptance  $1 <|\eta_\Delta| <1.6$. To accommodate the 2.76 TeV \pbpb\ data the SS 2D peak azimuth width $\sigma_{\phi_\Delta}$ is reduced from 0.65 (200 GeV) to 0.60 (2.76 TeV). The width reduction is consistent with the energy trend from 62 to 200 GeV~\cite{anomalous}, increasing the $v_3$ and $v_4$ trends slightly relative to $v_2$ [see Eq.~(\ref{fm})]. Given the SS 2D peak width adjustment all curves in   Fig.~17 of Ref.~\cite{multipoles} are scaled up by a common factor 1.3, the same factor attributed in Ref.~\cite{alicev2} to an increase in the inclusive spectrum mean $p_t$ at the higher energy.

Given those small adjustments the predictions derived from 2D angular correlations at 200 GeV describe the 2.76 TeV \pbpb\ data well. The $v_2\{2\}$ prediction {\em with} $\eta$ exclusion cut follows the $v_2\{2\}$ data, suggesting persistent presence of a sharp transition~\cite{anomalous} in the SS 2D peak $\eta$ width at the higher energy (see Fig.~\ref{quad2} and Ref.~\cite{anomalous}). 
The lack of $v_3$ and $v_4$ data for more-peripheral centralities is unfortunate because such data could provide a direct test of SS 2D peak systematics at the higher energy. Such peripheral data should fall to small values if the SS 2D peak is confined to $|\eta_\Delta| <1$ for more-peripheral collisions [see Fig.~\ref{soft} (left) for \pp\ collisions]. The LHC data are also consistent with the 200 GeV $v_2\{\text{2D}\}$ NJ quadrupole trend (solid curve) which shows no sensitivity to $\eta$ exclusion cuts because that correlation feature is uniform on $\eta_\Delta$ within the ALICE TPC $\eta$ acceptance.

It is notable that the centrality trends for $v_n\{\text{SS}\}$ and $v_2\{\text{2D}\}$ are dramatically different, consistent with independent physical mechanisms. All $v_n\{\text{SS}\}$ have the same functional form but differ in amplitude according to the Fourier components of a narrow Gaussian [i.e.\ the SS 2D peak projected onto 1D azimuth, see Eq.~(\ref{fm}) below]. 

That the SS 2D peak represents jets may actually be confirmed via $v_n\{\text{SS}\}$ trends with a simple argument (e.g.\ for $v_2\{\text{SS}\}$). $v_2(b)$ as defined is the square root of a pair ratio. The pair-ratio numerator represents correlated particle pairs while the denominator represents uncorrelated mixed pairs. If the source of correlations is jets the numerator should be proportional to $N_{bin} \approx (N_{part}/2)^{4/3}$ (per the Glauber model of \aa\ collisions) while the denominator is approximated by $\bar \rho_0^2 \propto (N_{part}/2)^2$. The $v_2$ trend for jets should then be approximately $v_2\{\text{SS}\} \propto \sqrt{(N_{part}/2)^{-2/3}} = (N_{part}/2)^{-1/3}$ and similarly for any other $v_n\{\text{SS}\}$ (higher harmonics). 

In Fig.~\ref{alice} the dash-dotted curves represent the predicted jet-related $v_2\{\text{SS}\}(b)$ trend. The lower curve is $0.038 (N_{part}/2)^{-1/3}$ (SS peak narrower on $\eta_\Delta$ for peripheral collisions) and the upper is $0.135 (N_{part}/2)^{-1/3}$  (SS peak broader on $\eta_\Delta$ for central collisions).  The change in amplitude can be related to the sharp transition in SS 2D peak properties near 50\% centrality (especially the peak width on $\eta_\Delta$). Note correspondence of the transition between two dash-dotted trends in Fig.~\ref{alice} and the sharp transition of the jet amplitude in Fig.~\ref{quad2} (left). The trend for $v_2\{2\}$ (solid dots) then actually confirms the presence of a sharp transition in the jet contribution from 2.76 TeV \pbpb\ collisions similar to 200 GeV \auau\ collisions~\cite{anomalous}. 

The same argument applied to the $v_2\{\text{2D}\}(b)$ trend is as follows: The NJ quadrupole correlated-pair number varies as $(N_{part}/2)N_{bin} \epsilon_{opt}^2 \approx (N_{part}/2)^{7/3} \epsilon_{opt}^2$~\cite{davidhq2} per Fig.~\ref{soft2} (right). The pair {\em ratio} is then $\propto (N_{part}/2)^{1/3} \epsilon_{opt}^2$ and $v_2\{2D\} \propto (N_{part}/2)^{1/6} \epsilon_{opt}$. That expression with coefficient 0.093 is shown as the top-most dashed curve which approximates the exact $v_2\{\text{2D}\}(b)$ expression (solid) well for more-central collisions. The fall-off for more-peripheral collisions results from the simplistic approximation $\bar \rho_0 \propto N_{part}$.  Because of the sharp transition in jet properties (see Fig.~\ref{quad2}, left) that approximation adjusted to accommodate more-central collisions then overestimates $\bar \rho_0$ for more-peripheral collisions, leading to the undershoot of the estimated $v_2\{\text{2D}\}(b)$ trend. The bold solid curve includes the exact TCM expression for $\bar \rho_0(b)$.

In summary, this exercise with LHC $v_n\{2\}(b)$ data demonstrates that deviations of $v_2\{2\} \approx v_2\{\text{EP}\}$ from $v_2\{\text{2D}\}$ data and any nonzero $v_n\{\text{2}\}$ data for $n > 2$ (higher harmonics) are predicted within point-to-point data uncertainties by measured MB dijet features, and there are no significant {\em nonjet} higher harmonics~\cite{multipoles}. Jet-related bias depends on the effective angular acceptance for a given detector, e.g.\ as determined by angular cuts on $(\eta_1,\eta_2)$ and the $v_n$ analysis method. This result suggests that $v_3$ data from small systems as reported in Ref.~\cite{nature} may actually represent MB dijet production in $x$-Au collisions and that $v_2$ may be significantly biased by jets.

\section{quadrupole $\bf p_t$ dependence} \label{quadrupole}

This section also responds to assumption (c) as noted in Sec.~\ref{assumptions}: Fourier amplitudes $v_n$ as conventionally defined actually measure final-state flows as opposed to some other phenomenon (e.g.\ jets).
Of central importance to claims of a perfect liquid or sQGP in \aa\ collisions and recent claims of QGP droplets in asymmetric $x$-A systems is the quality and interpretation(s) of \pt-differential $v_2(p_t)$ data. As noted in Sec.~\ref{vnmeasures} $v_2$ data may represent both nonjet and jet-related mechanisms depending on the preferred analysis method. The accuracy of $v_2$ data with respect to separation of underlying hadron production mechanisms is therefore critical for interpretation. For $v_2(p_t)$ data that do exclude any jet contribution further questions arise as to their correspondence with a flowing dense medium and hydro theory. 

\subsection{Quadrupole $\bf v_2(p_t)$ accuracy and MB dijets} \label{quadacc}

Assessing the accuracy of $v_2\{\text{method}\}(p_t,b)$ data is critical to establishing valid physical interpretations. As demonstrated below different $v_2$ statistical methods can produce dramatically different results from the same basic particle data. The equivalence $v_2\{2\} \approx v_2\{\text{EP}\}$ is established in  Sec.~IV-B of Ref.~\cite{njquad}. 
Other cumulant ranks (e.g.\ $v_2\{4\}$) are not relevant to the analysis in Ref.~\cite{nature}. The main issue for assessing measure accuracy is the response of any $v_2$ method to strong MB dijet contributions to 2D angular correlations. $v_2\{\text{2D}\}$ derived from model fits to 2D angular correlations explicitly excludes the dijet contribution [SS 2D peak on $(\eta,\phi)$ and AS 1D peak on $\phi$]. In contrast, $v_2\{\text{EP}\}$ is the quadrupole Fourier amplitude for {\em all} angular correlations projected onto 1D azimuth and must include any jet contribution to a selected $(\eta_1,\eta_2)$ acceptance. $v_2\{\text{EP}\}$ and $v_2\{\text{2D}\}$ methods for more-central 200 GeV \auau\ are compared below. 

Because the AS dipole is orthogonal to other multipoles the SS 2D peak and NJ quadrupole are the only significant contributors to total quadrupole $V_2^2\{\text{EP}\} \equiv \bar \rho_0 A_Q\{\text{EP}\}$ in more-central \aa\ collisions, leading to 
\bea \label{aqsum}
A_Q\{\text{EP}\} &=& A_Q\{\text{2D}\} + A_Q\{\text{SS}\}.
\eea
$A_Q\{\text{SS}\}$ can be derived from measured SS 2D peak properties as follows. The SS  peak per-particle quadrupole amplitude ($n = 2$ Fourier amplitude) is given by~\cite{multipoles}
\bea \label{sseq}
2 A_Q\{\text{SS}\} &=& F_2(\sigma_{\phi_\Delta}) G(\sigma_{\eta_\Delta};\Delta \eta) A_{\text{2D}},
\eea
where $A_{\text{2D}}$ is the per-particle amplitude of a fitted SS 2D Gaussian with r.m.s.\ widths $(\sigma_{\eta_\Delta},\sigma_{\phi_\Delta})$, $F_2$ is the $n=2$ Fourier amplitude of a unit-amplitude 1D Gaussian on azimuth with width $\sigma_{\phi_\Delta}$
\bea \label{fm}
2F_n(\sigma_{\phi_\Delta}) &=& \sqrt{2/\pi}\, \sigma_{\phi_\Delta} \exp\left( - n^2 \sigma_{\phi_\Delta}^2 / 2\right),
\eea
and $G(\sigma_{\eta_\Delta};\Delta \eta) \leq 1$ is a calculated $\text{2D} \rightarrow \text{1D}$ $\eta$ projection factor defined in Ref.~\cite{multipoles}. 
%
Jet-related quadrupole $A_Q\{\text{SS}\}$ can be identified with ``nonflow''~\cite{gluequad,tzyam,multipoles}. NJ quadrupole $A_Q\{\text{2D}\}$ would correspond to elliptic flow if that phenomenon were relevant. Equation~(\ref{sseq}) can be tested with results from previous 2D angular correlation analysis and published $v_2\{\text{method}\}$ data.

\begin{figure}[h]
	\includegraphics[width=1.65in,height=1.65in]{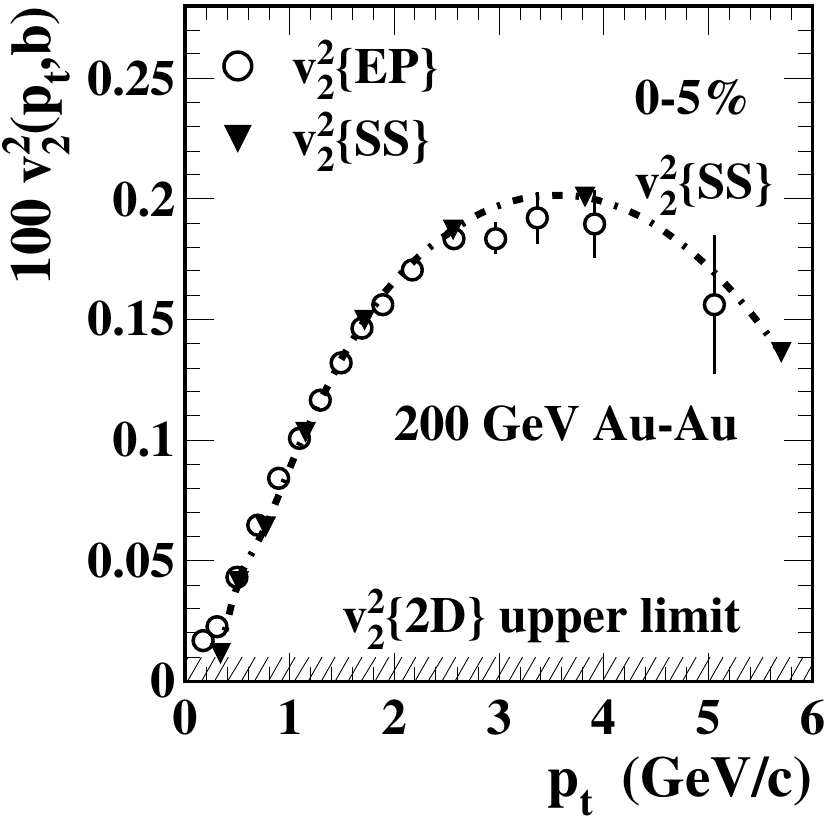}
	\includegraphics[width=1.65in,height=1.68in]{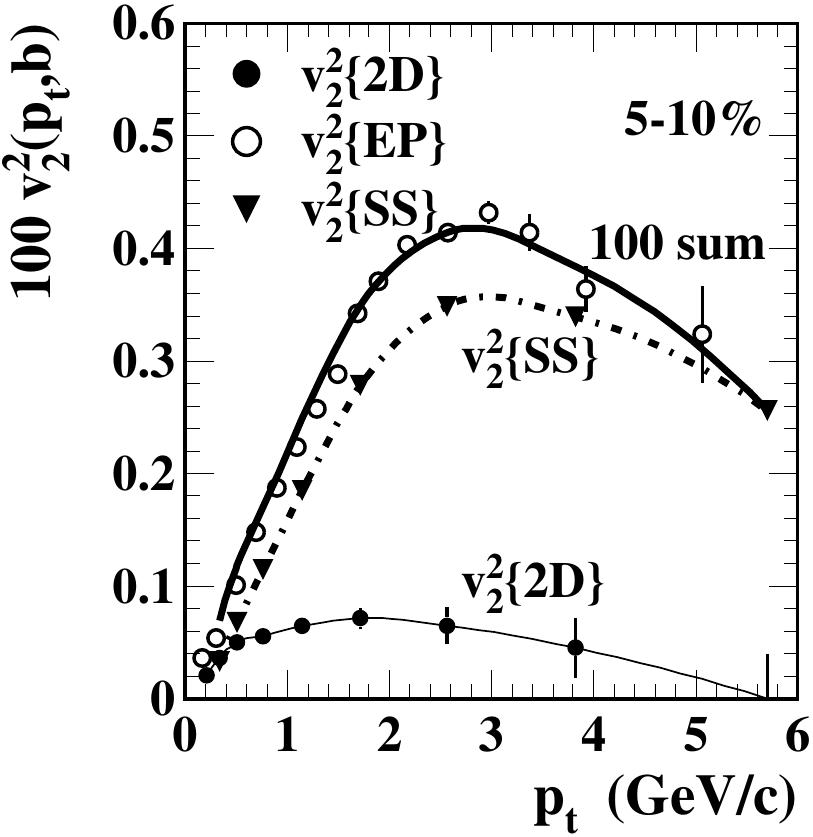}
	\caption{\label{ytcomp1}
		Left: Comparison of jet-related 2D angular correlations represented by $v_2^2\{\text{SS}\}(p_t,b)$ (dash-dotted) and nonjet $v_2^2\{\text{2D}\}(p_t,b)$ (hatched region, upper limit) with published $v_2^2\{\text{EP}\}(p_t,b)$ data (open circles) for 0-5\% central 200 GeV \auau\ collisions~\cite{2004}.
		Right: Similar comparison for 5-10\% central \auau\ collisions showing  close correspondence between the sum $v_2^2\{\text{2D}\}(p_t,b) + v_2^2\{\text{SS}\}(p_t,b)$ (bold solid) and published $v_2^2\{\text{EP}\}(p_t,b)$ data (open circles) as in Eq.~(\ref{aqsum}). 
	} 
\end{figure}

Figure~\ref{ytcomp1} shows published $B_Q\{\text{EP}\} \equiv v_2^2\{\text{EP}\}(p_t,b)$ data (open circles) for two centralities of 200 GeV \auau\ collisions~\cite{2004} compared to $v_2^2\{\text{2D}\}(p_t,b)$ data from Ref.~\cite{v2ptb} (hatched upper limit or solid dots) and ``nonflow'' prediction $v_2^2\{\text{SS}\}$ derived from characteristics of the SS 2D jet peak reported in Ref.~\cite{v2ptb}. $v_2^2\{\text{\text{EP}}\}(p_t,b)$ points are obtained by combining $v_2(b)$ and $v_2(p_t,b)$ data from Ref.~\cite{2004}. The $\eta$ acceptance for data from either method is $|\eta| < 1$. The hatched band at left denotes an upper limit for 0-5\% $v_2^2\{\text{2D}\}$ data. The bold solid curve at right is defined (without the factor 100) by $v_2^2\{\text{EP}\} = v_2^2\{\text{2D}\} + v_2^2\{\text{\text{SS}}\}$ per Eq.~(\ref{aqsum}).  $A_{\text{2D}}$ data that determine $v_2^2\{\text{\text{SS}}\}$ correspond to those in Fig.~\ref{quad2} (left).

Several conclusions can be drawn from these data: 
(i) Strong jet-related angular correlations extend down to at least 0.5 GeV/c whereas the NJ quadrupole trend (for non-central data) extends up to 5 GeV/c or more.
(ii) In the plotting format of Fig.~\ref{ytcomp1} jet-related and nonjet $v_2(p_t)$ trends appear similar whereas the $v_2(b)$ centrality trends are very different (See Fig.~\ref{alice}).
(iii) Measured $v_2^2\{\text{EP}\}(p_t,b)$ data (open circles) are predicted by a combination of nonjet $v_2^2\{\text{2D}\}(p_t,b)$ data (hatched band or solid dots) and a jet-related trend $v_2^2\{\text{SS}\}(p_t,b)$ (dash-dotted) representing the $m = 2$ Fourier component of the SS 2D jet peak when projected onto 1D azimuth as derived from SS 2D peak properties obtained from Refs.~\cite{anomalous,v2ptb}. The trend $v_2^2\{\text{EP}\} = v_2^2\{\text{SS}\} + v_2^2\{\text{2D}\}$ is confirmed for $p_t$-differential data based on the published $\eta$ dependence of 2D angular correlations. There has been no adjustment to accommodate the $v_2^2\{\text{EP}\}$ data.

Strategies have been adopted to reduce ``nonflow'' (mainly MB dijet) contributions to $v_2$ by excluding some parts of a nominal $(\eta_1,\eta_2)$ angular acceptance from $v_2\{\text{EP}\}$ calculations~\cite{multipoles} (e.g.\ Fig.~\ref{alice}). For instance, some $\eta_\Delta$ interval about zero may be excluded from projections onto $\phi_\Delta$ by determining the event plane with large-$\eta$ detectors~\cite{2004,cpanp,alicev2}.  The motivation is exclusion of ``short-range'' jet-related structure from azimuth projections $A_Q\{\text{EP}\}$ based on assumptions about the jet fragment distribution on $\eta$.
Such $\eta$ pair cuts may be less effective at distinguishing jet-related structure from a NJ quadrupole than 2D model fits applied directly within a more-limited central-detector $\eta$ acceptance. In more-central \auau\ collisions the SS 2D jet peak is strongly elongated on $\eta$~\cite{anomalous} and may develop non-Gaussian tails extending over a substantial $\eta$ interval~\cite{trigger}. The effects of $\eta$-exclusion cuts are then uncertain and may have little impact on jet-related biases for $v_2\{\text{method}\}$ data~\cite{multipoles,sextupole}. Such issues are discussed further in following subsections.

\subsection{NJ quadrupole $\bf p_t$ spectra from Au-Au collisions} \label{quadspecc}

Figure~\ref{quad3} shows $v_2(p_t)$ and inferred quadrupole spectra for three hadron species from 200 GeV \auau\ collisions averaged over centrality. Figure~\ref{ytcomp1} shows $v_2(p_t)$ data for unidentified hadrons from two centrality classes that reveal the extent of bias from MB dijets in some $v_2(p_t)$ data. This subsection considers centrality dependence of $v_2(p_t)$ data for unidentified hadrons as it relates to the question of source boost distributions and hydro models. The relevant chain of argument is this:  particle and energy densities should depend on \aa\ centrality, flow fields should depend on density gradients, inferred source boosts should then depend strongly on \aa\ centrality.


As defined, ratio measure $v_2\{\text{method}\}(y_t,b)$ includes the SP $y_t$ spectrum $\bar \rho_0(y_t,b)$ in its denominator. The SP spectrum has a strong jet contribution (spectrum hard component)~\cite{hardspec} that should not relate to hydro models and is therefore extraneous to the azimuth quadrupole problem. Depending on the $v_2$ method the numerator of $v_2(y_t,b)$ may also include significant contributions from jets in the form of a ``nonflow'' bias. To study quadrupole spectra in isolation jet contributions should be removed from numerator and denominator of $v_2(p_t,b)$ by focusing on NJ quadrupole pair amplitude $V_2^2\{\text{2D}\}(y_t,b) = \rho_0(b) \rho_0(y_t,b) B_Q\{\text{2D}\}(y_t,b)$.

It is desirable to define a quantity that reveals source boost evolution as inferred from $v_2\{\text{2D}\}(p_t,b)$ data in as differential a form as possible. Based on $B_Q(y_t,b)$ data in Refs.~\cite{v2ptb,quadspec} a {\em unit-normal} (on \yt) ratio can be defined
\bea \label{qb} 
Q(y_t,b) &\equiv &
\frac{(1/p_t) \bar \rho_0(y_t,b) v_2\{\text{2D}\}(y_t,b)}{\langle 1/p_t \rangle \bar  \rho_0(b) v_2\{\text{2D}\}(b) \times g(b)},
\eea
where 
 data from Ref.~\cite{v2ptb} are of the form $B_Q\{\text{2D}\}(y_t,b) = v_2\{\text{2D}\}(b) v_2\{\text{2D}\}(y_t,b)$. Quantity $v_2\{\text{2D}\}(b)$ corresponds exactly to the $A_Q\{\text{2D}\}(b)$ data shown in Fig.~\ref{soft2} (right).
The $O(1)$ factor $g(b)$ in the denominator is defined and discussed in Ref.~\cite{v2ptb}.
The main goal is inference of the centrality dependence of the quadrupole {\em spectrum shape} represented by $Q(y_t,b)$.

\begin{figure}[h]
	\includegraphics[width=1.65in]{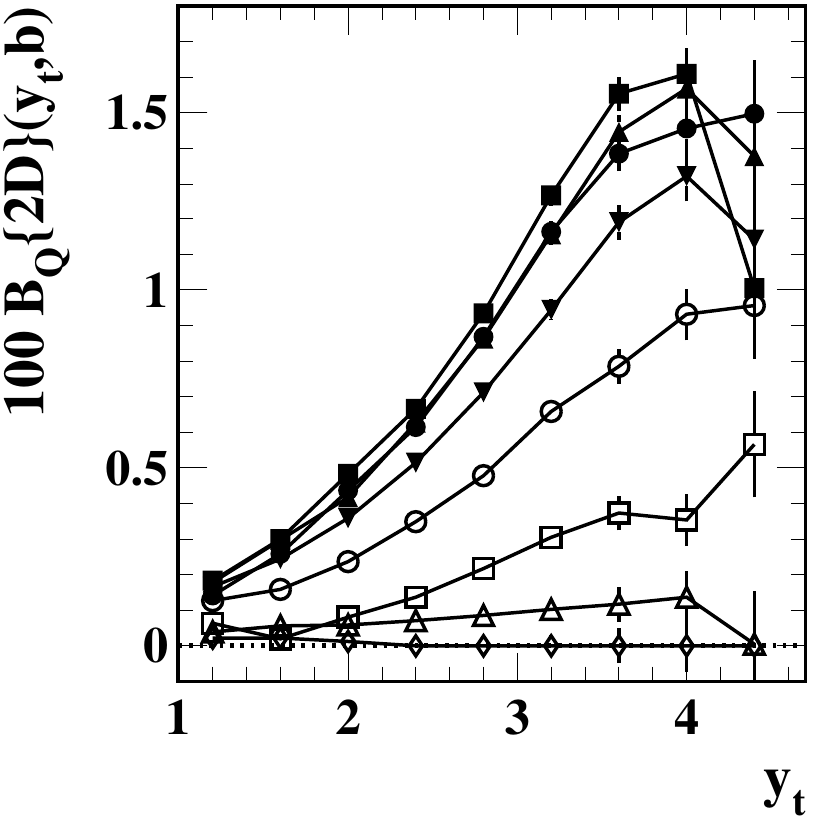}
	\includegraphics[width=1.65in]{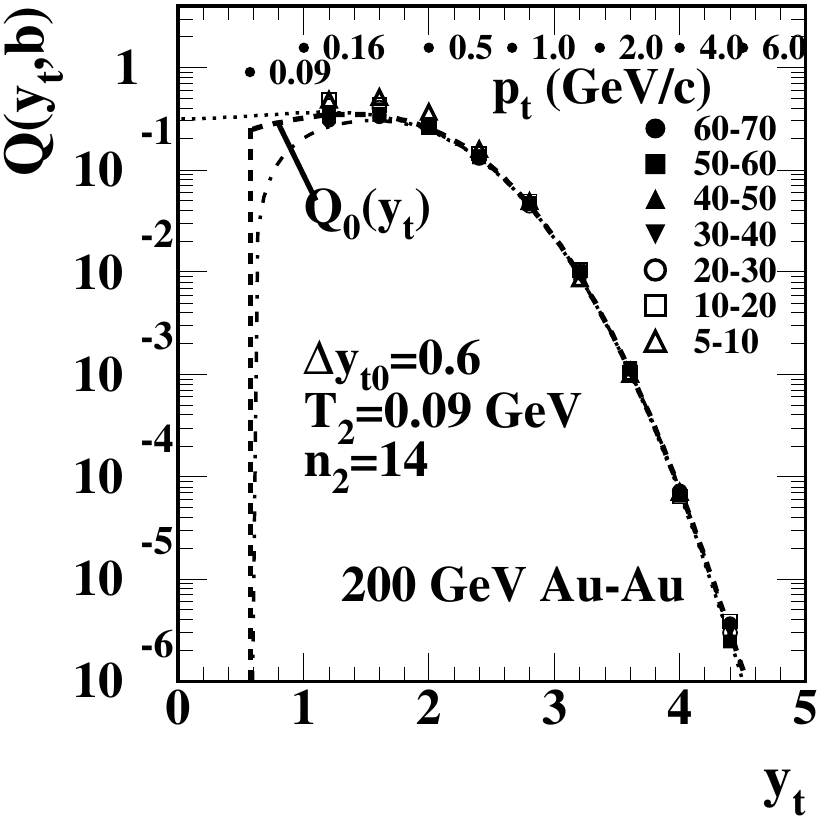}
	\caption{\label{quadspec}
		Left: NJ quadrupole data in the form $B_Q\{\text{2D}\}(y_t,b) = v_2\{\text{2D}\}(b) v_2\{\text{2D}\}(y_t,b)$ for seven centrality classes of 200 GeV \auau\ collisions obtained by model fits to 2D angular correlations~\cite{v2ptb}.
		Right: Unit-normal quadrupole spectra derived from Eq.~(\ref{qb}) and the 200 GeV \auau\ data in the left panel. The spectrum shapes are independent of \auau\ centrality -- all $Q(y_t,b)$ coincide within systematic uncertainties.  The common shape denoted by $Q_0(y_t)$ (dashed) is well described by a boosted L\'evy distribution~\cite{quadspec}.
	} 
\end{figure}

Figure~\ref{quadspec} (left) shows $B_Q(y_t,b)$ data from Ref.~\cite{v2ptb} for seven centrality classes of 200 GeV \auau\ collisions. The data symbols are defined in the right panel. The general trend of monotonic increase with \yt\ (or \pt) is determined in part by a factor $p_t(\text{boost})$ arising from the Cooper-Frye formalism for a boosted \pt\ spectrum with fixed boost (i.e.\ quadrupole spectrum in this case)~\cite{quadspec} and in part by the SP spectrum $\bar \rho_0(y_t,b)$ in the denominator of pair ratio $v_2\{\text{2D}\}(y_t,b)$. The turnover at larger \yt\ (or \pt) arises in part from the jet-related hard component of $ \bar \rho_0(y_t,b)$ in the denominator. Factors $(1/p_t) \bar \rho_0(y_t,b)$ in Eq.~(\ref{qb}) approximately remove those extraneous factors to reveal the sought-after quadrupole spectra.

Figure~\ref{quadspec} (right) shows $Q(y_t,b)$ data for seven centrality bins of 200 GeV \auau\ collisions derived from pair ratios $B_Q(y_t,b)$ in the left panel. 
The $B_Q(y_t,b)$ data are combined with SP spectra $\bar \rho_0(y_t,b)$ and yields $\bar \rho_0(b)$ from Ref.~\cite{hardspec} to compute ratio $Q(y_t,b)$. 
$Q(y_t,b)$ is undefined for the 0-5\% bin since $v_2\{\text{2D}\}(y_t,b)$ for that centrality is consistent with zero for $y_t > 2$ (see Fig.~\ref{ytcomp1}).  Within data uncertainties $y_t$-differential quadrupole spectra are observed to follow a universal shape represented by unit-normal quadrupole spectrum $Q_0(y_t)$ (dashed curve) in the form of a {\em boosted L\'evy distribution} derived previously from MB PID $v_2$ data~\cite{quadspec}. $Q_0(y_t)$ from Ref.~\cite{quadspec} is {\em not a fit} to more-recent $Q(y_t,b)$ data from Ref.~\cite{v2ptb}. 



The results from Ref.~\cite{quadspec} plus Figs.~\ref{quad3} and~\ref{quadspec} taken together suggest that all quadrupole spectra for any \auau\ centrality and for any hadron species follow universal $Q_0(m_t')$ in the boost frame (except for most-central \auau\ collisions where an additional reduction factor $g(b) < 1$ is required).  It is apparent that quadrupole source boost $\Delta y_{t0} \approx 0.6$ for unidentified hadrons (mainly pions) is {\em independent of \auau\ centrality} to good approximation.  The inferred constant source boost independent of \auau\ centrality suggests that the NJ quadrupole phenomenon does not  depend on energy or particle {\em densities}. In comparison with Fig.~\ref{quad2} there is no sensitivity to the sharp transition in jet properties reported in Ref.~\cite{anomalous} that might be attributed to a dense medium. If $v_2$ were indeed a measure of final-state rescattering~\cite{hydro} one must  conclude from these results that conjectured rescattering is independent of system size or particle/energy densities.


\section{Radial flow and jet quenching} \label{radial}

A critical test for claims of a flowing dense QCD medium in high-energy collisions is demonstration of clear evidence for radial flow in single-particle \pt\ spectra and quantitative determination of a boost distribution characterizing the corresponding velocity field. Since any {\em flow-related} azimuthal asymmetries are modulations of radial flow, radial flow inferred from \pt\ spectra should be quantitatively consistent with $v_n$ data inferred from angular correlations. Systematic variation of radial flow as a characteristic of a dense medium should then be quantitatively consistent with jet modifications (jet quenching) inferred from \pt\ spectra and angular correlations. Any viable hydro model should in turn describe radial flow and $v_n$ quantitatively and self-consistently. 
Any significant inconsistencies among the several elements could signal that the concept of a flowing dense medium is not valid for high-energy nuclear collisions. This section presents differential analysis of \pt\ spectra from \auau\ and \ppb\ collisions. The separate contributions from nucleon dissociation and jet production are isolated. Systematic variations with collision \nch\ or A-B centrality are discussed.

\subsection{200 GeV Au-Au PID spectrum data} \label{ppbpid2}

\begin{figure}[h]
	\includegraphics[width=1.65in]{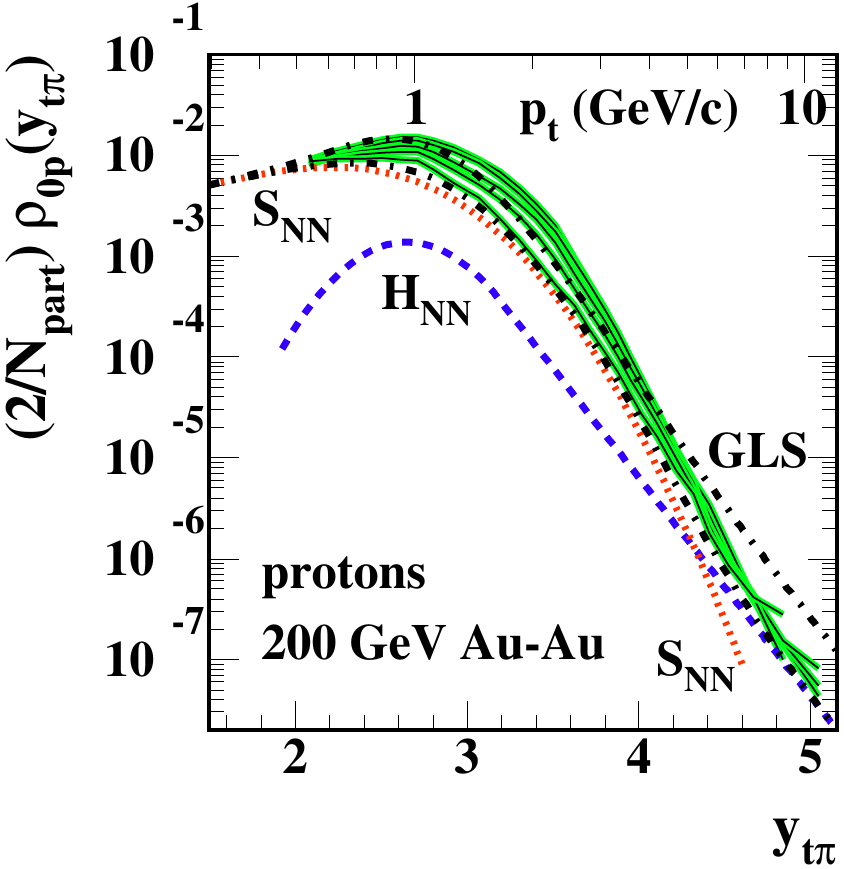}
	\includegraphics[width=1.65in]{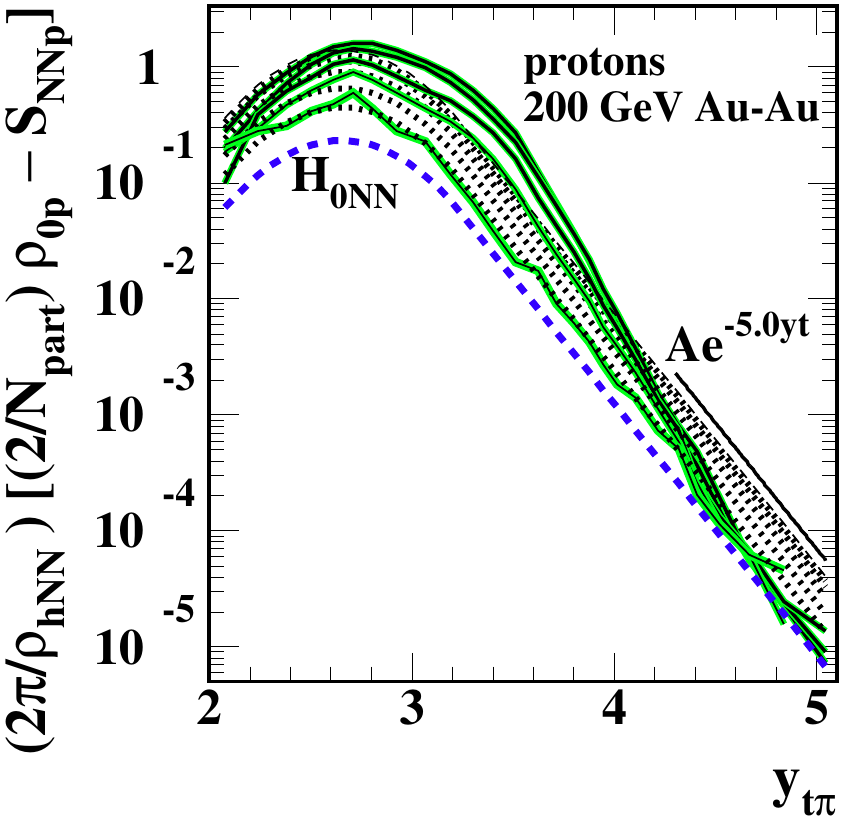}
	\caption{\label{auauspec} (Color online)
Left:  \yt\ spectra (green solid) for identified protons from five centrality classes  of 200 GeV \auau\ collisions normalized by the number of participant nucleon pairs $N_{part}/2$~\cite{hardspec}. The TCM soft-component model function is $S_{NN}$ (red dotted) and the hard-component model function is $H_{NN}$ (blue dashed).
Right: Spectrum hard components derived from data in the left panel (green solid). $H_{0NN}(y_t)$ (blue dashed) is the {\em unit-normal} TCM hard-component model function: a Gaussian with exponential (power-law on \pt) tail. Data suppression at larger \yt\ and enhancement at lower \yt\ are apparent relative to TCM GLS references (dotted curves).
	} 
\end{figure}

Figure~\ref{auauspec} (left) shows \yt\ spectra (green solid) for identified protons from five centrality classes (0-12\%, 10-20\%, 20-40\%, 40-60\% and 60-80\%) of 200 GeV \auau\ collisions normalized by the number of participant nucleon pairs $N_{part}/2$. According to the TCM normalized spectrum soft components should then coincide at low \yt, which is observed to be the case. $S_{NN} = \bar \rho_{sNN} \hat S_{0NN}(y_t)$ (red dotted) and $H_{NN}= \bar \rho_{hNN} \hat H_{0NN}(y_t)$ (blue dashed) are the TCM soft and hard model functions~\cite{hardspec}. The upper dash-dotted curve is a GLS reference for the most-central \auau\ data ($\nu \approx 5.5$). Data deviations from the GLS reference indicate jet modifications within more-central \auau\ collisions. Spectrum data indicate that the jet-related spectrum hard component dominates proton spectra above $y_t \approx 2.3$ ($p_t \approx 0.7$ GeV/c). The sharp transition in jet properties reported in Ref.~\cite{anomalous} occurs within the 40-60\% centrality interval. Parts of the same spectra appear in Fig.~\ref{gale1} (right) plotted on linear \pt.

Figure~\ref{auauspec} (right) shows spectrum hard components in the form $\nu H_{AA}(y_t,b) / \bar \rho_{hNN}$ inferred from data in  the left panel as indicated on the $y$-axis label. The factor $2\pi /\bar \rho_{hNN}$ (with $\bar \rho_{hNN} \equiv \alpha \bar \rho_{sNSD}^2$) makes the data consistent with unit-normal limiting case $\hat H_{0NN}$ (dashed) for peripheral collisions. $\nu H_{AA}(y_t) / \bar \rho_{hNN}$ for the most-peripheral centrality class is consistent with the corresponding GLS reference $\nu H_{0NN}(y_t)$  with $\nu \approx 2$. More-central data increasingly deviate from their GLS references (dotted): spectra are suppressed above 4 GeV/c (consistent with $R_{AA}$ results) but are {\em enhanced} in the interval $p_t \approx 2$-$3$ GeV/c ($y_t \approx 3.5$), a result concealed by the conventional $R_{AA}$ ratio. The jet-related lower-\pt\ enhancement for baryons corresponds to the baryon/meson ``puzzle''~\cite{puzzle} (see Sec.~IX of Ref.~\cite{hardspec} for relation to TCM hard components). Note that all proton data spectra are consistent with their GLS references below $p_t \approx 1$ GeV/c ($y_t \approx$ 2.7) corresponding to the proton mass. There is no evidence in these data for radial flow which would be indicated by soft components boosted on \yt. In particular, there is no reduction of spectra at lower \pt\ (compared to the GLS reference) that would indicate a significant particle source boost (see Sec.~\ref{hydrospec} and Fig.~\ref{nature1}, right).

\subsection{5 $\bf TeV$ $\bf p$-$\bf Pb$ PID spectrum data} \label{ppbpid1}

\begin{figure}[h]
	\includegraphics[width=3.3in]{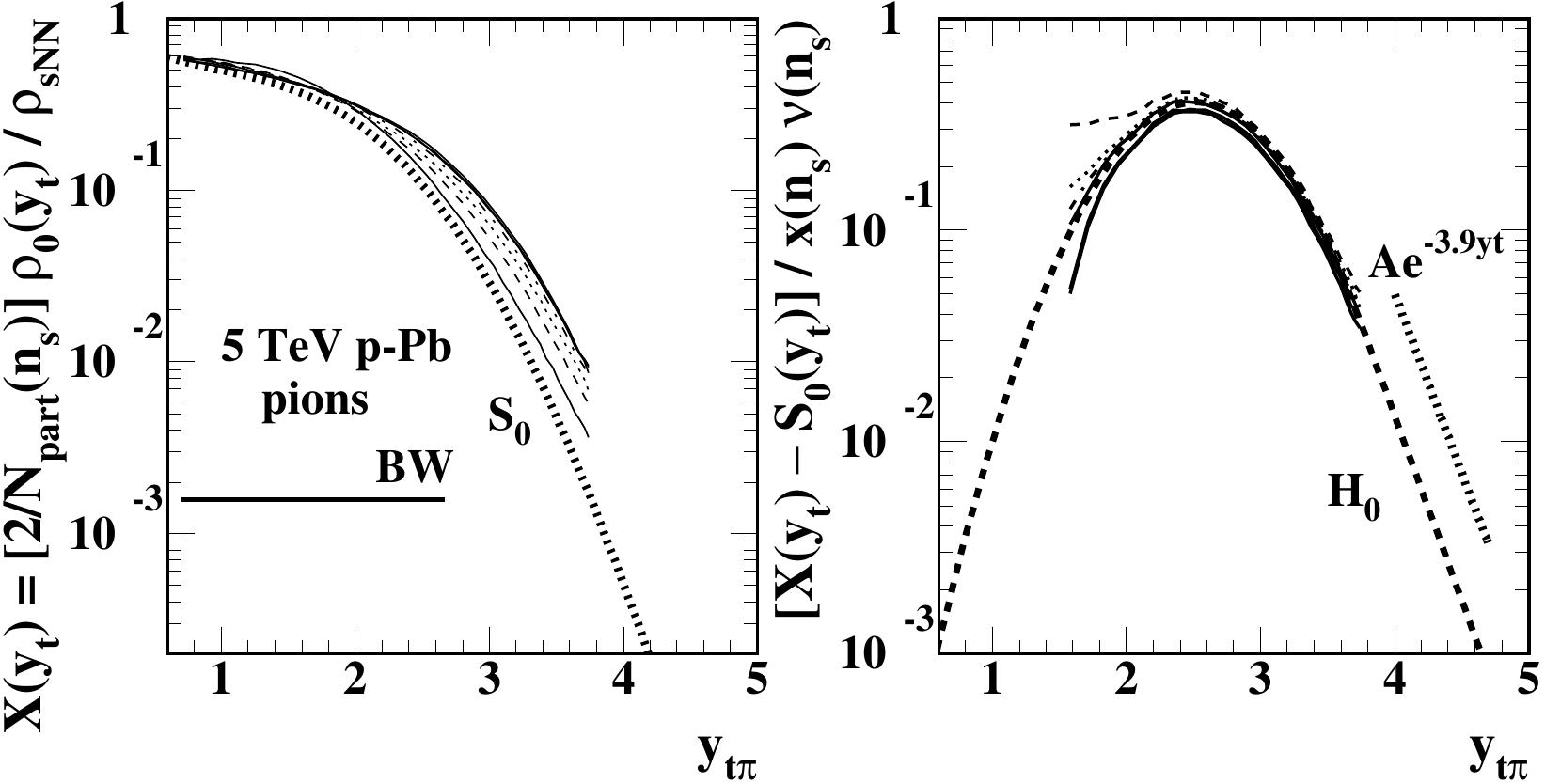}
	\caption{\label{pions}
		Left: Identified-pion spectra for 5 TeV \ppb\ collisions from Ref.~\cite{aliceppbpid} transformed to \yt\ with Jacobian $m_{t\pi} p_t / y_{t\pi}$ and normalized by TCM quantities from Ref.~\cite{ppbpid} (7 thinner curves of several styles). $\hat S_0(y_t)$ (bold dotted) is the TCM soft-component model.
		Right: Difference $X(y_t) - \hat S_0(y_t)$ normalized by $x(n_s) \nu(n_s) \approx \alpha \bar \rho_{sNN} \nu(n_s)$ using TCM values from Ref.~\cite{ppbpid} (thinner curves). The bold dashed curve is the TCM hard-component model $\hat H_0(y_t)$ with exponential tail.
	}  
\end{figure}

Figure~\ref{pions} (left) shows identified-pion \yt\ spectra for 5 TeV \ppb\ collisions  from a TCM analysis reported in Ref.~\cite{ppbpid}. Measured PID spectra from Ref.~\cite{aliceppbpid} have been multiplied by $2\pi$ to be consistent with the $\eta$ densities used in this study. The spectra are then normalized by soft-component density $\bar \rho_{s} = (N_{part}/2) \bar \rho_{sNN}$ with TCM parameters reported in Ref.~\cite{ppbpid}. The normalized spectra $X(y_t)$ are then compared with spectrum soft-component model $\hat S_0(y_{t\pi})$ (bold dotted) -- a L\'evy distribution defined on $m_{t\pi}$ (with parameters $T = 145$ MeV and $n = 8.5$) transformed to $y_{t\pi}$ that also describes unidentified hadrons from 5 TeV \pp\ collisions as reported in Ref.~\cite{alicetomspec}. The same parameter values are also used as a GLS reference for \pbpb\ spectra as in Fig.~\ref{nature1}. The solid line labeled BW marks the $y_{t\pi}$ (\pt) interval corresponding to blast-wave model fits as reported in Ref.~\cite{aliceppbpid}.

Figure~\ref{pions} (right) shows difference $X(y_t) - \hat S_0(y_t)$ normalized by $x(n_s) \nu(n_s)$ using TCM values as reported in Ref.~\cite{ppbpid}. The data are so normalized to be compatible with the plotting format in Fig.~\ref{auauspec} (right), with $x(n_s) = \alpha \bar \rho_s$. The result should be directly comparable to a \pp\ spectrum unit-normal hard-component model in the form $\hat H_0(y_t)$ [or $H_{0NN}(y_t)$ in \aa]. The bold dashed curve is $\hat H_0(y_t)$ with model parameters  $(\bar y_t,\sigma_{y_t},q)$ for pions. 
Any deviations from $\hat H_0(y_t)$ in the right panel would indicate the ``fit'' residuals for the model, but the TCM is not a free fit to individual spectra; it is highly constrained with only a few adjustable parameters (see Sec.~\ref{predict2}). 

\begin{figure}[h]
	\includegraphics[width=3.3in]{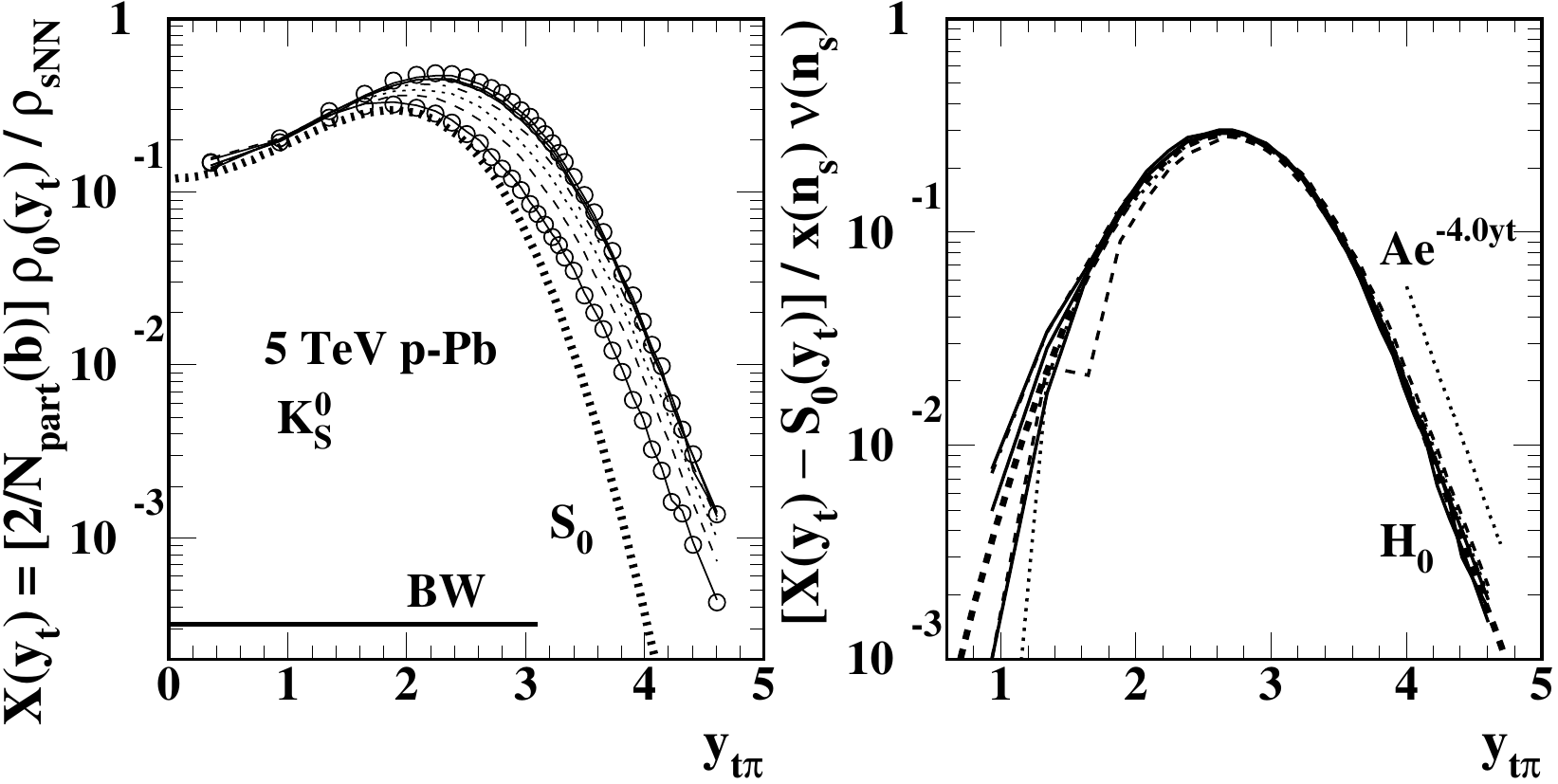}
	\caption{\label{kch}  
	Left:	Identified-$K_S^0$ spectra for 5 TeV \ppb\ collisions from Ref.~\cite{aliceppbpid} transformed to $y_{t\pi}$. $\hat S_0(y_t)$ (bold dotted) is the TCM soft-component model.
	Right:  Difference $X(y_t) - \hat S_0(y_t)$ normalized by $x(n_s) \nu(n_s) \approx \alpha \bar \rho_{sNN} \nu(n_s)$ using TCM values from Ref.~\cite{ppbpid} (thinner curves). The bold dashed curve is the TCM hard-component model $\hat H_0(y_t)$ with exponential tail.
	}  
\end{figure}

Figure~\ref{kch} shows neutral $K^0_s$  spectra from Ref.~\cite{aliceppbpid} as processed in Ref.~\cite{ppbpid} in the same manner as for charged pions.
$K^\pm$ spectra are consistent with  $K^0_S$ spectra within data uncertainties as reported in Ref.~\cite{aliceppbpid}. The $K^0_S$ data subtend the spectacular interval $p_t \in [0,7]$ GeV/c.  The data below $y_{t\pi} \approx 1.2$ ($\approx 0.2$ GeV/c) demonstrate that only a {\em fixed} soft component independent of \ppb\ centrality contributes in that interval, and a L\'evy distribution on $m_{tK}$ (bold dotted) describes the data well. Individual data points (open circles) are shown for the lowest and highest $\bar n_{ch}$ classes. Data for the other classes are plotted as joined line segments of varied styles. A usable estimate for $\hat H_0(y_t)$ obtained down to $y_t = 1.2$ ($p_t \approx 0.2$ GeV/c) confirms that the TCM hard component drops off sharply below its mode. These  $K^0_S$ data thereby strongly support a MB jet-spectrum lower bound near 3 GeV~\cite{fragevo}. 

\begin{figure}[h]
	\includegraphics[width=3.3in]{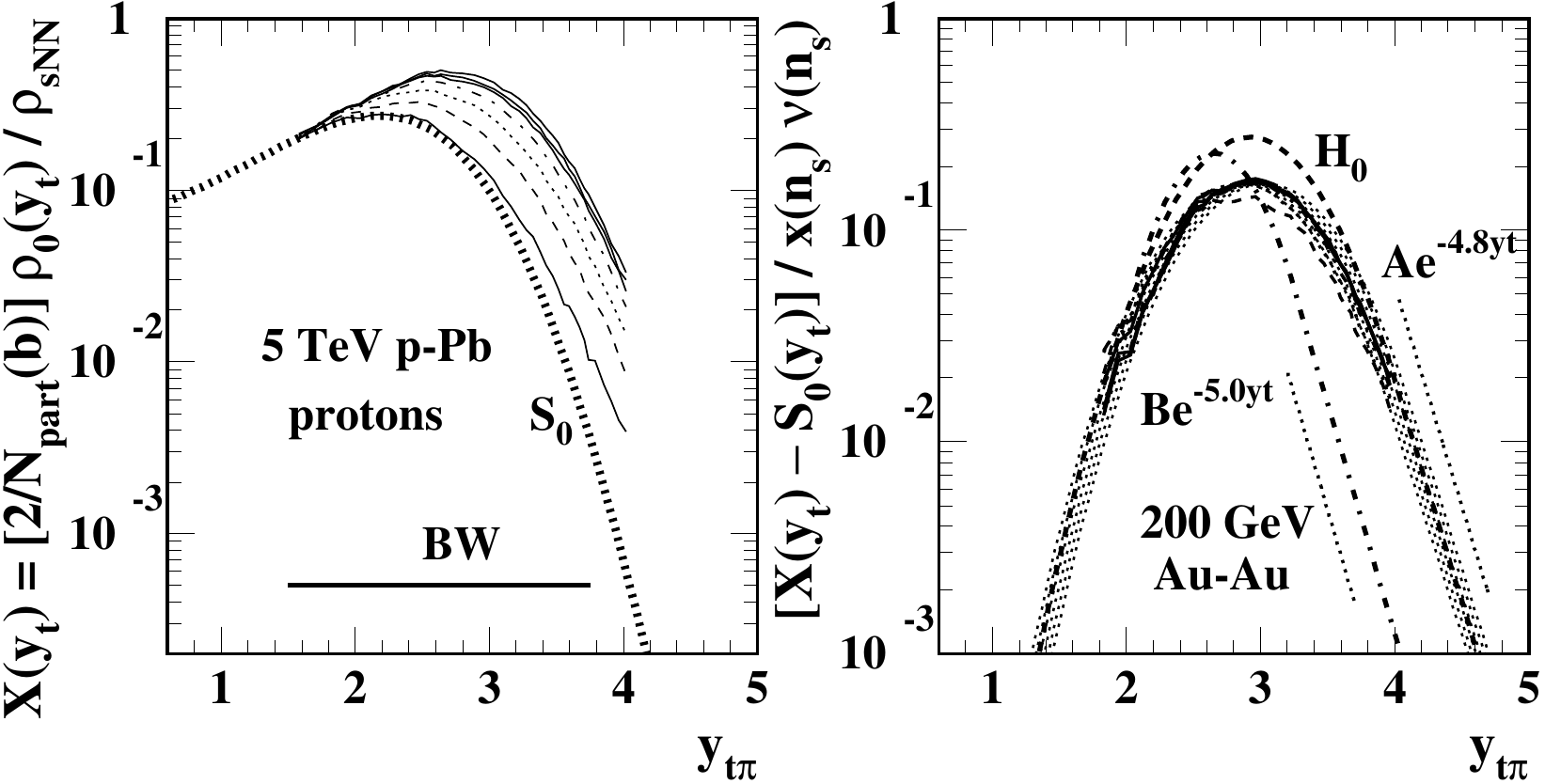}
	\caption{\label{protons}
	Left:	Identified-proton spectra for 5 TeV \ppb\ collisions from Ref.~\cite{aliceppbpid} transformed to $y_{t\pi}$. $\hat S_0(y_t)$ (bold dotted) is the TCM soft-component model.
Right:  Difference $X(y_t) - \hat S_0(y_t)$ normalized by $x(n_s) \nu(n_s) \approx \alpha \bar \rho_{sNN} \nu(n_s)$ using TCM values from Ref.~\cite{ppbpid} (thinner curves). The bold dashed curve is the TCM hard-component model $\hat H_0(y_t)$ with exponential tail. The dash-dotted curve is the TCM hard component for proton spectra from 200 GeV \auau\ collisions.
	}   
\end{figure}

Figure~\ref{protons} (left) shows identified-proton spectra from 5 TeV \ppb\ collisions corresponding to the previous two figures. The soft-component model $\hat S_0(y_t)$ (bold dotted) is similar to that for $K_S^0$. These proton spectra make clear the dominance of jet fragments scaling $\propto \bar \rho_{hNN}N_{bin}$ compared to $\hat S_0(y_t)$ at and above $y_t = 2$ ($p_t \approx 0.5$ GeV/c).

Figure~\ref{protons} (right) shows hard components inferred from spectrum data in the left panel. The expected proton hard-component model function $\hat H_0(y_t)$ for 5 TeV \pp\ collisions is given by the bold dashed curve. \ppb\ data are systematically suppressed (for protons only) near the mode as reported in Ref.~\cite{ppbpid}. The reason is not known. Spectrum data for pions, kaons and Lambdas are consistent with TCM predictions, e.g.\ Figs.~\ref{pions} and \ref{kch} (right).

The dash-dotted curve in the right panel is the TCM spectrum hard-component model for 200 GeV \auau\ collisions as shown (dashed) in Fig.~\ref{auauspec} (right). The difference above the mode is expected based on evolution of \pp\ spectrum hard components~\cite{alicetomspec} (i.e.\ underlying jet spectra) with \nn\ collision energy~\cite{jetspec2}. The properties of \ppb\ TCM spectrum hard components are quantitatively consistent with expectations for MB jet fragmentation.

\subsection{Conclusions: radial flow and jet quenching}

TCM analysis of \yt\ spectra from \pa\ and \aa\ collisions relates to questions of radial flow and jet modification in high-energy nuclear collisions. In the conventional narrative both conjectured phenomena should be coupled via the presence of a dense and flowing QCD medium. Their manifestations should therefore be closely correlated.

Radial flow is typically inferred from blast-wave (BW) model fits to spectra~\cite{blastwave} within a limited \pt\ interval defined by data availability and/or fit quality (see horizontal lines labeled BW in figures above). Increased ``flattening'' of spectra (with hadron mass, collision centrality or collision energy) is interpreted as an indicator for radial flow~\cite{hydro}. Inferred flow values in the form $\beta_t$ range from 0.25 for \pp\ collisions to 0.60 for central \auau\ collisions~\cite{betatauau}. BW fit exercises typically do not acknowledge significant jet contributions to fitted spectra.

Differential analysis of spectrum structure (see examples in this section) reveals no evidence for radial flow in any collision system. Radial flow should be manifested as systematic boost (shift to higher \yt) of the nonjet contribution to spectra (i.e.\ the TCM soft component). In fact, spectrum soft components are remarkably stable. Note that in Fig.~\ref{protons} (left) the blast-wave fit interval (BW straight line) is centered on the hard-component jet peak that dominates proton spectra within that interval.

Evolution of \ppb\ spectrum structure with \nch\ or centrality also reveals no changes in spectrum hard-component {\em shapes} within data uncertainties. There is no evidence for jet modification (jet quenching) in \ppb\ collisions, in contrast to clear evidence for jet modification in \auau\ collisions as in Fig.~\ref{auauspec} (right). Increase of dijet production proportional to number of \nn\ binary collisions accounts for \pt\ spectrum ``flattening'' conventionally attributed to radial flow. In Fig.~\ref{protons} (right) evolution of the spectrum hard component with collision energy is consistent with evolution of the underlying jet spectrum as inferred from event-wise reconstructed jets~\cite{jetspec2}. Those conclusions are consistent with the message inferred early in the RHIC program from \dau\ collision data interpreted as a {\em control experiment}~\cite{daufinalstate}

The combination of no radial flow and no jet modification in \ppb\ collisions implies no dense QCD medium, no QGP formation, in small asymmetric collision systems. The absence of radial flow also implies no significant {\em nonjet} $v_3$ or $v_4$ as azimuthal modulations of radial flow. \ppb\ spectrum data thus rule out the possibility of QGP droplet formation in $x$-Au as claimed in Ref.~\cite{nature}.

\section{Transport: Hydro models} \label{hydro}

This section responds to assumption (d) as noted in Sec.~\ref{assumptions}: Hydrodynamic models assuming a low-viscosity flowing dense QCD medium, including ``plasma droplets'' in small systems, have some relation to real collision dynamics. There are several issues: What can hydro models truly predict? Are hydro models simply fitted to {\em selected} data features  by specific model choices and parameter variations? What data features are actually compared with nominal hydro predictions? Do some claimed flow manifestations arise from MB dijets?

\subsection{Conventional hydro narrative}

The general structure of hydro models is summarized in Refs.~\cite{hydro,galehydro}. From a set of initial conditions (matter-density and temperature distributions) a locally-equilibrated ``hot and dense'' partonic medium evolves in response to pressure gradients such that a flow field develops. The dense QCD medium (QGP) expands and cools until a freezeout condition is met whereupon the medium hadronizes via a phase transition. Collective expansion may continue until the hadron density is low enough to achieve kinetic decoupling and free streaming. Detected hadrons provide information on hydro evolution and the state of the velocity field upon hadron decoupling.

A hydro model includes several complex parts: Initial conditions (particle and energy densities, temperature distribution) are based on assumptions about pQCD parton scattering~\cite{hydro} or a saturation model~\cite{glasmafluc}. Hydro evolution (ideal or viscous) requires an EoS that may change with decreasing system density (e.g.\ from QGP to hadron resonance gas). A phase transition from partonic to hadronic matter must be modeled, and possible hadronic rescattering (``afterburner'') may be included.

Hydro model outputs may be compared to hadron spectra and two-particle correlations. \pt\ spectra are assumed to consist of two components divided by a transition point near 3 GeV/c: a soft exponential component at lower \pt\ representing a flowing medium and a hard power-law component at higher \pt\ representing jet production~\cite{hydro}. That description should not be confused with the TCM wherein hard and soft spectrum components may overlap extensively on \pt. So-called blast-wave fits over a limited interval at lower \pt\ are used to infer a temperature at kinetic decoupling $T_{kin}$ and a mean flow velocity $\beta_t$~\cite{blastwave}. The same partition is assumed for experimental analysis as in Ref.~\cite{phenixflow}: ``For low momentum particles ($p_T \le 3$ GeV/c), this [azimuthal] anisotropy is understood to result from hydrodynamically driven flow of the Quark-Gluon Plasma....'' That assumption is one basis for measurement of ``higher-order flow harmonics....''

Hydro comparisons to two-particle correlations emphasize ``azimuthal asymmetry'' in pair distributions on azimuth difference $\Delta \phi = \phi_1 - \phi_2$. It is assumed that particle production in primary [\nn] collisions is isotropic (inconsistent with \pp\ data as demonstrated in Sec.~\ref{ppquadd}) so azimuth asymmetries require final-state rescattering~\cite{hydro}. 
Azimuthal anisotropy is measured by Fourier coefficients $v_n$ (see Sec.~\ref{vnmeasures}), and it is asserted that if hydro models describe $v_n$ data the QGP must be a fluid. If ideal hydro is a good approximation to $v_2$ data it is concluded that a perfect liquid is formed~\cite{perfect}.

\subsection{What should be expected of hydro models}

A viable hydro model for flows in high-energy nuclear collisions should simultaneously and self-consistently describe all nominal flow-related data features for a given collision system comprehensively and with a single parametrization. The same model should address evidence for radial flow in \pt\ spectra and evidence for elliptic flow (e.g.\ $v_2$ data) and possibly other flow-related azimuthal asymmetries in angular correlations. 

The predicted velocity field of an expanding dense medium at kinetic decoupling should be described via an azimuth-dependent radial-boost distribution directly comparable with data through spectrum and angular-correlation features~\cite{quadspec}. A candidate hydro model should be compared only with those data features in spectra and two-particle correlations that are accurately distinguished from MB dijet contributions. 

If a hydro model is invoked for smaller collision systems (e.g.\ \pp\ or $x$-A) the predicted data trends should be consistent with extrapolation from larger \aa\ systems. In particular, the same model invoked for $x$-A and \aa\ collisions should be required to describe the \nch\ trend for NJ quadrupole $v_2\{\text{2D}\}$ in 200 GeV \pp\ collisions~\cite{ppquad}.

\subsection{Complex hydro models and data fitting}

According to Ref.~\cite{hydro} a hydro theory calculation may be composed of several parts: initial conditions, hydro evolution following one or more equations of state,  a freezeout surface determined by a condition such as local temperature, hadron production from a hydro-evolved stress-energy tensor and a possible ``afterburner'' that models hadronic rescattering. For each model component multiple choices are available for implementation, and for each implementation a range of parameter values may compete for data description. There are thus many possibilities for constructing a hydro theory.

A stated goal in the formulation of a hydro model is ``the extraction of properties of the created quark-gluon plasma and to learn about the initial state and its fluctuations''~\cite{galehydro}. The process depends on adjusting a theory Monte Carlo to accommodate selected data:  ``Comparisons of experimental data to hydrodynamical simulations thus allows to extract properties of the matter created in heavy-ion collisions...''~\cite{galehydro}.
 An example is provided by this description in Ref.~\cite{hydro}: ``...a stiffer EoS and a lower freeze-out temperature lead to larger $p_T$-averaged flow if nothing else in the model is changed. This also changes the single particle distributions. If these are still required to fit the data, additional changes are required. For example, a stiffer EoS usually necessitates a higher freeze-out temperature. The combined effect largely cancels and the final $p_T$-averaged anisotropy is almost unchanged in semi-peripheral collisions in which a hydrodynamical description works best.'' The quoted text describes a process of fitting competing complex models to selected data, consistent with further descriptions below.

``Recently, event-by-event hydrodynamic calculations have produced a plethora of predictions and explanations of a wide range of experimental data, for which the inclusion of initial state fluctuations is the essential ingredient''~\cite{galehydro}. In connection with results appearing in Fig.~\ref{gale2} of the present study: ``This agreement [between theory Ref.~\cite{galehydro2} and data Ref.~\cite{alice}] indicates that initial state fluctuations...are the main ingredient to explain the measured flow coefficients''~\cite{galehydro}. Whereas that theory result may be {\em sufficient} to describe selected data it is not demonstrated that hydro theory is {\em necessary}, or even able at all, to describe more-comprehensive data in comparison with possibly simpler models as in the present study.

Significant uncertainties are admitted: e.g.\ for hadron production~\cite{galehydro} (a) the choice between ``a simple Cooper-Frye description...or coupling to a hadronic rescattering simulation'' and (b) ``the prescription of the conversion of energy densities to particle degrees of freedom.'' Uncertainty (b) applies in reverse to (c) defining initial conditions -- the conversion of particle degrees of freedom to energy densities: ``Every wounded nucleon...is then assigned an energy or entropy density...''~\cite{galehydro} ``...either given by physical arguments or fixed by {\em comparing...the results with experimental data} [emphasis added]''~\cite{hydro}.

The fitting procedure is described explicitly: ``Some effort has been directed at developing a framework for precision determination of different parameters...in a detailed multi-parameter $\chi^2$ fit to multiple experimental quantities''~\cite{galehydro}.  ``...a suitable choice of freeze-out surface allows one to fit the $p_T$ spectra even if the kinetic freeze-out temperature is taken to be the same...as the chemical freeze-out temperature''~\cite{hydro}.  ``Studies exploring the effects of initial time, the shape of initial distributions, and the value of $T_{dec,chem}$ while using two separate freeze-outs are needed to settle the issue [of decoupling]''~\cite{hydro}. Such fitting programs are considered a success: ``The current state-of-the-art viscous hydrodynamic calculations with the most advanced models for the fluctuating initial state show remarkable agreement with {\em flow measurements} [emphasis added] from both RHIC and LHC [20]''~\cite{galehydro}. But the assumption that $v_n(p_t,b)$ data represent flows can be questioned, as demonstrated in the present study.  Note that ``[20]'' in the quoted text denotes Ref.~\cite{galehydro2} of the present study from which figures in Sec.~\ref{predict} are taken.

\subsection{Assumptions about hydro-related data} \label{assumptions2}

Comparisons of hydro theory with data tend to include selected subsets of data plotted in preferred formats based on critical assumptions about hadron production mechanisms.  \pt\ spectra are conventionally assumed to have a ``two-component'' (not to be confused with TCM nonjet and jet-related components) structure relative to some boundary \pt\ value. A typical spectrum model is described in Ref.~\cite{hydro}: ``At RHIC, the transition from steep exponential to a shallower power behavior takes place at $p_T \approx 3$ GeV[/c].'' The fraction with $p_t > 3$ GeV/c is said to be small and to originate from ``fragmentation of high-energy partons'' whereas ``low-energy partons...{\em are assumed to thermalize} [emphasis added].'' ``The range where hydrodynamics can be used to describe the hadron spectra [e.g.\ BW model fits below 3 GeV/c] is indicated clearly in Figure 7 [of Ref.~\cite{hydro}],'' approximated there by an exponential suggesting thermal equilibrium.  That description can be contrasted with the discussion of Fig.~\ref{nature1} below, especially the exponential function there.

Hydro theory comparisons with two-particle angular correlations emphasize projection of 2D angular correlations on $(\eta,\phi)$ onto 1D azimuth $\phi$, resulting structure (denoted as azimuthal anisotropy or asymmetry) conventionally measured by amplitudes $v_n$ of a Fourier series. In Ref.~\cite{hydro} the following observations are made: ``...the anisotropy of the final particle distribution [in effect including all correlation structure] is a measurement of the frequency of rescatterings during the dense phase of the collision.'' ``The first and second coefficient..., $v_1$ and $v_2$ are usually referred to as directed and elliptic flow....'' The last statement is formally incorrect (see Sec.~\ref{auauquad}).

So-called mass ordering of $v_2(p_t,b)$ data at lower \pt\ is seen as an indicator that collective motion (flows) must play a role in collisions.  ``...measurements of $v_2$ for different particle species [show] a clear mass-splitting, predicted by hydrodynamics and due to the emergence of all particles from a single velocity field''~\cite{galehydro}. ``In the extreme case of a {\em thin expanding shell} [emphasis added], this reduction [of $v_2$ with hadron mass for fixed \pt] can be so strong that it reverse the sign of the anisotropy and $v_2$ becomes negative. ...a more realistic velocity distribution weakens the reduction of $v_2$ at low $p_T$, but the mass ordering...remains''~\cite{hydro}.  

However, detailed {\em quantitative} analysis of the data feature referred to qualitatively as mass ordering or mass splitting reveals a boost distribution inconsistent with the hydro narrative, as described in Sec.~\ref{interpretations}. And $v_2(p_t,b)$ data for baryons actually provide strong evidence for the thin expanding shell scenario.  In Fig.~10 of Ref.~\cite{2004} $v_2(p_t)$ Lambda data (solid blue dots) are compared to a hydro result (solid blue curve) that exhibits the monotonic approach to zero corresponding to ``...a more realistic velocity distribution.''  The earlier Lambda data already suggest a fixed boost. More-recent Lambda data are reported in Fig.~12 (d) (note the inset) of Ref.~\cite{2008}.

\begin{figure}[h]
	\includegraphics[width=1.65in,height=1.68in]{boost1anewer}
	\includegraphics[width=1.65in,height=1.65in]{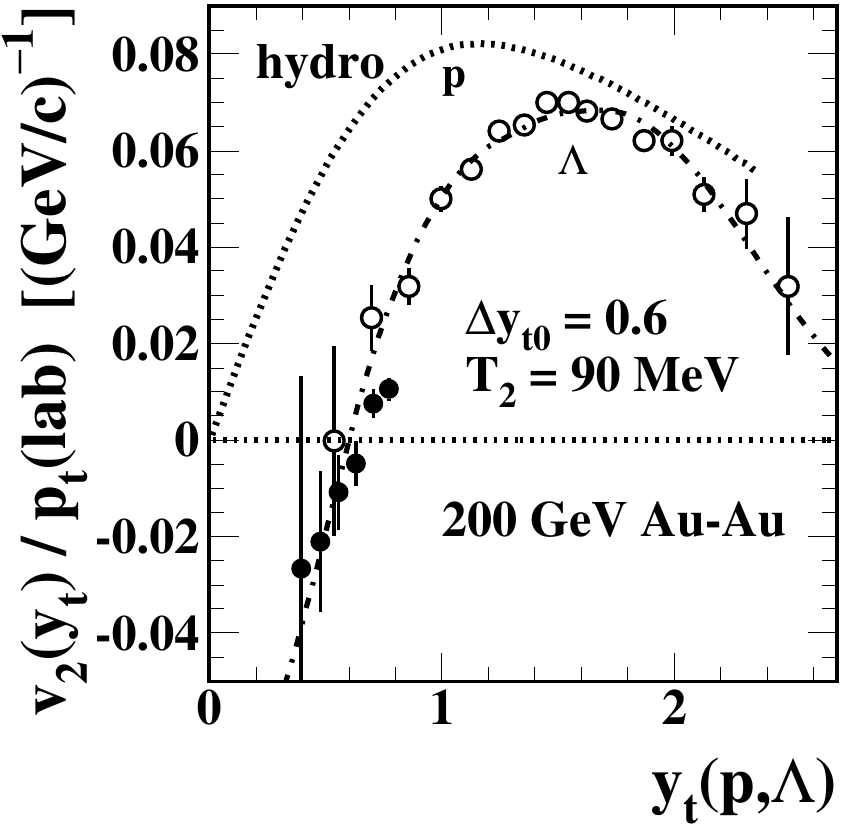}
	\caption{\label{shell}
		Left: Published PID $v_2(p_t)$ data for three hadron species from Refs.~\cite{v2pions,v2strange} plotted in the conventional format. The curves are transformed from a universal quadrupole spectrum described in Sec.~\ref{interpretations}.
		Right:  $\Lambda$ data from the left panel (open points) are replotted in the form $v_2(p_t)/p_t$ vs transverse rapidity $y_t$ calculated with the Lambda mass. The solid points are 0-10\% Lambda data taken from the inset in Fig.~12 (d) of Ref.~\cite{2008} treated similarly. The dotted curve is a viscous hydro prediction~\cite{rom}. The dash-dotted curve is the common quadrupole spectrum on $m_{t}(\text{boost})$ (solid curve) in Fig.~\ref{quad3} (right) back-transformed to $y_{t\Lambda}(\text{lab})$ with fixed $\Delta y_{t0} = 0.6$.
	}  
\end{figure}

Figure~\ref{shell} compares PID $v_2(p_t)$ vs \pt\ data on the left (repeat of Fig.~\ref{quad3}, left) with Lambda data reported in Ref.~\cite{2008} plotted here on the right as $v_2(p_t) / p_t(\text{lab})$ vs $y_{t\Lambda}$, an intermediate step in derivation of a quadrupole spectrum (as described in Sec.~\ref{interpretations}). The newer Lambda data (solid dots) demonstrate that $v_2(p_t)$ for baryons is consistent with a negative-going trend as for a thin shell~\cite{hydro}. The Lambda data clearly demonstrate the extent of the negative trend because the hadron \pt\ acceptance lower bound is located on \yt\ (for larger masses) well below the fixed source boost $\Delta y_{t0} = 0.6$. The boost distribution indicated by Lambda $v_2(p_t)$ data is thus inconsistent with Hubble expansion of a dense bulk medium~\cite{v2ptb}. The dash-dotted curve is a {\em prediction} derived assuming a L\'evy quadrupole spectrum with slope parameter $T \approx 90$ MeV and fixed source boost $\Delta y_{t0} = 0.6$ characteristic of a thin expanding shell (see Fig.~\ref{quad3}). The dotted curve is a viscous-hydro prediction for proton $v_2(p_t)$~\cite{rom} transformed to \yt\ in the same manner with factor $1/p_t$(lab). The zero intercept for centrality-averaged Lambda data (open points) is consistent with the 0-10\% data (solid dots), which agreement is in turn consistent with Fig.~\ref{quadspec} (right) for unidentified hadrons.

\subsection{What data should test hydro Monte Carlos?}

Just as there are many choices for construction of a complex hydro  Monte Carlo there are many choices for data selection, analysis methods and plot formats. A single most-central (e.g.\ 0-5\%) A-B \pt\ spectrum may be presented as in Ref.~\cite{nature} when full centrality coverage down to 95\% is accessible from available particle data and would provide much more information on centrality evolution of {\em multiple contributions} to spectrum structure. Spectra may be presented conventionally as semilog plots on linear \pt\ whereas plots on logarithmic transverse rapidity \yt~\cite{ppprd,ppquad} provide  much better visual access to the low-\pt\ region where crucial issues of  interpretation are localized. See Secs.~\ref{ppbpid1} and \ref{hydrospec} for examples.

2D angular correlations are conventionally projected onto 1D azimuth difference $\phi_\Delta$, thereby discarding much information conveyed within full 2D angular correlations on $(\eta_\Delta,\phi_\Delta)$. Given the 1D projection, instead of measuring the azimuth distribution of {\em correlated pairs} on $\phi_\Delta$ any Fourier-series amplitude $v_n$ as defined represents the square root of a pair ratio. Quantity $v_2(p_t)$ is then effectively a {\em ratio of single-particle spectra} and includes in its denominator the single-particle spectrum $\bar \rho_0(p_t)$ from which jet quenching is inferred via ratio $R_{AA}$. Yet the $v_2(p_t)$ trend on \pt\ near and above its mode (where $R_{AA}$ shows dramatic suppression) is interpreted to provide estimates for viscous-hydro parameter $\eta/s$ from which claims of ``perfect liquid'' are published~\cite{perfect}.

If the $v_2(p_t)$ numerator {\em is} isolated the associated {\em quadrupole spectrum}~\cite{quadspec,v2ptb} provides information on  a source boost distribution that hydro theory should confront, but those data conflict with the standard flow narrative for \aa\ collisions (in which Hubble expansion of a dense medium is expected). The spectrum $\bar \rho_2(p_t)$ for quadrupole-correlated hadrons is dramatically different from spectrum $\bar \rho_0(p_t)$ that describes most hadrons. Thus, $v_2(p_t)$ data invoked as the main basis for claims of QGP formation in \aa\ collisions and small asymmetric $x$-A systems actually conflict with the flow narrative.

Any conjectured hadron production mechanism should be tested against full 2D angular correlations rather than 1D projections. Based on centrality or multiplicity trends for A-B collisions nonjet and jet-related contributions can be clearly distinguished. In particular, Fourier components of the jet-related SS 2D peak, including a jet-related quadrupole component, should be distinguished from a NJ quadrupole component. Interpretation of any aspect of nonjet angular correlations as representing flows should in turn be consistent with evidence (or not) for radial flow in nonjet hadron spectrum structure. If a hydro model appears to describe data from a collision system in which radial flow has been excluded (e.g.\ by nonjet spectrum structure) the hydro interpretation is unlikely.

\subsection{The TCM as a nonhydro predictive model} \label{predict2}

The TCM describes hadron production mechanisms that could be complementary to hydro flows, if they exist, and should therefore furnish a context for any hydro model. The TCM also provides an example of a predictive model in contrast to current hydro models as summarized above.
The TCM is predictive in the following sense: it is determined by a small number of parameters that are required to describe accurately a broad array of A-B collision systems over the currently accessible range of collision energies and that relate directly to fundamental QCD physics. For instance, the \pp\ TCM reported in Ref.~\cite{alicetomspec} includes six parameters: $T$ and $n$ for the soft component, $\bar y_t$, $\sigma_{y_t}$ and $q$ for the hard component and  $\alpha$ to relate hard and soft components. $T$, $\bar y_t$ and $\sigma_{y_t}$ are independent of collision energy or vary quite slowly with $\log(\sqrt{s})$. Energy evolution of $n$ relates to the splitting cascade within projectile nucleons, evolution of $q$ relates to the underlying MB dijet spectrum~\cite{jetspec2} and evolution of $\alpha$ relates directly to MB dijet production measurements.

Examples are provided in Secs.~\ref{ppbpid1} and \ref{hydrospec}. In Sec.~\ref{ppbpid1} the pion spectrum TCM for 5 TeV \ppb\ collisions as reported in Ref.~\cite{ppbpid} is predicted based on the \pp\ TCM. The TCMs for kaons and protons are adjusted to accommodate data with the pion TCM as starting point, and also with reference to the \auau\ PID TCM reported in Ref.~\cite{hardspec} so as to achieve a self-consistent system.  In Fig.~\ref{protons} (right) the dashed curve is a TCM prediction for the proton hard component based on spectrum details at lower \pt, that is, based on the spectrum soft component. Because of the relation $\bar \rho_h \approx \alpha \bar \rho_s^2$, and because $\alpha(\sqrt{s})$ is determined by a fixed parametrization covering 17 GeV to 13 TeV, there is no flexibility in the model to accommodate the suppressed spectrum hard-component data. The proton data fall 40\% below the TCM prediction and that disagreement is currently unexplained.

In Sec.~\ref{hydrospec} a GLS reference for the pion spectrum from 2.76 TeV \pbpb\ collisions is determined by the 7 TeV \pp\ (for unidentified hadrons and pions) and 5 TeV \ppb\ (for kaons and protons) TCMs. The spectrum hard-component models are simple Gaussians without exponential tails because of the limited \pt\ range of the data, thus eliminating TCM parameter $q$. In more detail, soft component $S_{NN}$ includes parameters $T = 145$ MeV (universal for all pion and unidentified-hadron spectra) and $n = 8.8$ for 5 TeV \pp\ collisions (interpolated via energy-dependent TCM) with $\bar \rho_s = 4.55$ from Ref.~\cite{ppbpid}. Hard component $H_{NN}$ includes parameters $\bar y_t = 2.63$, $\sigma_{y_t} = 0.56$ and $\bar \rho_h = \alpha \bar \rho_s^2 = 0.012 \times 4.55^2$. In Eq.~(\ref{pbpbtcm}) the TCM model functions are multiplied by factor 0.8 representing the fraction of pions within hadrons and with $\nu \approx 6$ for central \pbpb.  There is no further adjustment of TCM parameters to accommodate the \pbpb\ data. 



\section{Hydro models $\bf vs$ selected data} \label{predict}

Given the principles set forth in Sec.~\ref{hydro} it is instructive to consider data descriptions from some prominent recent hydro models in comparison with several forms of data. Hydro models typically include a Monte Carlo Glauber~\cite{naglesonic} or Glasma-based~\cite{glasmafluc} estimate of initial conditions followed by (viscous or nonviscous) hydro evolution and in some cases a hadronic-cascade afterburner. 
  
Reference~\cite{galehydro}, a general hydro description cited as ``[17]'' in Ref.~\cite{nature}, provides a context for hydro theory. The hydro model therein consists of components IP-Glasma + Music where IP-Glasma~\cite{glasmafluc} models initial conditions and MUSIC~\cite{music} models viscous hydro evolution. The EoS is determined from ``lattice QCD results and a hadron resonance gas model.'' Kinetic freezeout occurs at $T_\text{FO} = 120$ MeV which can be compared with TCM slope parameters for pions (145 MeV), kaons (200 MeV) and protons (210 MeV) as in Ref.~\cite{ppbpid} and Sec.~\ref{radial}. 

The hydro model actually utilized in Ref.~\cite{nature} comparisons with $x$-Au $v_n\{\text{EP}\}$ data is based on the SONIC hydro model. According to Refs.~\cite{naglesonic,naglesonic2} SONIC combines  Monte-Carlo Glauber initial conditions with a 2+1 viscous hydrodynamics evolution and hadronic cascade afterburner. A recent update is super-SONIC, an event-by-event generalization of the SONIC model including pre-equilibrium flow, viscous hydrodynamics and hadronic cascade afterburner~\cite{romatsonic}. Super-SONIC is described in Ref.~\cite{romatsonic} as the ``most realistic description currently available.'' Comparisons with CMS and ATLAS \ppb\ $v_n$ data show good agreement between theory and data.  

The data descriptions provided by the two hydro models cited in Ref.~\cite{nature} are similar in their correspondence with data. For the present study specific hydro theory trends compared to data are taken from Ref.~\cite{galehydro2} (IP-Glasma + MUSIC) with its broader selection of results. 

\subsection{Hydro vs single-particle $\bf p_t$ spectra} \label{hydrospec}

\begin{figure}[h]
	\includegraphics[width=1.65in,height=1.68in]{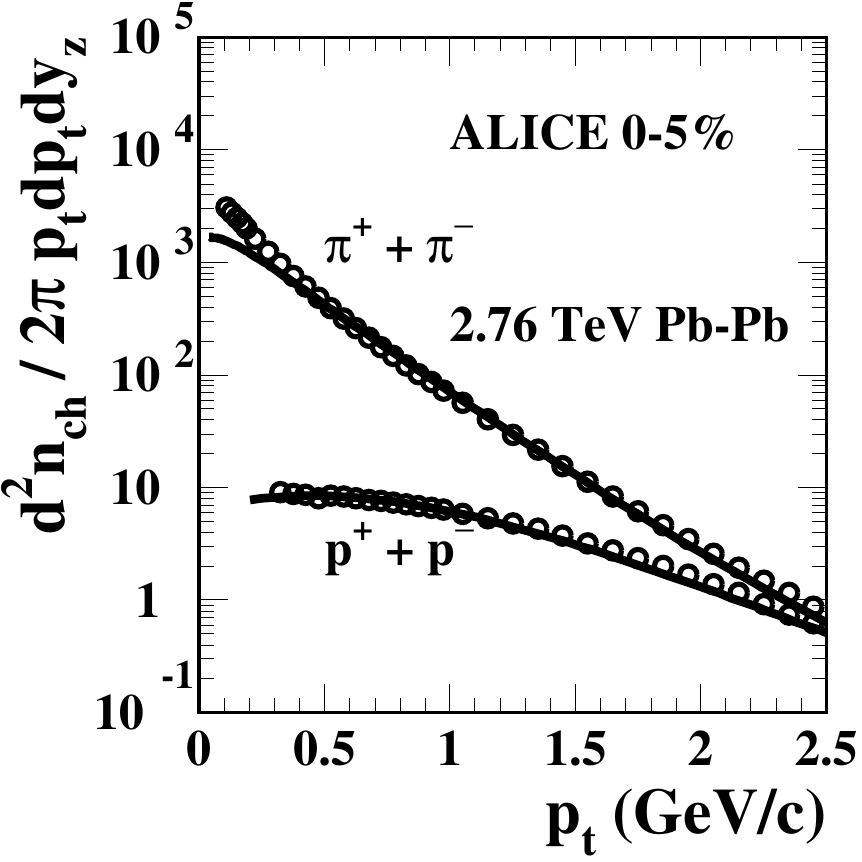}
	\includegraphics[width=1.65in,height=1.65in]{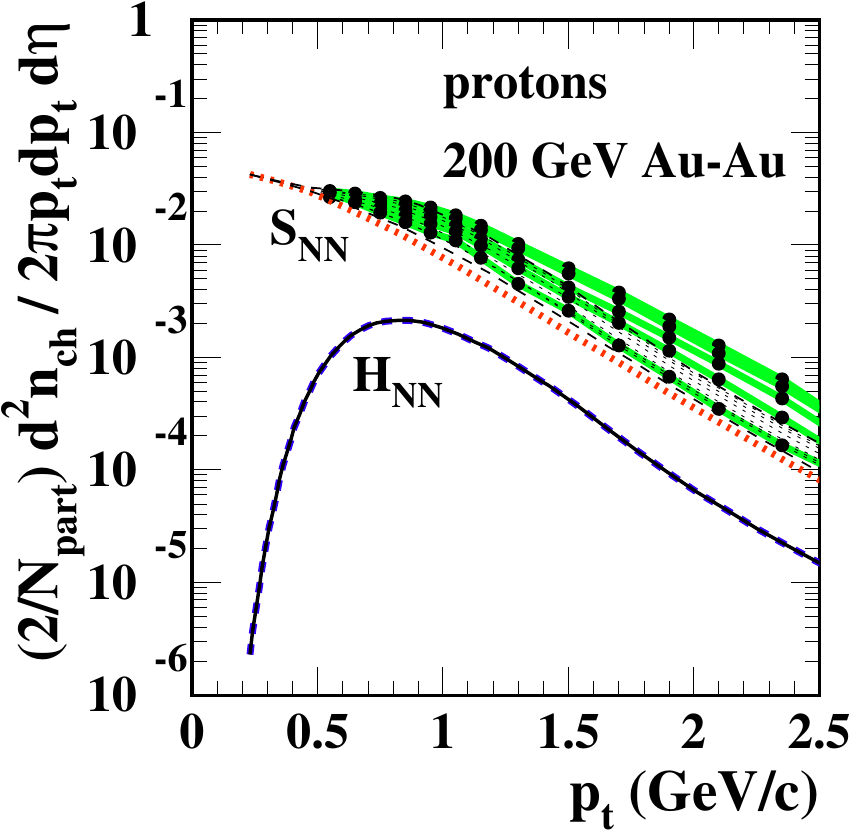}
	\caption{\label{gale1} (Color online)
		Left: Comparison of hydro theory (curves) from Ref.~\cite{galehydro2} with PID spectrum data for 0-5\% central 2.76 TeV \pbpb\ collisions as reported in Ref.~\cite{alicepid} (points).
		Right:	TCM analysis of proton spectra from five centrality classes of 200 GeV \auau\ collisions (points and green solid curves) as reported in Ref.~\cite{hardspec}. The bold dotted  and dashed curves are TCM model functions. Those spectra have been normalized by participant pair number $N_{part}/2$.
	} 
\end{figure}

Figure~\ref{gale1} (left) shows hydro theory (curves) compared to 
PID \pt\ spectra for charged pions and protons from 0-5\% central 2.76 TeV \pbpb\ collisions (points) within a conventional semilog plot vs \pt. The figure is adopted from Ref.~\cite{galehydro2} reporting the theory results. Agreement seems good in that format, especially for protons. But there is no distinction made between soft processes that might be amenable to a hydro description and substantial MB dijet contributions to PID spectra.

Figure~\ref{gale1} (right) shows proton spectra from 200 GeV \auau\ collisions to illustrate TCM analysis of PID spectrum evolution with \aa\ centrality~\cite{hardspec}. The points and solid (green) curves correspond to five \auau\ centrality classes: 0-12\%, 10-20\%, 20-40\%, 40-60\% and 60-80\%. The spectra have been rescaled by the number of participant pairs $N_{part}/2$. The full \pt\ spectra extend to 12 GeV/c (see Fig.~\ref{auauspec}) but are truncated here to match the left panel from Ref.~\cite{galehydro2}. The dotted $S_{NN}$ (soft) and dashed $H_{NN}$ (hard) model functions result from a TCM analysis of pion and proton spectra in Ref.~\cite{hardspec}. Spectrum evolution with centrality indicates that the data soft component (what scales with $N_{part}$) remains unchanged, and the data hard component (what scales with $N_{bin}$) also remains unchanged for more-peripheral collisions. For more-central collisions (above a sharp transition in jet characteristics, see Fig.~\ref{quad2}) the hard component is suppressed above 4 GeV/c (``jet quenching'') but {\em strongly enhanced} just above 1 GeV/c relative to a \nn\ linear superposition reference (thin dotted curves). Below 1 GeV/c the proton hard component {\em remains unmodified}.

This comparison suggests that the hydro description in the left panel mainly accommodates the MB dijet contribution to spectra.  In more-central \auau\ collisions proton spectra are dominated by the contribution from MB dijets scaling with $N_{bin}$, with a five-fold hard/soft ratio near 1 GeV/c for central collisions as in the right panel. Although the \pbpb\ data in the left panel correspond to higher collision energy spectrum trends are typically slowly varying, as $\log(\sqrt{s})$~\cite{jetspec2,alicetomspec}. At higher collision energies jet-related structure should be substantially greater compared to the soft component. If one examines the \pbpb\ proton spectrum on the left end near 0.5 GeV/c the upturn in data there corresponds well with the detailed spectrum structure in the right panel. 

The \pbpb\ PID data in Fig.~\ref{gale1} (left) can be analyzed in far greater detail within the context of the TCM. The spectrum TCM for A-B collisions is~\cite{hardspec,ppbpid}
\bea \label{pbpbtcm}
\bar \rho_0(p_t) ~= && \hspace{-.2in} (N_{part}/2)S_{NN}(p_t) + N_{bin}H_{NN}(p_t)~~~~
 \\ \nonumber
(2/N_{part}) \bar \rho_0(p_t) &=& \bar \rho_{sNN} \hat S_{0NN}(p_t) + \nu \bar \rho_{hNN} \hat H_{0NN}(p_t),
\eea
where $\bar \rho_0(p_t) \equiv d^2 n_{ch} / p_t dp_t d\eta$ and $\hat S_0(p_t)$ and $\hat H_0(p_t)$ are unit-normal TCM model functions~\cite{hardspec}. For spectra included in this subsection the hard-component model $\hat H_0(p_t)$ consists of a simple Gaussian (on \yt) without exponential tail because of the limited \pt\ range of the data.

\begin{figure}[h]
	\includegraphics[width=3.3in]{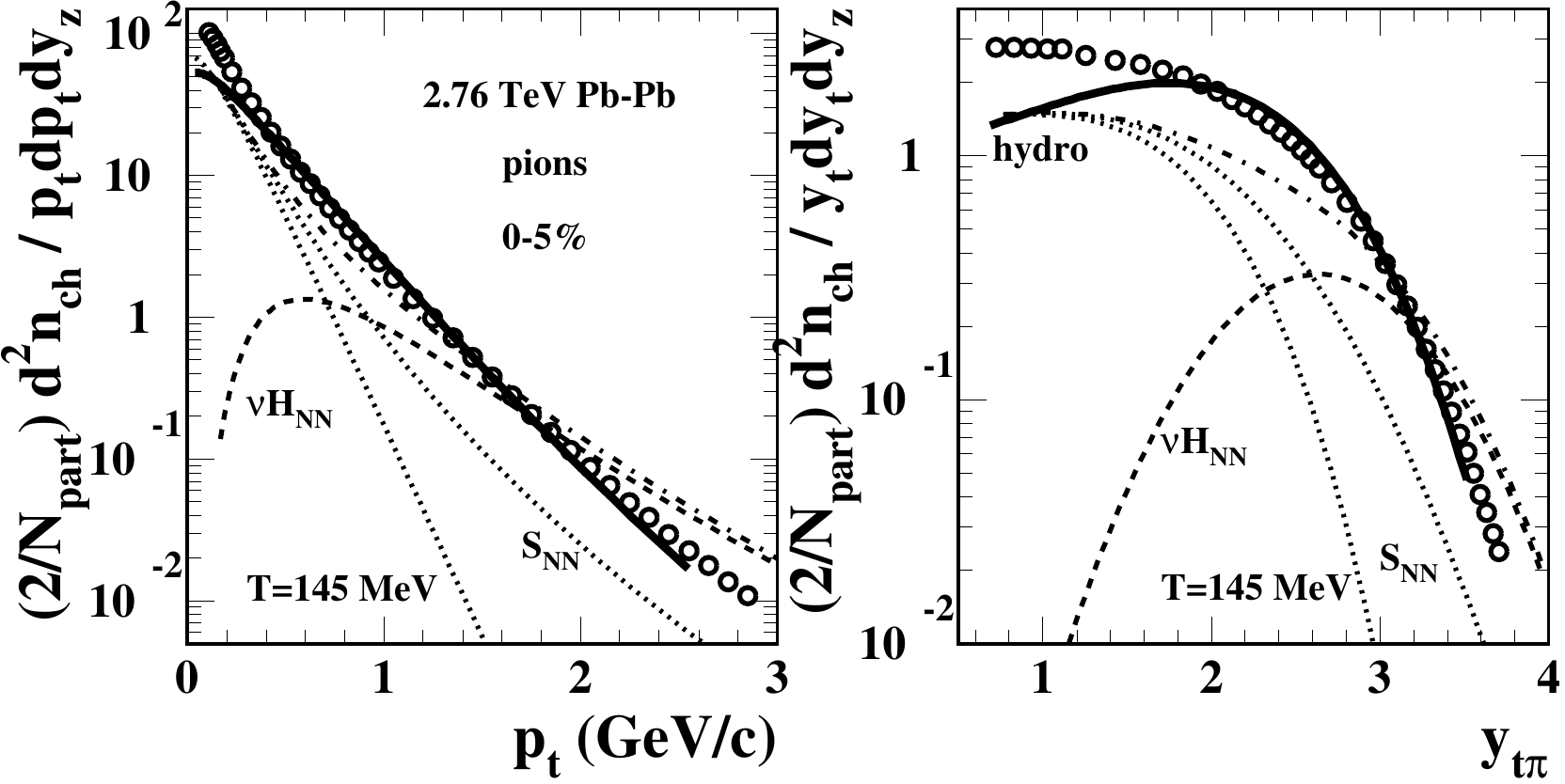}
	\caption{\label{nature1}
		Left: Hydro theory curve for pions (bold solid) from Fig.~\ref{gale1} (left) compared to the full pion spectrum from 0-5\% central \pbpb\ collisions (open points) plotted on linear \pt\ as reported in Ref.~\cite{alicepid}. The dashed and dotted curves are model functions derived from TCMs for 7 TeV \pp\ and 5 TeV \ppb\ spectrum data~\cite{alicetomspec,ppbpid} with no adjustment for this comparison.
		Right:	Curves and data in the left panel transformed to transverse rapidity \yt\ via Jacobian factor $m_{t\pi} p_t / y_{t\pi}$.
	} 
\end{figure}

Figure~\ref{nature1} (left) shows the full \pbpb\ pion spectrum from Ref.~\cite{alicepid} (points), part of which is shown in Fig.~\ref{gale1} (left) taken from Ref.~\cite{galehydro2}. The published spectrum data have been multiplied by $2\pi$ to correspond with $\eta$ densities used in this study and divided by participant pair number $N_{part}/2 \approx 187$ corresponding to 0-5\% central \pbpb\ collisions. The bold solid curve is the hydro result from Ref.~\cite{galehydro2} and Fig.~\ref{gale1} (left) terminating at $p_t = 2.5$ GeV/c. 

The other curves in Fig.~\ref{nature1} (left) are TCM soft component $S_{NN}(p_t)$ (upper dotted), hard component $H_{NN}(p_t)$ (dashed) and an exponential with slope parameter $T = 145$ MeV (lower dotted). The exponential is included in response to claims that data tend to follow an exponential trend below 3 GeV/c~\cite{hydro}. With slope parameter determined by the large fraction of (soft) hadrons lying below 0.5 GeV/c the exponential demonstrates that spectrum data are nowhere consistent with that model function. The dash-dotted curve is the sum $(2/N_{part})\bar \rho_0(p_t)$ of TCM soft and hard components as in Eq.~(\ref{pbpbtcm}) (second line). The \pbpb\ TCM model parameters are adopted from the spectrum TCM for 7 TeV \pp\ and  5 TeV \ppb\  collisions~\cite{alicetomspec,ppbpid} (e.g.\ see figures in Sec.~\ref{ppbpid1}) and therefore constitute a {\em prediction} for \pbpb\ assuming linear superposition of \nn\ collisions, i.e.\ no jet modification. There is no adjustment of TCM parameters to accommodate the \pbpb\ data. Details of the TCM parametrization are presented in Sec.~\ref{predict2}. The conventional semilog format on linear \pt\ suppresses the detailed spectrum structure at lower \pt\ in relation to model trends.

Figure~\ref{nature1} (right) shows the same data and curves transformed to pion transverse rapidity $y_{t\pi}$ via Jacobian factor $m_{t\pi} p_t / y_{t\pi}$. Compare with Fig.~\ref{pions} (left) for \ppb\ data. The detailed relations among data and model curves are more accessible within this plot format. The spectrum TCM (dash-dotted), serving as a GLS reference for data and other models, falls well below the data for smaller \yt\ and well above the data for larger \yt, a result consistent with trends for PID spectra from 200 GeV \auau\ collisions~\cite{hardspec}  (see Fig.~\ref{specfrag}). The measured pion spectrum (points) is suppressed at larger \yt\ compared to the GLS reference (as also indicated by spectrum ratio $R_{AA}$) but {\em strongly enhanced} at smaller \yt\ suggesting approximate energy conservation within modified jets~\cite{hardspec}. The same trend is shown by 200 GeV \auau\ data in Fig.~\ref{specfrag} (right) where the low-\yt\ enhancement is accurately described by a pQCD convolution integral (solid curve).  Such low-\yt\ enhancements are concealed by ratio $R_{AA}$.

The main feature of the right panel is comparison of the hydro result (bold solid) with data (points). The hydro curve falls well below the data at smaller \yt\ where the great majority of hadrons are located but is consistent with data at larger \yt\ where the spectrum is dominated by the jet-related hard component {\em including the effect of jet suppression}. The hydro trend is consistent with a large-amplitude and broad source-boost distribution (e.g.\ radial flow) as illustrated by Fig.~3 of Ref.~\cite{hydro}. The decrease at lower \yt\ is generally not observed in spectrum data, and the hydro increase at higher \yt\ ($\propto N_{bin}$) is consistent with MB dijet production. See Fig.~\ref{specfrag} (right) for a pQCD description (solid curve) of the pion hard component from 0-5\% central 200 GeV \auau\ collisions.

\begin{figure}[h]
	\includegraphics[width=3.3in]{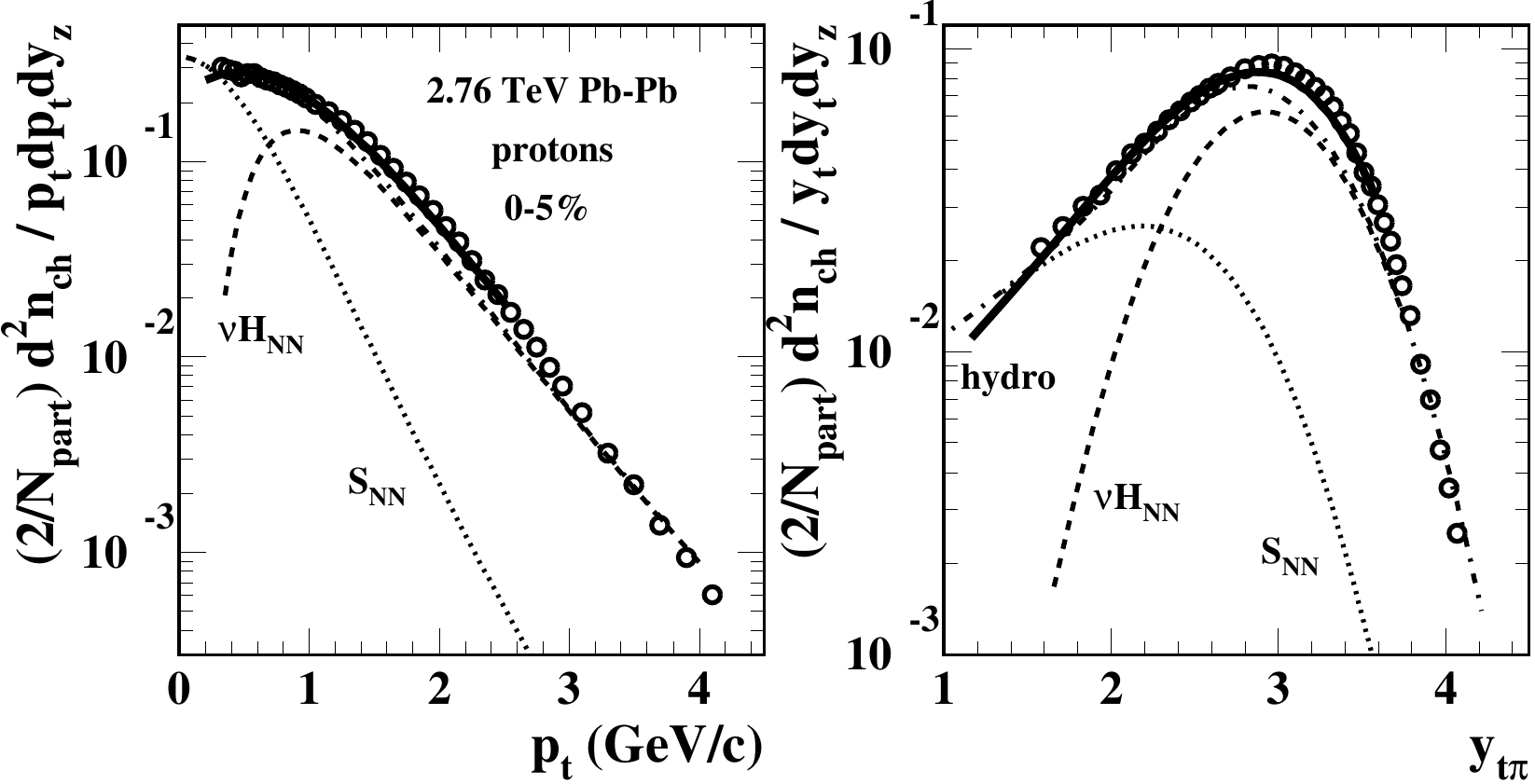}
	\caption{\label{nature2}
		Left: Hydro theory curve for protons (bold solid) from Fig.~\ref{gale1} (left) compared to the full proton spectrum from 0-5\% central \pbpb\ collisions (open points) plotted on linear \pt\ as reported in Ref.~\cite{alicepid}. The dashed and dotted curves are model functions derived from a TCM for 5 TeV \ppb\ proton spectrum data~\cite{alicetomspec,ppbpid} with no adjustment for this comparison.
Right:	Curves and data in the left panel transformed to transverse rapidity \yt\ via Jacobian factor $m_{t\pi} p_t / y_{t\pi}$.
	} 
\end{figure}

Figure~\ref{nature2} (left) shows the full 2.76 TeV \pbpb\ proton spectrum from Ref.~\cite{alicepid} (points).  The bold solid curve is the hydro result from Ref.~\cite{galehydro2} terminating at 2.5 GeV/c as in Fig.~\ref{gale1} (left). The other curves represent the proton spectrum TCM with line styles as described for Fig.~\ref{nature1}, with parameter values as discussed in Sec.~\ref{predict2}. This panel corresponding to 0-5\% central \pbpb\ can be compared with Fig.~\ref{gale1} (right) and Fig.~\ref{auauspec} (left) showing proton spectra for 200 GeV \auau\ collisions taken from Ref.~\cite{hardspec} including a range of \auau\ centralities.

Figure~\ref{nature2} (right) shows  the same data and curves transformed to pion transverse rapidity $y_{t\pi}$ via Jacobian factor $m_{t\pi} p_t / y_{t\pi}$. The choice of pion rather than proton \yt\ is motivated by the underlying MB jet spectrum as determining the properties of spectrum hard components~\cite{hardspec}. $y_{t\pi}$ simply serves as a logarithmic measure of \pt\ for any hadron species~\cite{hardspec}. In each case the hard-component mode then appears near $p_t = 1$ GeV/c or $y_{t\pi} \approx 2.7$, consistent with the MB jet spectrum having an effective lower bound near 3 GeV~\cite{jetspec2,alicetomspec}.
This panel can be compared with  Fig.~\ref{auauspec} (left) taken  from Ref.~\cite{hardspec} that shows spectrum evolution with \auau\ centrality.

Figure~\ref{nature2} (right) confirms a result from Ref.~\cite{hardspec}: For proton spectra the jet-related hard component strongly dominates the soft component (dashed vs dotted curves). Almost all protons for $p_t > 0.5$ GeV/c ($y_t > 2$) arise from jet production scaling with the number of \nn\ binary collisions. The dominance of MB dijet production is especially evident for more-central \aa\ collisions or large-\nch\ \pp\ or \pa\ collisions (e.g.\ Fig.~\ref{protons}). The effects of jet modification are more subtle for protons than for pions, but the onset of suppression (relative to GLS dash-dotted) is apparent near \yt\ = 4, and enhancement just above \yt\ = 3 is also visible. The enhancement of protons within 1-2 GeV/c (as opposed to pion enhancement well below that interval as in Fig.~\ref{nature1}) is consistent with results from 200 GeV \auau\ collisions as reported in Ref.~\cite{hardspec}.

The hydro result (bold solid) is remarkable. The curve closely follows data at higher \yt\ dominated by jet production but at lower \yt\ falls below the TCM soft component that might be associated with hydro phenomena. As noted in Sec.~\ref{radial} spectrum data at lower \pt, within an interval that includes most hadrons, do not exhibit the reduction with centrality that is expected within the hydro model as a manifestation of radial flow~\cite{hydro}. The relation of hydro descriptions to MB jet manifestations represents a major issue addressed by the present study.

\subsection{Hydro vs $\bf v_n(b)$ centrality dependence}

A conventional method for azimuth correlation analysis in the context of a flow/QGP narrative is two-particle cumulants denoted by $v_n\{2\}$ representing the $n^{th}$ Fourier amplitude of the single-particle azimuth distribution, actually derived from the pair distribution on $\phi_\Delta$~\cite{njquad,v2ptb}. In attempts to reduce ``nonflow'' (e.g.\ MB dijet contributions to azimuth correlations) cuts on the pair space $(\eta_1,\eta_2)$ may be imposed based on the assumption that jet angular correlations are localized near the origin of difference variable $\eta_\Delta$ (so-called ``short-range'' correlations, e.g.\ Fig.~\ref{soft}, left). But in more-central \aa\ collisions the relevant jet contribution (i.e.\ the SS 2D jet peak) may broaden on $\eta_\Delta$~\cite{anomalous} (e.g.\ Fig.~\ref{unique}, right) resulting in strong MB dijet contributions to $v_n\{2\}$. A hydro Monte Carlo may be applied to $v_n\{2\}$ data assuming that they are exclusively flow related when in fact they may be biased by or even dominated by MB dijet contributions. This subsection addresses hydro descriptions of the \aa\ centrality dependence of  \pt-integral $v_2(b)$ data while the next subsection addresses issues for \pt-differential $v_2(p_t)$ data.

\begin{figure}[h]
	\includegraphics[width=1.7in,height=1.65in]{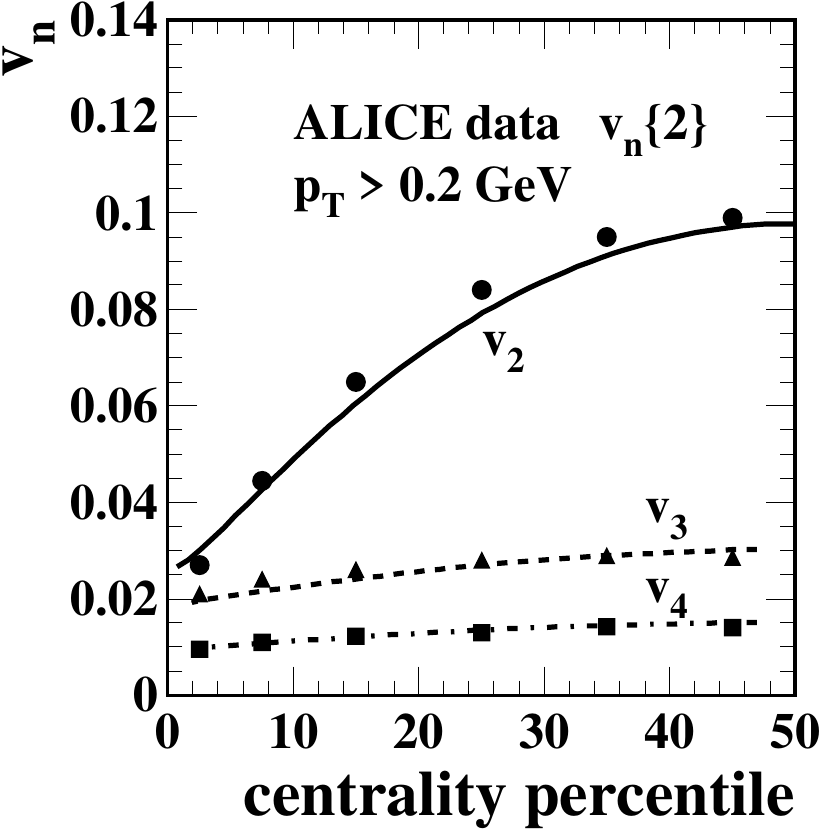}
	\includegraphics[width=1.65in,height=1.65in]{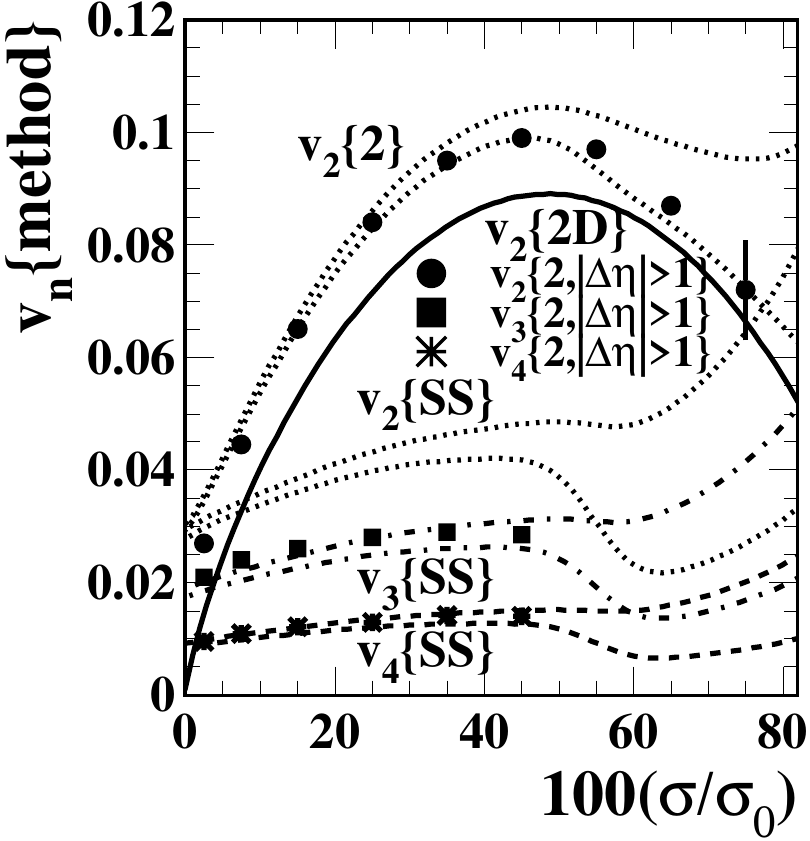}
	\caption{\label{gale2}
		Left: Hydro theory results (curves) from Ref.~\cite{galehydro2}  compared to \pbpb\ $v_n\{2\}(b)$ data as  reported in Ref.~\cite{alice} (points).
		Right: A similar comparison but with curves derived from model fits to 2D angular correlations. The data points are again from Ref.~\cite{alice}. The several curves are obtained from Ref.~\cite{multipoles} based on 200 GeV \auau\ data but with $\sigma_{\phi_\Delta}$ reduced from 0.65 to 0.60 and with an overall multiplier 1.3. The relative magnitudes and centrality variations agree closely between data and curves, suggesting that the \pbpb\ SS 2D peak at 2.76 TeV has similar properties to the \auau\ peak at 200 GeV, as also noted in Ref.~\cite{ppridge}.
	} 
\end{figure}

Figure~\ref{gale2} (left) shows hydro theory (curves) from Ref.~\cite{galehydro2} vs 
$v_n\{\text{2}\}(b)$ data (points) as reported in Ref.~\cite{alice} showing the centrality dependence for 2.76 TeV \pbpb\ collisions. The agreement seems good but no distinction is made between soft processes, what might be amenable to hydro descriptions, and MB dijet contributions.

Figure~\ref{gale2} (right)  repeats Fig.~\ref{alice} with the same ALICE data as in the left panel, except all the $v_2\{\text{2}\}$ data from Ref.~\cite{alice} are presented in this panel.  The solid curve is equivalent to the straight line in Fig.~\ref{soft2} (right). It was noted in Sec.~\ref{nonjethigher} that the ALICE data for $v_3\{2\}$ and $v_4\{2\}$ are entirely accounted for by corresponding Fourier components $v_n\{\text{SS}\}$ of the jet-related SS 2D peak appearing in Fig.~\ref{unique}, and $v_2\{2\}$ is accurately described by a combination (in quadrature) of nonjet $v_2\{\text{2D}\}$ and jet-related $v_2\{\text{SS}\}$. The $v_n\{\text{SS}\}$ all have the same slowly-varying functional form on centrality, and their relative amplitudes are consistent with the Fourier spectrum of a narrow Gaussian. The multiple Fourier components $v_n\{\text{SS}\}$ are thus equivalent to a single SS Gaussian (as in Fig.~\ref{unique}) when projected onto 1D $\phi_\Delta$. As pointed out in relation to Fig.~\ref{alice}, the dotted and dashed $v_n\{\text{SS}\}$ trends are consistent with dijet production scaling as $\propto N_{bin}$ in combination with jet modification (``jet quenching'') varying with centrality according to a sharp transition as reported in Ref.~\cite{anomalous} and illustrated in Fig.~\ref{quad2} (left).

Given the large MB dijet contribution to the $v_n\{\text{2}\}(b)$ data the close correspondence of a hydro description (curves) to $v_n$ data (points) in the left panel is questionable. It is also unexplained why more-peripheral $v_2\{\text{2}\}$ points have been omitted from that figure, and why more-peripheral points for $v_3\{\text{2}\}$ and $v_4\{\text{2}\}$ were not provided in Ref.~\cite{alice}. The availability of more-peripheral $v_n\{2\}$ data could make possible more-stringent tests of the hypothesis that $v_n\{2\}$ data for $n > 2$ are dominated by MB dijet contributions rather than hydro-related flows.

\subsection{Hydro vs $\bf v_n(p_t)$ $\bf p_t$ dependence}

\begin{figure}[h]
	\includegraphics[width=1.7in,height=1.65in]{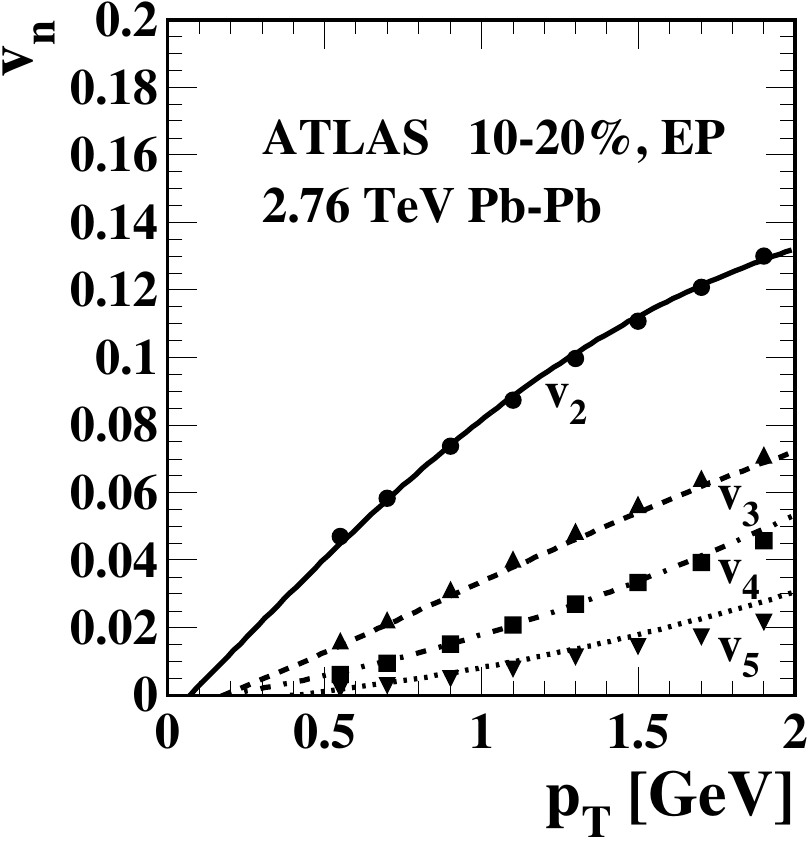}
	\includegraphics[width=1.65in,height=1.65in]{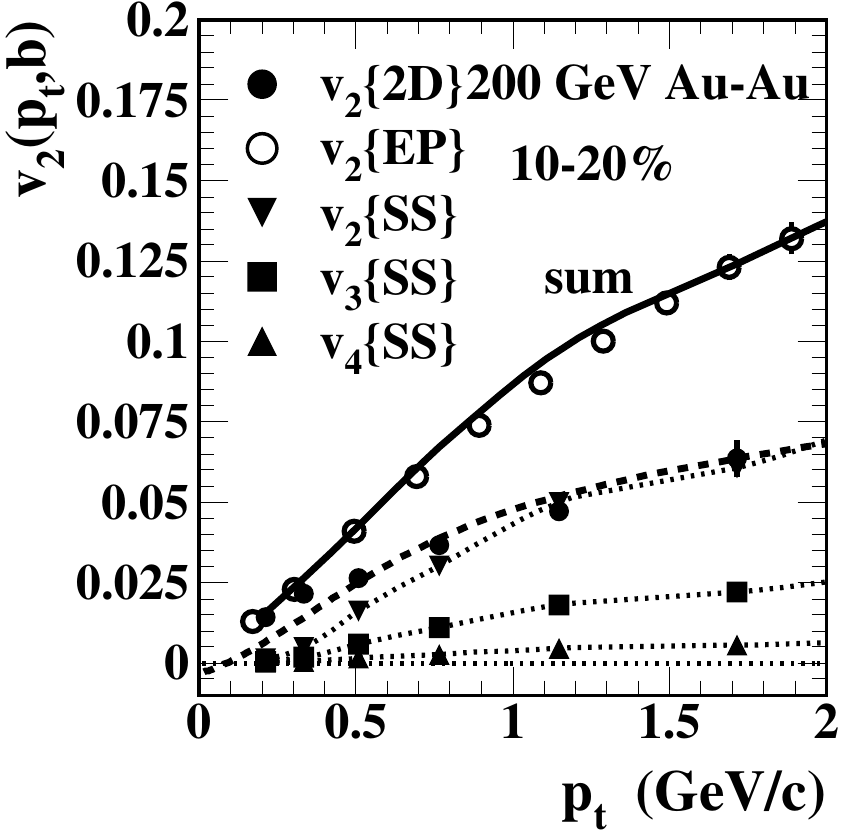}
	\caption{\label{gale3}
Left: Hydro theory results (curves) from Ref.~\cite{galehydro2} compared to $v_n\{\text{EP}\}(p_t)$ data for 10-20\% central 2.76 TeV \pbpb\ collisions (points) as reported in Ref.~\cite{atlasflow}.
Right: A similar comparison but with  $v_2\{\text{EP}\}(p_t)$ data from 10-20\% central 200 GeV \auau\ collisions as reported in Ref.~\cite{2004} compared with $v_2\{\text{2D}\}(p_t)$ and $v_2\{\text{SS}\}(p_t)$ data derived from model fits to 2D angular correlations from the same collision system as reported in Ref.~\cite{v2ptb}. The bold solid curve represents a combination of $v_2\{\text{2D}\}(p_t)$ and $v_2\{\text{SS}\}(p_t)$ data as in Eq.~(\ref{aqsum}). The dashed curve is  the solid pion curve in Fig.~\ref{quad3} (left) (divided by two) derived from the universal quadrupole spectrum in Fig.~\ref{quad3} (right) and is equivalent to model curve $Q_0(y_t)$ in  Fig.~\ref{quadspec} (right).
	} 
\end{figure}

Figure~\ref{gale3} (left) shows  hydro theory (curves) from Ref.~\cite{galehydro2} compared with $v_n\{\text{EP}\}(p_t)$ data for 10-20\% central 2.76 TeV \pbpb\ collisions from Ref.~\cite{atlasflow} (points). Again the hydro trends describe the data well but there is no distinction between jet-related azimuth structure and nonjet contributions that might be amenable to a hydro description. It is notable that the hydro curve for $v_2$ is consistent with a zero intercept near 0.09 GeV/c expected for pions (and unidentified hadrons) given a fixed boost $\Delta y_{t0} = 0.6$. One should also contrast the limited \pt\ interval for this plot with those for Figs.~\ref{quad3} (left), \ref{ytcomp1} and \ref{quadspec} where data are described accurately out to 6 GeV/c by the combination of a fixed quadrupole spectrum and jets.
 
Figure~\ref{gale3} (right) shows $v_2\{\text{EP}\}(p_t)$ data from 10-20\% central 200 GeV \auau\ collisions (open points)~\cite{v2ptb} compared to $v_2\{\text{SS}\}$ and $v_2\{\text{2D}\}$ data.  See the discussion of Fig.~\ref{ytcomp1}. Although the published \auau\ $v_2\{\text{EP}\}$ data extend over a broader \pt\ range as in Fig.~\ref{ytcomp1} the right panel matches the interval presented at  left and in Ref.~\cite{galehydro2}. Given the trend for $v_2\{\text{SS}\}$ derived from measured SS 2D peak characteristics (the points are derived from 2D model fits reported in Ref.~\cite{v2ptb}) the corresponding trends for the ``higher harmonics'' $v_3\{\text{SS}\}$ and $v_4\{\text{SS}\}$ are determined exactly by the SS 2D jet peak azimuth width and Eq.~(\ref{fm}). The dotted lines serve to guide the eye. 

The solid dots represent NJ quadrupole $v_2\{\text{2D}\}$ data inferred from the same 2D model fits. The dashed curve is the solid pion curve in Fig.~\ref{quad3} (left) derived by back-transforming the universal quadrupole spectrum in Fig.~\ref{quad3} (right). The minimum-bias result from  Fig.~\ref{quad3} is divided by 2 to correspond with 10-20\% central collisions in Fig.~\ref{gale3}. The curve has a zero intercept at 0.09 GeV/c (for pions) corresponding to fixed source boost $\Delta y_{t0} \approx 0.6$ reported in Refs.~\cite{quadspec,v2ptb}. Compare with the solid hydro theory curve at left. The good agreement between the MB trend and 10-20\% $v_2\{\text{2D}\}$ data is consistent with the result in Fig.~\ref{quadspec} (right) that the \pt\ dependence of the NJ quadrupole is independent of \auau\ centrality within data uncertainties. Note that $Q_0(y_t)$ in  Fig.~\ref{quadspec} (right) is equivalent to the dashed curve in Figure~\ref{gale3} (right).

The solid curve, combining the $v_2\{\text{2D}\}$ and $v_2\{\text{SS}\}$ data per Eq.~(\ref{aqsum}), agrees well with the published $v_2\{\text{EP}\}$ data. From the data trends within this limited \pt\ acceptance and this \auau\ centrality 2D and SS contributions have apparently similar \pt\ dependences. But the definition of $v_n$ as the square root of a pair ratio serves to confuse two distinct physical phenomena that have dramatically different centrality dependences as demonstrated in Fig.~\ref{alice}. The jet-related $v_n\{\text{SS}\}(p_t)$ in effect represent a ratio of the spectrum hard component (jets) to the sum of soft and hard components (total hadron spectrum). The hard-component shape depends strongly on \aa\ centrality per Ref.~\cite{hardspec} such that in more-central \auau\ collisions large excess at lower \pt\ and suppression at higher \pt\ appear, as shown in Fig.~\ref{specfrag}. Thus, the form of jet-related $v_n\{\text{SS}\}(p_t)$ should exhibit strong centrality dependence whereas  the form of NJ quadrupole $v_2\{\text{2D}\}(p_t)$ as measured remains independent of centrality as shown in Fig.~\ref{quadspec} (right). The validity of a candidate hydro Monte Carlo in describing $v_n(p_t)$ data should then be tested against a full range of \aa\ centralities.

Given the large MB dijet contribution to the $v_n\{\text{EP}\}(p_t) \approx v_n\{\text{2}\}(p_t)$ data demonstrated in the right panel the close correspondence of a hydro theory description (curves) to $v_n$ data (points) in the left panel is again questionable. This example also illustrates the necessity for full exploration of \pt\ and centrality or \nch\ dependence for any theory-data comparison in any collision system.

\subsection{Conclusions: hydro models vs data}

The several examples of hydro model descriptions of selected spectrum and correlation data presented in this section suggest that to a significant extent hydro models are employed to fit jet-related data features. No clear distinction is made between nonjet data features that might relate to hydro theory and jet-related features that should be expected based on accurate measurements of jet properties in a broad array of collision systems. The complexity of typical hydro models, indicated by a large number of degrees of freedom (model configurations, parameter values), is applied to data having only a few degrees of freedom. In contrast, the TCM describes several collision systems simultaneously and self-consistently with a few fixed parameters that accurately reflect the true degrees of freedom in high-energy collision systems. One can then question any role for hydro theory in descriptions of high-energy nuclear collisions~\cite{nohydro}.

\section{Summary}\label{summ}

A claim of ``QGP droplet'' formation in small collision systems recently published in the journal Nature is based on certain assumptions about correlation data, geometry simulations and hydro theory as applied to or inferred from \aa, \pa\ and $x$-Au collisions ($x$ represents protons, deuterons or helions). This article reports a review of previous analysis results that conflict with those assumptions and interpretations of certain $x$-Au data as evidence for QGP formation in small collision systems.

The supporting argument appearing in the Nature letter is summarized as follows: The initial-state geometry of each of three $x$-Au collision systems is estimated via Glauber Monte Carlo simulations and characterized by eccentricities $\epsilon_2$ (ellipticity) and $\epsilon_3$ (triangularity). The simulated geometries are used as initial conditions for hydro theory evolution of a conjectured flowing medium (QGP) to final-state azimuth distributions. The final state of each collision system is measured by \pt-differential azimuth Fourier amplitudes $v_2(p_t)$ (elliptic flow) and $v_3(p_t)$ (triangular flow). The general magnitudes of $v_2$ and $v_3$ for the three systems are found to follow the trends for $\epsilon_2$ and $\epsilon_3$, and the hydro descriptions are also found to correspond with the $v_n(p_t)$ data. It has been argued for results from \aa\ collisions that if hydro theory is able to describe $v_2(p_t)$ data the agreement is interpreted to support formation of a QGP in \aa\ collisions. Arguing by analogy, if the same phenomena and theory-data correspondence are observed in small asymmetric collision systems QGP must also appear in those systems. That argument can be questioned as follows.

\pt\ spectrum and angular-correlation data (especially $v_2$ interpreted as a measure of elliptic flow) from nucleus-nucleus (\aa) collisions actually conflict with hydro theory expectations. Centrality trends observed for jet modification (``jet quenching'') in more-central \aa\ collisions conflict with the centrality trend for nonjet azimuth quadrupole measure $v_2(b)$, inconsistent with presence of a flowing QCD medium as common to both phenomena.

A Glauber Monte Carlo simulation conventionally used to estimate initial-state collision geometries for A-B collisions is found to overestimate nucleon participant number $N_{part}$ by up to a factor 3 in \ppb\ collisions based on comparisons with ensemble-mean \mmpt\ data, thus casting doubt on application of a Glauber MC to asymmetric $x$-Au collisions as a critical element of that analysis.

Fourier amplitudes $v_n(p_t,b)$, conventionally interpreted to represent azimuthally asymmetric flow patterns in A-B collisions, are found to be strongly biased by or entirely due to minimum-bias jet contributions to angular correlations, in particular a two-dimensional peak on the angular space $(\eta,\phi)$ expected for intrajet correlations.

Differential study of transverse-momentum \pt\ spectra for identified hadrons from \pp, \pa\ and \aa\ collisions has established that radial flow, which should manifest as a systematic boost of the nonjet contribution to \pt\ spectra, is not present to any detectable degree in any collision system. Since Fourier amplitudes $v_n$ nominally represent modulations of radial flow, its absence rules out any true flow origin for measured $v_n$ which must then arise from other mechanisms.

Differential comparisons of hydro model descriptions (e.g.\ so-called blast-wave fits, hydro Monte Carlos) with \pt\ spectrum and angular-correlation data reveal that the models tend to describe data features established independently as arising from minimum-bias jet production. Such results suggest that theory models with many degrees of freedom are able to describe data with a few degrees of freedom despite no fundamental correspondence of assumed and actual underlying physical mechanisms.

The present study thus demonstrates that formation of QGP droplets in $x$-Au collision systems, argued on the basis of certain assumptions, theory calculations and  correlation data, is unlikely based on the larger context of data derived from an array of A-B collision systems. 
 

\end{document}